\documentclass[fleqn,subeqn,12pt]{article}
\usepackage{amsmath,amsfonts,epsfig,mathrsfs}

\numberwithin{equation}{section}

\def\pa{\partial} 
\def\pat{\tilde \partial} 

\def\L{\mathbb{L}}
\def\M {\mathbb{M}}

\def\T {\mathbb{T}}
\def\cT {\cal T}
\def\PT {\mathbb{PT}}
\def\cPT {\mathbb{\cal PT}}
\def\Z {\mathbb{Z}}
\def\R {\mathbb{R}}
\def\C {\mathbb{C}}
\def\CM {\mathbb{CM}}
\def\CP {\mathbb{CP}}
\def\RP {\mathbb{RP}}

\def\ti{\tilde}

\def\be{\begin{equation}} \def\eq{\begin{equation}}
\def\ee{\end{equation}} \def\eqe{\end{equation}}

\def\bea{\begin{eqnarray}} \def\eqa{\begin{eqnarray}}
\def\ena{\end{eqnarray}} \def\eea{\end{eqnarray}}
                                      \def\eqae{\end{eqnarray}}

\def\a{\alpha}
\def\b{\beta}
\def\c{\gamma}
\def\d{\delta}
\def\e{\epsilon} 
\def\f{\phi} 
\def\g{\gamma}
\def\h{\eta}
\def\i{\iota}

\def\k{\kappa} 
\def\l{\lambda}
\def\m{\mu}
\def\n{\nu}
 \def\w{\omega}
\def\p{\pi} 
\def\r{\rho} 
\def\s{\sigma} 
\def\t{\tau}

\def\F{\Phi}

\def\L{\Lambda}

\def\U{\Upsilon}

\def\del{\partial} 

\newcommand{\caO}{{\cal O}}

\newcommand{\rd}{\mathrm d}
\newcommand{\Mc}{\mathcal M}
\newcommand{\CMc}{\C\mathcal M}
\newcommand{\PTc}{P\mathcal T}
\newcommand{\Tc}{\mathcal T}
\newcommand{\dbar}{\bar\partial}

\ifx\ltimes\undefined
\newcommand{\ltimes}{{\kern3pt\hbox{\vrule width 0.4pt height 5.30pt
depth .0pt}\kern-1.76pt\times\kern1pt}} \fi

\begin{document}

\begin{titlepage}

\begin{flushright}
hep-th/yymmnnn\\
  Imperial/TP/06/CH/06 \\
  \end{flushright}
  \vskip 0.8cm

\begin{center}

{\bf \large Einstein Supergravity and New Twistor String Theories}

\vspace{.5cm}

Mohab Abou-Zeid\footnote{mohab.abouzeid@cern.ch}, Christopher M.\
Hull\footnote{c.hull@imperial.ac.uk} and Lionel J.\
Mason\footnote{lmason@maths.ox.ac.uk}

\vspace{.5cm} $^1${\small \em Theoretische Natuurkunde, Vrije
Universiteit Brussel \& The International Solvay
Institutes,\\Pleinlaan 2, 1050 Brussels, Belgium}

\vspace{.5cm} $^2${\small \em Theoretical Physics Group, The
Blackett Laboratory, Imperial College London,\\ Prince Consort
Road, London SW7 2BW, United Kingdom}

\vspace{.5cm} {\small \em $^2$The Institute for Mathematical
Sciences, Imperial College London,\\53 Prince's Gate, London SW7
2PG, United Kingdom}

\vspace{.5cm} $^3${\small \em The Mathematical Institute,
University of Oxford,\\ 24-29 St Giles, Oxford OX1 3LB, United
Kingdom}

\vspace{1cm}

\begin{abstract}
A family of new twistor string theories is constructed and shown
to be free from world-sheet anomalies. The spectra in space-time
are calculated and shown to give   Einstein supergravities with
second order field equations   instead of the  higher derivative
conformal supergravities that arose from earlier twistor strings.
The theories include one with the spectrum of $N=8$ supergravity,
another with the spectrum of $N=4$ supergravity coupled to $N=4$
super-Yang-Mills, and a family with $N\ge 0$ supersymmetries with
the  spectra   of self-dual supergravity coupled to self-dual
super-Yang-Mills. The  non-supersymmetric string with $N=0$ gives
self-dual gravity coupled to self-dual Yang-Mills and a scalar. A
three-graviton amplitude is calculated for the $N=8$ and $N=4$
theories and  shown to give  a result consistent with the cubic
interaction of Einstein supergravity.
\end{abstract}

\end{center}
\vspace{1cm}

\end{titlepage}

\section{Introduction}

\label{intro}

The string theories in twistor space proposed by Witten and by
Berkovits ~\cite{Witten2003,NB,BM} give a formulation of $N=4$
supersymmetric Yang-Mills theory coupled to conformal
supergravity. They provide an elegant derivation of a number of
remarkable properties exhibited by the scattering amplitudes of
these theories, giving important results for super-Yang-Mills tree
amplitudes in particular~\cite{PT,BG}.  However, in these theories
the conformal supergravity is inextricably mixed in with the gauge
theory so that, in computations of gauge theory loop amplitudes,
conformal supergravity modes propagate on internal
lines~\cite{BWsc}. There appears to be no decoupling limit giving
pure super-Yang-Mills amplitudes, and although there has been
considerable progress in studying the twistor-space Yang-Mills
amplitudes at loops (see e.~g.~\cite{CachSv} and references
therein), the results do not follow from the known twistor
strings. A twistor string that gave Einstein supergravity coupled
to super-Yang-Mills would be much more useful, and might be
expected to  have a limit in which the gravity could be decoupled
to give pure gauge theory amplitudes. (By Einstein supergravity,
we mean a supergravity with 2nd order field equations for the
graviton, in contrast to conformal supergravity which has 4th
order field equations.) Indeed, it is known that MHV amplitudes
for Einstein (super) gravity~\cite{BGK} have an elegant
formulation in twistor space~\cite{Witten2003,Gio,Nair2,Bjeretal},
and it is natural to ask whether these can have a twistor string
origin. In this paper, we propose  new twistor string models which
give Einstein (super) gravity coupled to Yang-Mills.

The new theories are constructed by gauging certain symmetries of
the Berkovits twistor string.  The structure of the theory is very
similar to that of the Berkovits model, but the gauging adds new
terms to the BRST operator so that the vertex operators have new
constraints and gauge invariances.  In this paper we construct a
family of theories for which the world-sheet anomalies cancel, and
find their spectra. We postpone a detailed discussion of the
interactions and scattering amplitudes to a subsequent paper, but
do show that there is a non-trivial cubic graviton interaction for
two of the theories, so that at least these theories are
non-trivial.  The theories of ~\cite{Witten2003,NB,BM} give target
space theories that are anomalous in general, with the anomalies
canceling only for 4-dimensional gauge groups.  It is to be
expected that these anomalies should arise from inconsistencies in
the corresponding twistor string model, but the mechanism for this
is as yet unknown~\cite{BWsc}.  If there are such inconsistencies
in the Berkovits twistor string that only cancel in special cases,
there should be similar problems for our theories, and this may
rule out some of the models we construct, or restrict the choice
of gauge group.

We find two classes of anomaly-free theories.  The first is
formulated in $N=4$ super-twistor space. Gauging a symmetry of the
string theory generated by one bosonic and four fermionic currents
gives a theory with the spectrum of $N=4$ Einstein supergravity
coupled to $N=4$ super-Yang-Mills with arbitrary gauge group,
while gauging a single bosonic current gives a theory with the
spectrum of $N=8$ Einstein supergravity, provided the number of
$N=4$ vector multiplets is six. In the Yang-Mills sector, the
string theory is identical to that of Berkovits, so that it gives
the same tree level Yang-Mills amplitudes. Both theories have the
MHV 3-graviton interaction (with two positive helicity gravitons
and one negative helicity one) of Einstein gravity.

The gauging introduces new ghost sectors into our twistor string
theories, and in the second family of string theories, gauging
different numbers of bosonic and fermionic symmetries allows
anomalies to be cancelled against ghost contributions for strings
in twistor spaces with $3$ complex bosonic dimensions and any
number $N$ of complex fermionic dimensions, corresponding to
theories in four-dimensional space-time with $N$ supersymmetries.
We then find the spectrum of states arising from ghost-independent
vertex operators.  For $N=0$, we find a theory with the bosonic
spectrum of self-dual gravity together with self-dual Yang-Mills
and a scalar, and for $N<4$ we find supersymmetric versions of
this self-dual theory. As twistor theory has been particularly
successful in formulating self-dual gravity~\cite{Penrose1} and
self-dual Yang-Mills~\cite{Ward}, it seems fitting that these
theories should emerge from twistor string theory.  With $N=4$, we
find a theory whose spectrum is that of $N=4$ Einstein
supergravity coupled to $N=4$ super-Yang-Mills with arbitrary
gauge group. It is intriguing that some of the theories we find
have similar structure to ${\cal N}=2$ string theories~\cite{OV}.

One of the  achievements of twistor theory was to give a general
solution of the self-dual and conformally self-dual Einstein
equations. Penrose's non-linear graviton
construction~\cite{Penrose1} provides an equivalence between
4-dimensional space-times $\cal M$ with self-dual Weyl curvature
and certain complex 3-folds, the curved projective twistor spaces
$\PTc$, providing an implicit construction of general conformally
self-dual space-times. For flat space-time, the corresponding
twistor space $\PT$ is $\CP^3$. In Euclidean signature, there is
an elegant realisation of the twistor space $\PTc$ corresponding
to a space $\cal M$ with signature $++++$ as the projective primed
spin-bundle over $\cal M$, the bundle of primed spinors $\pi_{A'}$
on $\cal M$ identified under complex scalings $\pi_{A'}\sim t
\pi_{A'}$, so that it is a $\CP^1$ bundle over $\cal
M$~\cite{AHS}. For other signatures, the construction of  curved
twistor space $\PTc$ is not quite so straightforward, and will be
reviewed in section \ref{nonlineargrav}.

New twistor spaces, and hence new conformally self-dual
space-times, can be constructed by deforming the complex structure
of a suitable region of a given twistor space $\PTc_0$ (such as a
neighbourhood $\PT_0$ of a projective line in $\CP^3$).  The
complex structure of a space can be specified by a (1,1) tensor
field $J$ satisfying $J^2=-1$ that is integrable, so that the
Nijenhuis tensor $N(J)$ vanishes. Given the complex structure
$J_0$ of $\PT_0$, one can construct a new complex structure
\begin{equation}
J=J_0+ \lambda J_1 + \lambda ^2 J_2 +\dots
\end{equation}
as a power series in a parameter $\lambda$, imposing the
conditions $J^2=-1$ and $N(J)=0$. In holomorphic coordinates for
$J_0$, $J^2=-1$ implies that $J_1$ decomposes into a section $j$ of
$\L ^{(0,1)}\otimes T^{(1,0)}$ and its complex conjugate on
$\PTc_0$. The linearised condition $N(J)=0$ is equivalent to
$\dbar j=0$.  Furthermore, $j$ represents an infinitesimal
diffeomorphism if $j=\dbar \alpha$ for some section $\alpha$ of
$T^{(1,0)}$. Thus a deformation corresponds to an element of the
first Dolbeault cohomology group on twistor space with values in
the holomorphic tangent bundle. Moreover, the linearised
deformations $J_1$ are unobstructed to all orders and determine
the tangent space to the moduli space of complex structures if
certain second cohomology groups vanish, which they do when
$\PT_0$ is a small enough neighbourhood of a line.

Witten's twistor string~\cite{Witten2003} is a topological string
theory on (super-)twistor space and has physical states
corresponding to deformations of the complex structure of the
target space $\PTc_0$. The corresponding vertex operator
constructed from $J_1$ is physical precisely when $j$ represents
an element of $H^1_{\bar
\partial} (\PT_0)$.  The twistor space string field theory action for
Witten's theory has a term with a Lagrange multiplier imposing
$N(J)=0$~\cite{BWsc} and the corresponding term in the space-time
action is
\begin{equation}
\label{sdW} \int d^4 x \sqrt g U^{ABCD}W_{ABCD} ,
\end{equation}
where $W_{ABCD}$ is the anti-self-dual part of the Weyl tensor. If
this were the complete gravity action, then $U^{ABCD}$ would be a
Lagrange multiplier imposing the vanishing of $W_{ABCD}$, so that
the Weyl tensor would be self-dual. However, in addition there is
a term $\int U^2$, which arises from D-instantons in Witten's
topological B-model~\cite{BWsc, Mason3}. Integrating out $U$ gives
the conformal gravity action $\int W^2$.

In split $++--$ space-time signature, there is a three real
dimensional submanifold $\PTc_\R$ of complex twistor space $\PTc$.
In the flat case, $\PT_\R\subset\PT$ is the standard embedding of
$\RP^3\subset \CP^3$, and the information about deformations of
the complex structure is encoded in an analytic vector field $f$
on $\PTc_\R$.
       It was shown in
       \cite{LeBrunMason} that conformally self-dual space-times in split
       signature can also be constructed
by deforming the embedding of  $\PT_\R$ to some $\PTc_\R$ in $\PT$
instead of
        deforming the complex structure of some region in $\PT$ to give $\PTc$.
       The deformations of the
anti-self-dual conformal structure   correspond to deformations of
the embedding of $\PTc_\R$ in $\CP^3$ and are determined at first
order by a vector field $f$ on $\PTc_\R$, or more precisely by  a
section of the normal bundle to $\PTc_\R\subset\CP^3$.

Berkovits' twistor string~\cite{NB,BM} has open strings with
boundaries on the   real twistor space  $\PT_\R$, and (conformal)
supergravity physical states are created by  an open string vertex
operator constructed from a vector field $f$ defined on $\PT_\R$,
corresponding to deformations of the embedding of $\PT_\R$ in
$\PT$.

There is an important variant of the Penrose construction that
applies to the Ricci-flat case (in fact, this is the original
non-linear graviton construction). A special case of the
conformally self-dual spaces are those that are Ricci-flat, so
that the full Riemann tensor is self-dual. The corresponding
twistor spaces $\PTc$ then have extra structure, as will be
discussed in section 3. In particular, they have a fibration
$\PTc\rightarrow\CP^1$. The holomorphic one-form on $\CP^1$ pulls
back to give a holomorphic one-form on $\PTc$ which takes the form
$I_{\a\b}Z^\a dZ^\b$ in homogeneous coordinates $Z^\a$, for some
$I_{\a\b}(Z)=-I_{\b\a}$ (Z) (which are the components of a closed
2-form on the non-projective twistor space $\cT$). The dual
bi-vector $I^{\alpha\beta}=\frac 12 \e^{\a\b\g\d}I_{\g\d}$ defines
a Poisson structure  and is called the {\it infinity twistor}.

Consider for example flat space-time ${\cal M}=\R^4$ in signature
$++++$, which has conformal compactification $S^4$. The twistor
space is $\CP^3$, which is a $\CP^1$ bundle over $S^4$: it is the
projective primed spin bundle over the conformal compactification
of ${\cal M}$. If conformal invariance is broken, then there is a
distinguished point at infinity. Removing the point at infinity
from $S^4$ to leave $\R^4$ amounts to  removing the fibre over
this point in the twistor space, leaving $\PT '= \CP^3- \CP^1$,
the projective primed spin bundle over $\R^4$. However, $\PT '$ is
also a bundle over $\CP^1$ with fibres $\C^2$, the planes through
the missing $\CP^1$. A projective line joining two points
$X^\alpha$ and $Y^\beta$ in twistor space can be represented by a
bivector $X^{[\alpha}Y^{\beta]}$, and the infinity twistor is the
bivector corresponding to the projective line over the point at
infinity in $S^4$. Choosing a point at infinity, or an infinity
twistor, breaks the conformal group down to the Poincar\'e group.
For Minkowski space,  the infinity twistor determines the
light-cone at infinity in the conformal compactification. A
similar situation obtains more generally: the infinity twistor
breaks conformal invariance.

Self-dual space-times are obtained by seeking deformations of the
complex structure of twistor space as before, but now
Ricci-flatness in space-time places further restrictions on the
deformations allowed. In the split signature picture, the vector
field $f$ on $\RP^3$ is required to be a Hamiltonian vector field
with respect to the infinity twistor, so that in homogeneous
coordinates we can write
\begin{equation}
\label{trans1} f^\a = I^{\alpha\beta} \frac{\partial h} {\partial
Z^\b}
\end{equation}
for some function $h$ of homogeneity degree 2 on $\RP^3$. In the
linearised theory, such a function $h$ corresponds to a
positive-helicity graviton in space-time via the Penrose
transform, and the non-linear graviton construction gives the
generalisation of this to the non-linear theory. In the Dolbeault
picture, the tensor $J_1$ is given by a $(0,1)$-form  $j^\a$ of
the form
\begin{equation}
\label{einsteindrep} j ^\a= I^{\alpha\beta} \frac{\partial
h}{\partial
          Z^\b} \end{equation}
where   $h$ is a $(0,1)$-form  representing an element of
$H^1(\PT',\caO(2))$.

This suggests seeking a twistor string that is a modification of
either the Berkovits or the Witten string theories  which
introduces explicit dependence on the infinity twistor, such that
there are extra constraints on the vertex operators imposing that
the deformation of the complex structure be of the
form~(\ref{trans1}) or~(\ref{einsteindrep}). Then the leading term
in the action analogous to~(\ref{sdW}) should
        have a multiplier imposing self-duality, not just conformal
self-duality, and further terms quadratic in the multiplier (from
instantons in Witten's approach) could then give Einstein gravity.
A formulation of Einstein gravity of just this form was discussed
in~\cite{AH1}.

We will present such a modification of the Berkovits twistor
string here. The key ingredient is that the one-form corresponding
to the infinity twistor is used to construct a current, and the
corresponding symmetry is gauged. The resulting gauge-fixed theory
is given by the Berkovits twistor string theory plus some extra
ghosts, and there are extra terms in the BRST operator involving
these ghosts. The dynamics and vertex operators are of the same
form as for the Berkovits twistor string, but the extra terms in
the BRST charge give extra constraints and gauge invariances for
the vertex operators, including the constraint~(\ref{trans1}) that
takes us from conformal gravity to Einstein gravity. Variants of
the theory are obtained by also gauging some fermionic currents.
The case of $N=4$ is particularly interesting as in that case the
spectrum is parity invariant and is that of  $N=4$ Einstein
supergravity (together with $N=4$ Yang-Mills).  We expect that
similar refinements of Witten's
       twistor string should also be possible.

A key difference between our models and the twistor strings of
refs.~\cite{Witten2003,NB,BM} is that space-time conformal
invariance is broken. The magnitude of the infinity twistor
defines a length scale in space-time, and so determines the
gravitational coupling $\kappa$. The theory has two independent
coupling constants: the gravitational coupling $\kappa$, determined by
the magnitude of the infinity twistor, and the Yang-Mills coupling
$g_{YM}$, arising as in~\cite{BWsc}.  Then for the $N=4$ theory
there is a limit in which $\kappa \rightarrow 0$ and supergravity
decouples from the super-Yang-Mills, so that, if the twistor
string theory is consistent at loops, it will have a decoupling
limit that gives $N=4$ super-Yang-Mills loop amplitudes.

The plan of the paper is as follows.  In
section~\ref{twistorspace}, relevant aspects of twistor theory are
reviewed, including special features of different space-time
signatures, super-twistor space, the Penrose transform and the
infinity twistor.  In section~\ref{nonlineargrav}, the non-linear
graviton construction of Penrose is reviewed, and its
generalisations to bosonic spaces of split signature and to
super-twistor spaces are given. In particular, we adapt
\cite{LeBrunMason} to the Ricci-flat case. In
section~\ref{Berkovitsstring}, the Berkovits twistor string theory
is reviewed. In section~\ref{betagamma}, the gauging of symmetries
of so-called beta-gamma systems is studied.  In
section~\ref{GaugedBerkovits}, this analysis is applied to the
Berkovits twistor string, gauging various symmetry groups of the
theory and calculating the world-sheet anomalies. In
section~\ref{Anomalies}, the conditions for anomaly cancellation
are solved, and a number of anomaly-free bosonic and
supersymmetric models is found. The spectra of these models are
found in section~\ref{spectra}, where they are compared to known
(super)gravity theories.  In section~\ref{3pointN4}, we give a
sample calculation of a nontrivial three point function in the
theory giving $N=4$ supergravity coupled to $N=4$
super-Yang-Mills. Finally, in section~\ref{Discuss} we discuss our
results and the space-time theories that might emerge from our
twistor strings.

Our conventions are those of Penrose, see for example
\cite{HuggTod}, apart from our choice of sign of the helicity,
which  is opposite to that of Penrose.


\section{Twistor space and the infinity twistor}

\label{twistorspace}

\subsection{Twistor space for flat complex space-time}

\label{flatspace}

We start by considering complexified flat space-time $\C^4$, and
postpone the discussion of the real slices giving space-times of
signature $(4,0)$, $(3,1)$ or $(2,2)$ to the next subsection.  The
twistor space $\T$ corresponding to flat complex space-time is
also $\C^4$, with coordinates $Z^\a$, $\a=0,1,2,3$.  We also use
$Z^\a$ as homogeneous coordinates on projective twistor space
$\PT=\CP ^3$, which is obtained by identifying $Z^\alpha
\sim\lambda Z^\a$ for complex $\l \ne 0$.  The $Z^\a$ transform as
a {\bf 4} under the complexified conformal group\footnote{Strictly
speaking, the
       complexified conformal group is $PGL(4,\C)=SL(4,\C)/\mathbb{Z}_4$,
       as the centre $\Z_4$ acts trivially, but this $\Z_4$ will not
       play a role in this paper.}  $SL(4,\C)$ and decompose into
two-component spinors under the complexified Lorentz group
$SL(2,\C)\times SL(2,\C)$:
\begin{equation*}
Z^\a=(\omega^A,\pi_{A'})\, ,
\end{equation*}
where $A=0,1$  and $A'=0',1'$ are spinor indices for the two
$SL(2,\C)$ factors.
        Spinor indices are raised and lowered with
$\epsilon_{AB}=\epsilon_{[AB]}$, $\epsilon_{01}=1$, and its dual
and primed counterparts.

Complex flat space-time $\CM$ is $\C^4$ with complex coordinates
$x^{AA'}$ and complex-valued metric
\begin{equation}
\label{metrics} ds ^2 = \epsilon_{AB}
\epsilon_{A'B'}dx^{AA'}dx^{BB'} .
\end{equation}
A point $ x^{AA'}$ in
         $\CM$ corresponds to a two dimensional linear subspace of $\T$
         given by the incidence relation
\begin{equation}\label{incidence}
\omega^A= x^{AA'}\pi_{A'}\, .
\end{equation}
In the projective twistor space $\PT$, these two-dimensional
subspaces determine projective lines (i.e. $\CP^1$'s), so that
each point  $ x^{AA'}$ in
         $\CM$ corresponds to a   $\CP^1$ in $\PT$.

However, some  two-dimensional subspaces in $ \T$ cannot  be
expressed in this way, and these correspond to \lq points at
infinity' in the  conformal compactification $\widetilde \CM$ of
$\CM$. The conformal compactification  is obtained by adding a
light cone at infinity $\mathscr{I}$ to $\CM$ ~\cite{HuggTod}. The
vertex $i$ of the lightcone $\mathscr{I}$ at infinity corresponds
to the subspace $\pi_{A'}=0$, and other points of $\mathscr{I}$
correspond to two-dimensional subspaces lying in the three-spaces
$\alpha^{A'}\pi_{A'}=0$ in which one linear combination of the two
components of $\p$ vanishes. There is then a one-to-one
correspondence between points in compactified space-time
$\widetilde \CM$
        and  two dimensional linear subspaces of $\T$, or projective lines in $\CP^3$.

A  two dimensional linear subspace of $\T$ is determined by two
vectors  $X^\a, Y^\a$ that lie in it, or equivalently by a simple
bi-vector, that is a bi-vector $P^{\a\b}=- P^{\b\a}$ satisfying
the simplicity condition
\begin{equation}
P^{[\alpha\beta}P^{\gamma\delta]}=0
\end{equation}
which implies $P^{\a\b}=X^{[\a}Y^{\b]}$ for some $X,Y$. Then a
point in compactified space-time corresponds to the linear
subspace in $\T$  determined  by a simple bi-vector $P^{\a\b}$.
        As $P^{\a\b}$ and $\l P^{\a\b}$  ($\l \ne 0$) determine the same linear space, we are only interested in
equivalence classes under scaling, so that the 6-dimensional space
of bivectors $P^{\a\b}$ is reduced to the space $\CP ^5 $ of
scaling equivalence classes, and the simplicity condition selects
a quadric in $\CP ^5 $. In this way,  the conformal
compactification $\widetilde \CM$ is represented as a complex
4-quadric in $\CP^5$ ~\cite{HuggTod}. Instead of using a simple
bi-vector, one can equivalently use the
         simple 2-form $P_{\a\b}=\frac 12 \epsilon_{\a\b\c\d}P^{\c\d}$ in $\T$ (where a simple 2-form is one satisfying
         $P_{[\alpha\beta}P_{\gamma\delta]}=0$).

A point  $Z^\alpha$ in twistor  space corresponds to an
`$\alpha$-plane' in $\CM$, which is a    totally null self-dual
2-plane. This can be seen by regarding the incidence relation
(\ref{incidence}) as a condition on $x^{AA'}$ for fixed $Z^\a$,
the general solution of which is
$x^{AA'}=x_0^{AA'}+\lambda^A\pi^{A'}$; this describes a 2-plane
        parametrised by
$\lambda^A$. The two-form orthogonal to the two-plane is given by
the symmetric bi-spinor $\pi_{A'}\pi_{B'}$, and is null and
self-dual. In this way, the twistor space $\PT$ can be defined as
the space of $\a$-planes in $\CM$, and this formulation is useful
as it generalises to curved space-times.

A standard tool for studying twistor correspondences is the double
fibration of the  bundle of primed spinors $\mathbb{S}$ over
space-time and over twistor space
\begin{equation}\label{doublefibration}
\begin{array}{rlcrl}
&&\mathbb S&&\cr &q\swarrow&&\searrow r&\cr \CM&&&&\T
\end{array}
\end{equation}
Using coordinates $(x,\pi_{A'})$ on the spin bundle, $q$ is the
projection $q(x^{AA'},\pi_{B'})=x^{AA'}$, whose fibre at $x^{AA'}$
is the spin space at $x^{AA'}$. The other projection $r$ takes
$(x^{AA'},\pi_{A'})\in \mathbb{S}$ to the point
$(\w^{A'},\pi_{B'})=(x^{AA'}\pi_{A'},\pi_{B'})\in \T$. The fibre
at
    $Z^\a=(x^{AA'}\pi_{A'},\pi_{B'})$ is the set of all $(x,\pi_{A'})\in \mathbb{S}$
with $Z^\a=(x^{AA'}\pi_{A'},\pi_{B'})$, which is the 2-surface
$(x^{AA'}+\lambda^A\pi^{A'},\pi_{A'})$ parameterised by
$\lambda^A$; this surface
    is the lift to the spin bundle of the $\alpha$-plane corresponding to
$Z^\a$ with tangent spinor $\pi_{A'}$.  There is clearly a
corresponding double fibration of the projective spin bundle
$\mathbb{PS}$, but now over projective twistor space $\PT$. The
Penrose transform can be understood in terms of this double
fibration as   pulling back objects from twistor space using $r^*$
and then pushing them down to space-time using $q_*$.

The space $\T$ has various canonical structures. The space $\T-0$
has a natural fibration over  $\PT$ and the
     Euler
homogeneity operator
\begin{equation}
\Upsilon = Z^\a \frac{\del}{\del Z^\a} \label{Euler}
\end{equation}
is a vector field which points up the fibres of the line bundle
$\{ \T-0\} \rightarrow \PT$. We will represent objects on $\PT$ by
their pull-backs to $\T$. Thus functions on $\PT$ are given by
functions on $\T$ that are annihilated by $\Upsilon$. The line
bundle $\caO(n)$ over $\PT$ has sections that are
         functions on   $\T$ that are homogeneous of degree $n$, i.~e.\
$\Upsilon f=nf$. Similarly, a form $\alpha$ on $\PT$ with values
in $\caO(n)$ pulls back to a form on $\T$ (which we will also
denote by $\alpha$) satisfying
\begin{equation}
\i (\Upsilon)\alpha=\i (\bar\Upsilon) \alpha=0, \hspace{1cm}
\mathcal L_{\bar \Upsilon}\alpha=0, \hspace{1cm} \mathcal
L_\Upsilon\alpha=n\alpha ,
\end{equation}
where $\i (\Upsilon)$ denotes the interior product (i.~e.\
contraction) with $\U$. We will denote the space of  $p$-forms on
$\PT$ with values in $\caO(n)$ as $\L ^p(n)$.

We define the 3-form
\begin{equation}
\Omega=\frac16 \epsilon_{\a\b\c\d}Z^\a \rd Z^\b \wedge \rd Z^\c
\wedge \rd Z^\d\, , \qquad
\epsilon_{\a\b\c\d}=\epsilon_{[\a\b\c\d]} \, , \qquad
\epsilon_{0123}=1\, . \label{defmeas}
\end{equation}
This annihilates $\Upsilon$ (i.e. $\i (\Upsilon) \Omega=0$),
        but it does not descend to $\PT$, since
it has homogeneity degree 4. However, it does so descend when
multiplied by functions that are of homogeneity degree $-4$, and
gives an isomorphism $\L^{(3,0)}(\PT)\simeq \caO(-4)$ (or
alternatively defines a holomorphic section of $\L^{(3,0)}(4)$).
This also determines the holomorphic volume form $\rd \Omega$ on
$\T$:
     \begin{equation}
d\Omega=\frac16 \epsilon_{\a\b\c\d}\rd Z^\a \wedge \rd Z^\b \wedge
\rd Z^\c \wedge \rd Z^\d\,  . \label{defmeasa}
\end{equation}

\subsection{The infinity twistor}

The conformal compactification $\widetilde{\CM}$ of space-time is
invariant under the full conformal group. In order to break
conformal invariance to conformal Poincar\'e invariance (i.~e. the
Poincar\'e group together with dilations), we choose a point in
$\widetilde{\CM}$ to be the point
        $i$ at infinity, and the complexified conformal Poincar\'e group is the subgroup of $SL(4,\C)$ preserving this point.
        In particular,
with a further choice of an origin $0$ in $\widetilde{\CM}$, this
chooses a Lorentz subgroup  $SL(2,\C)\times SL(2,\C)\subset
        SL(4,\C)$, and different choices of $i,0$ lead to different conjugate Lorentz subgroups.

         The point
        $i$ at infinity in  $\widetilde{\CM}$  corresponds to a bi-vector
        $I^{\a\b}$ up to scale which is simple,
        \begin{equation}
I^{[\alpha\beta}I^{\gamma\delta]}=0,
\end{equation}
and which is called  the  {\em infinity twistor}.
        The infinity twistor can also be represented by the 2-form $\tau$
on $\T$ defined by
\begin{equation*}
        \tau = \frac{1}{2} I_{\a\b} \rd Z^\a\wedge \rd Z^\b ,
\end{equation*}
        where $I^{\a\b} = \frac{1}{2}
\varepsilon^{\a\b\c\d} I_{\c\d}$.
     Choosing a point  $0$ in $\widetilde{\CM}$ to be the   origin
     $x^\m=0$ corresponds to choosing a second two-form $\mu$ (dual to a
     simple bi-vector), and this can be chosen so that
\footnote{If no choice of origin is made, the two-form $\mu$ is
defined by (\ref{factmeas}) up to the addition of multiples of
$\rd\pi_{A'}$.}
     \begin{equation}
\label{factmeas} d\Omega = 4 \m \wedge \t .
\end{equation}
The choice of $i,0$ in $\widetilde{\CM}$
      selects an $SL(2,\C)\times SL(2,\C)$ subgroup of $SL(4,\C)$
that preserves $\m$ and $\t$ separately, and this is the double
cover of the rotation group $SO(4,\C)$ preserving the origin $x=0$
and the point at infinity in $\widetilde{\CM}$.
     It is natural to use
2-component spinor notation for this $SL(2,\C)\times SL(2,\C)$
subgroup, with $ Z^\a=(\omega^A,\pi_{A'}) $. Then
\begin{equation}
\label{}\t =\frac{R}{2}\epsilon^{A'B'}\rd \pi_{A'}\wedge
\rd\pi_{B'} , \qquad \m= \frac{1}{2R}\epsilon_{AB}\rd
\omega^{A}\wedge \rd\omega^{B}  \end{equation}
        for some $R$.
        The corresponding space-time metric is
        \begin{equation}
\label{metricst} ds ^2 =R^2  \epsilon_{AB}
\epsilon_{A'B'}dx^{AA'}dx^{BB'},
\end{equation}
so that scaling the infinity twistor by $R$ leads to a conformal
scaling of the
        metric by $R^2$, and the scale of the infinity twistor determines the length scale in space-time.
        For the rest of the paper, we will set $R=1$.

The infinity twistor determines the projective line $I$ in $\PT$
corresponding to $i$
         by
\begin{equation*}
Z^\alpha I_{\alpha\beta}= 0 ,\end{equation*} which in adapted
coordinates is the line $\p_{A'}=0$,
     while the origin $x=0$ corresponds to the line $\m^{A}=0$.
      Removing the light-cone at
infinity $\mathscr{I}$ from $\widetilde{\CM}$ leaves complex
space-time $\CM$ while removing the line $I$ in $\PT$
corresponding to the infinity twistor gives the twistor space
$\PT'=\PT-I$. As $I$ is the $\CP^1\subset \PT$ given by
$\p_{A'}=0$, $\PT'$ consists of points $Z^\alpha =
(\omega^A,\pi_{A'}) $ in which at least one component of $\p$ is
non-zero.  For non-conformal theories, it is natural to use
$\PT'$, and this (and its curved generalisations) is the twistor
space that will be used in our constructions.

The infinity twistor determines a projection $\T\rightarrow
\mathbb{S}_{A'}$ to $\mathbb{S}_{A'}$, the dual primed spinor
space, given by $Z^\alpha = (\omega^A,\pi_{A'}) \rightarrow
\pi_{A'}$. Projectively, this projection determines a fibration
$\PT'\rightarrow \CP^1$.  The infinity twistor $I^{\a\b}$ defines
a Poisson structure of homogeneity $-2$ by
\begin{equation*}
\{f,g\}_I: = I^{\a\b}\frac{\del f}{\del Z^\a} \frac{\del g}{\del
Z^\b} = \epsilon^{AB} \frac{\del f }{\del \omega^A}\frac{\del
g}{\omega^B} \,  .
\end{equation*}
We further define the one-form
\begin{equation}\label{forms1}
         k= I_{\a\b} Z^\a dZ^\b = \epsilon^{A'B'} \pi_{A'}
\rd\pi_{B'} ,
         \end{equation}
for which $\tau = \frac{1}{2}dk= \frac{1}{2}\epsilon^{A'B'}\rd
\pi_{A'}\wedge \rd\pi_{B'} $;  $k$ is the pull-back of a
holomorphic one-form on $\CP^1$ with weight $2$ and will play a
central role in our construction.

\subsection{Twistor spaces for real space-times}
        \label{realspace}

We can choose a real slice $\M\subset \CM$ in such a way that the
metric has signature $(p,4-p)$ for $p=0,1,2$, and the subgroup of
the complexified conformal group that preserves the real slice is
a real form of $SL(4,\C)$.  For Euclidean signature, Lorentzian
signature, or split signature $(2,2)$,  the real  conformal groups
are $SU^*(4) = SL(2,\mathbb{H})=Spin(5,1)$, $SU(2,2)=Spin (4,2)$
and $SL(4,\R)=Spin (3,3)$ respectively, where $\mathbb{H}$ denotes
the
quaternions.%
\footnote{Again, we are ignoring factors of $\Z_4$ here.}

The conformal group acts on the twistor space $\T=\C^4$, with
$Z^\a$ transforming as a complex Weyl spinor for $SO(6,\C)$. For
split signature, this representation is reducible: it decomposes
into the direct sum of two copies of the real Majorana-Weyl
representations of $Spin (3,3)$, and it is possible to impose a
reality condition on the twistors, giving the real twistor space
$\RP^3$. However, for the other two signatures, the Weyl
representation is irreducible so that twistors are necessarily
complex.

We can characterise the real slices $\M$ of $\CM$ as fixed points
of a complex conjugation $\tau:\CM\to\CM$ which,  in local
coordinates that are real on the appropriate real slice, are given
by standard complex conjugation, $\tau(x^\m)= (x^\m)^*$.  A point
$x^\m$ in $\CM$ is represented by a complex matrix $x^{AB'}$.  The
different conjugations can be expressed on this matrix as follows.
For space-time of split signature, $\tau(x^{AB'})= ({x^{AB'}})^*$
is the entry-by-entry complex conjugate, for Lorentzian signature
$\tau(x^{AB'})$ is
     the hermitian conjugate $\tau(x)=x^\dagger$, while for Euclidean signature
$\tau(x^{AB'})=\hat x^{AB'}$, where $\hat x= \e  x^* \e$ with $\e$
the real anti-symmetric $2\times 2 $ matrix (given in terms of the
Pauli matrix $\s_2$ by $\e=i\s_2$).\footnote{Note that in this
      definition, neither the map $x\to \bar x$ nor $x\to\e x \e$ are
      invariant under the $SO(4)$ rotation group, only the composition
      $x\to\e \bar x \e$ is.}

Complex conjugation $x \to \tau x$ in $\CM$ leads to a map on
twistor space. In split signature and in Euclidean signature,
$\tau$ sends $\alpha$-planes to $\alpha$-planes, but in Lorentz
signature it sends $\alpha$-planes to $\beta$-planes where
$\b$-planes are totally null 2-planes in $\CM$ that are
anti-self-dual.  The space of such $\b$-planes together with
tangent spinor $\l_A$, is dual twistor space $\T^*$ with
coordinates $W_\a = (\l_A , \m^{A'})$; a point in $\T^*$
corresponds to the $\b$-plane in $\CM$ defined by the dual
incidence relation $ \m^{A'}= x^{AA'}\l_A $.  The complex
conjugation $\tau$ on $\CM$ therefore induces a complex
conjugation $\tau:\T\to\T$ in split signature and Euclidean
signature, but in Lorentz signature, it determines an
anti-holomorphic map $\tau:\T\to\T^*$.

We have the complex conjugate twistor space $\bar \T$ (i.e. $\T$
with the opposite complex structure) with coordinates $\bar
Z^{\bar
      \a}=({Z^\a})^*$ on twistor space, and their counterparts on
dual twistor space $\T^\dagger$ with coordinates $\bar W_{\bar
      \a}=({W_\a}
)^*$.  For the real and split signature complex structure, $\tau$
is an isomorphism from $\bar \T$ to $\T$ and in the Lorentzian
case it is a natural map from $\bar \T$ to $\T^*$, and this can be
used to express conjugate twistors in $\bar \T$ in terms of
twistors in $\T$ or $\T^*$, so that conjugate twistor indices are
never needed explicitly.  We now describe features of twistor
geometry appropriate to each signature in more detail.

\subsubsection{Lorentzian signature}

In the case of Lorentzian signature, the conformal group $SU(2,2)$
preserves a Hermitian metric $\Sigma _{\a \bar \b}$, and this
defines the map $\tau: \bar \T\to \T^*$ under which $ \bar Z^{\bar
      \a}=(Z^\a)^* \to \Sigma _{\a \bar \b} \bar Z^{\bar \b}$, so that
each conjugate twistor can be identified with a dual twistor.
Complex conjugation on $\CM$ leads to an anti-holomorphic map
$Z^\a\rightarrow\bar Z_\a= \Sigma _{\a \bar \b} \bar Z^{\bar \b}$
from $\T\rightarrow \T^*$. The real Minkowski space-time $\M$ is
the subspace of $\CM$ in which $x^{AB'}$ is Hermitian and is
preserved by this conjugation. This is the standard case,
discussed in detail in e.~g.~\cite{HuggTod}.

\subsubsection{Split signature}
\label{splsign}

For extensive discussions of the twistor correspondences in split
signature see~\cite{Mason4,LeBrunMason}. Here we give a summary of
the main ideas.

For split signature, the real space-time $\M$ is the subspace of
$\CM$ with $x^{AB'}$ real. The ordinary complex conjugation on
$\CM$ that preserves $\M$ is
         represented by the ordinary component-by-component complex
conjugation on $\T$, viz.\ $Z^\a\rightarrow ( Z^\a)^*$, that fixes
the real slice $\T_\R=\R^4\subset \C^4 =\T$ and hence
$\PT_\R=\RP^3\subset \PT$. Points of this real slice correspond to
totally real $\alpha$-planes in $\M$ and there is  a totally real
version of the twistor correspondence in which points in
$\widetilde\M$ correspond to real projective lines (i.e. $\RP^1$s)
in $\PT_\R$ via the incidence relation $\omega^A=x^{AA'}\pi_{A'}$
where now $\omega^A$, $\pi_{A'}$ and $x^{AA'}$ are all real. Here
$\widetilde\M$ is the conformal compactification of $\M$, which is
$\widetilde \M=S^2\times S^2/\Z_2$.

In order to use deformed twistor correspondences in split
signature, we will also need to use the correspondence between
$\M$ and the complex twistor space $\PT$. Each point $x\in\M$
corresponds to a complex line $L_x=\CP^1$ in $\PT$ that intersects
the real slice $\PT_\R$ in a real line $L_{\R x}=\RP^1$. This real
line divides $L_x$ into two discs $D^\pm_x$, each with boundary
$L_{\R x}\subset \PT_\R$. The space of such discs naturally
defines a double cover $\widetilde {\widetilde\M}$ of conformally
compactified Minkowski space $\widetilde \M$ (which is the space
of all $L_{\R x}\subset \PT_\R$). In fact $\widetilde
{\widetilde\M}= S^2\times S^2$ with the conformal structure that
is determined by the split signature  product metric
\begin{equation*}
g= \pi_1^* h -\pi_2^* h ,
\end{equation*}
where $h$ is the standard round metric on $S^2$ and
$\pi_1,\pi_2:S^2\times S^2\rightarrow S^2$ are the two factor
projections. The conformal compactification  $\widetilde
\M=S^2\times S^2/\Z_2$ is obtained from the double cover
$\widetilde {\widetilde\M}$ by identifying under the
        $\Z_2$ that acts as  the joint antipodal map on both $S^2$ factors.

$\widetilde {\widetilde\M}$ can be thought of as two copies
$\M^\pm$ of $\M$ glued together across the double cover of the
lightcone at infinity $\mathscr I$. With the choice of the
infinity twistor, we have the fibration $\PT'=\PT-I\rightarrow
\CP^1$ as above. The condition that $i\pi_{A'}\bar\pi^{A'}$ be
positive, negative or zero defines $\PT_\pm$ and $\PT_0$. The
holomorphic discs in $\PT_\pm$ project to $\pm
i\pi_{A'}\bar\pi^{A'}>0$ in $\CP^1$ and correspond respectively to
points of $\M^\pm$, whereas the holomorphic discs in $\PT_0$
correspond to points of the double cover $\widetilde {\mathscr I}$
of $\mathscr I$.  This will be important later for the Berkovits
string, where the open string world-sheets are holomorphic discs.
The moduli space of discs in twistor space gives $\widetilde
{\widetilde\M}$ with two copies of space-time $\M$, and to get
just one copy, the theory must be restricted to one in which the
world-sheets are discs in one half of twistor space, say  in
$\PT_+$.

\subsubsection{Euclidean signature}

The anti-linear map $\tau:\T \to \T$ is given by the conjugation
$Z^\alpha\rightarrow \hat Z^\a$ where, if $Z^\a=(\w^A,\p_{A'})$,
then  $\hat Z^\a=(\hat\w^A,\hat\p_{A'})$, with $\hat\w^A=(\bar
\w^1,-\bar \w^0)$ and $\hat\p_{A'}=(\bar \p_{1'},-\bar\p_{0'})$.
The conjugation extends to multi-spinors and the real Euclidean
space-time $\M$ is the subspace of $\CM$ preserved by this,
$x^{AB'}=\hat x^{AB'}$. The conjugation $Z^\alpha\rightarrow \hat
Z^\a$ is then the lift of the complex conjugation $x^\m \to
(x^\m)^*$ on $\CM$ preserving real Euclidean slices. The
conjugation $Z^\alpha\rightarrow \hat Z^\a$ is quaternionic in the
sense that $\hat{\hat Z}^\a=-Z^\a$ so that it defines a complex
structure that anticommutes with the standard one. It therefore
has no fixed points (as $Z^\a=\hat Z^\a $ implies $Z^\a=-Z^\a$),
and it is induced by the standard quaternionic conjugation on
spinors: $\hat\pi_{A'}=(\bar\pi_{1'},- \bar\pi_{0'})$ and
similarly for $\omega^A$.

The conformal compactification $\widetilde M$ of Euclidean  $\R^4$
is given by adding a single point $i$ at infinity to give $S^4$.
The Euclidean signature correspondence is particularly
straightforward since we have a fibration $\PT=\CP^3 \rightarrow
S^4$ given by sending $Z^\a$ to the point in Euclidean space
corresponding to the projective line through $Z^\a$ and $\hat
Z^\a$ (this includes a line at infinity corresponding to
$\pi_{A'}=0$). The fibre over any point $x^{AA'}$ in $S^4$ is a
$\CP^1$ with projective coordinates $  \pi_{A'}$, and the
corresponding point in $\PT$ is
\begin{equation}
\label{dfghgdf}(\omega^A,\pi_{A'})=(x^{AA'}\pi_{A'},\pi_{A'}) .
\end{equation}
Conversely, a point  in $\PT$ with holomorphic coordinates
$(\omega^A,\pi_{A'})$ is represented in  local non-holomorphic
coordinates $(x^{AA'},\pi_{A'})$ by
\begin{equation}
\label{hjkgjk}(x^{AA'},\pi_{A'})=\left(\frac{\omega^A\hat
         \pi^{A'}-\hat\omega^A\pi^{A'}}{\pi_{A'}\hat\pi^{A'}} ,
         \pi_{A'}\right).\end{equation}
         The $\CP^1$ fibre at each point is the space of primed spinors $\p_{A'}$, identified under scaling, so that
          $\PT$ is the projective primed spin bundle over
$S^4$. Similarly,   $\T -0$ is   the bundle of primed spinors
minus the zero section, and we can again use the formulae
(\ref{dfghgdf}),(\ref{hjkgjk}).

To obtain $\M=\R^4$, we choose a point $i$ on $S^4$ to be the
point at infinity, and this corresponds to an infinity twistor
$I$, specifying the $\CP^1$ fibre over $i$. Then the twistor space
for $\R^4$ is given by removing this $\CP^1$, so that $\PT'=\PT -
\CP^1$ is the projective spin bundle over $\R^4$.  Choosing an
infinity twistor and an origin chooses a subgroup $SU(2)\times
SU(2)\subset SU^*(4)$ and a decomposition of $Z^\a$ into
holomorphic coordinates $(\omega^A,\pi_{A'})$ transforming under
this $SU(2)\times SU(2)$; in this frame, the twistor
correspondence is given by (\ref{dfghgdf}),(\ref{hjkgjk}) on
$\T'=\T-\{\pi_{A'}=0\}$ so that the point at infinity is
$x^{AA'}=\infty$, corresponding to the 2-plane in $\T$ (or $\CP^1$
in $\PT$) given by $\{\pi_{A'}=0\}$.

\subsection{The Penrose transform}

\label{Penrosetransf}

The Penrose transform identifies fields of helicity $-n/2$
satisfying the massless wave equation on a suitable region
$U\subset\CM$ with the cohomology group $H^1(\PT(U),\caO(n-2))$
for $\PT(U)$ the corresponding subset of $\PT$. A Dolbeault
representative of this group is a $(0,1)$-form $\alpha$ with
values in $\caO(n-2)$ such that $\dbar\alpha=0$, where $\alpha$ is
defined modulo $\dbar g$ with $g$   a smooth section of
$\caO(n-2)$. The corresponding massless space-time field of
helicity $|n|/2$ for $n\leq 0$ is given by the integral formula
\begin{equation}
\label{massl1}
        \phi_{A_1'\ldots
A'_{-n}}(x)=\int_{\omega^A=x^{AA'}\pi_{A'}}
\pi_{A'_1}\ldots\pi_{A'_{-n}}\alpha\wedge \pi_{C'}\rd\pi^{C'} .
\end{equation}
        For $n\geq 0$, the massless space-time field of helicity $-n/2$
is given by
\begin{equation}
\label{massl2}
        \phi_{A_1\ldots
A_{n}}(x)=\int_{\omega^A=x^{AA'}\pi_{A'}} \left( \frac{\del}{\del
\omega^{A_1}} \ldots \frac{\del}{\omega^{A_n}} \alpha \right)
\wedge \pi_{C'}\rd\pi^{C'} \, .
\end{equation}
Alternatively, a \v Cech representative can be chosen for the
cohomology class, and the space-time fields are then given by a
contour integral formula. This can be implemented simply when it
is possible to cover $\PT(U)$ by two open sets, $V_0$ and $V_1$
(this is the case for $\PT'$, for which we can take
$V_0=\{\pi_{0'}\neq 0\}$ and $V_1=\{\pi_{1'}\neq 0\}$). Then the
\v Cech cohomology class can be represented by a holomorphic
function $f$ of homogeneity $n-2$ on $V_0\cap V_1$. The analogues
of the above formulae are then, for $n\leq 0$,
     \be \label{zrmpos} \phi_{A_1'\ldots A'_{-n}}(x)=\oint_\Gamma
\pi_{A'_1}\ldots\pi_{A'_{-n}}f \; \pi_{C'}\rd\pi^{C'} \ee and, for
$n\geq 0$, \be\label{zrmneg} \phi_{A_1\ldots
A_{n}}(x)=\oint_{\Gamma} \frac{\del}{\del \omega^{A_1}} \ldots
\frac{\del}{\omega^{A_n}} f \;\pi_{C'}\rd\pi^{C'} \, . \ee In
both~(\ref{zrmpos}) and~(\ref{zrmneg}) the contour $\Gamma$ is a
suitable circle in $V_0\cap V_1 \cap
\{\omega^A=x^{AA'}\pi_{A'}\}$.

In split signature, instead of considering cohomology classes, we
can consider smooth functions defined on $\PT_\R$ that are
homogeneous of degree $n-2$ and apply the integral formulae
(\ref{zrmpos}) and (\ref{zrmneg}), where now $\Gamma$ is taken to
be the real line $\{\omega^A=x^{AA'}\pi_{A'}\}$ in $\PT_\R$ for
$x^{AA'}$ a point in real split signature Minkowski space. In the
case of $n=0$ this is known as the X-ray transform, and it is a
classic theorem that these formulae define an isomorphism from
functions on $\PT_\R$ to solutions of the ultrahyperbolic wave
equation on $\M$~\cite{John}. The close relationship between the
Penrose transform and the X-ray transform was observed by Atiyah
\cite{AtiyahYM}. The connection between the X-ray transform and
the Penrose transform can be understood naively by requiring $f$
to be analytic, extending it to some complex neighbourhood of
$\PT_\R$ and reinterpreting it as a \v Cech cohomology class.
However there are a number of issues that this approach does not
deal with; a full treatment of the relationship between the X-ray
and Penrose transforms is given in~\cite{BEGM,BEast}.
    For the most part, it is this X-ray transform version of the Penrose
    transform that is used by Witten and Berkovits in~\cite{Witten2003,NB}.

\subsection{Super-twistor space}

\label{superspace}

The superspace with $N$ supersymmetries has
        space-time coordinates $x^{AA'}$
        and anti-commuting coordinates $\theta_a^A,\ti \theta^{aA'}$
        where $a,b=1,....,N$.
        The latter  are space-time spinors and transform as an
        $N$-dimensional representation of an R-symmetry group, which is
        $U(N)$ or $SU(N)$ for Lorentzian signature,
        $GL(N,\R)$ or $SL(N,\R)$ for split signature and
$U^*(N)$ or $SU^*(N)$ for Euclidean signature.

The complexified superconformal group is $SL(4|N;\C)$ and its real
forms are $SU(2,2|N)$  for  Lorentzian signature,
         $SL(4|N; \R)$ for split signature and
        $SU^*(4|N)$ for Euclidean signature.
The group  $SL(4|N;\C)$ is realised on
        the space  $ \C^{4|N}$ with coordinates $Z^I=(Z^\a,\psi^a)\in \C^{4|N}$,
consisting of the usual   commuting coordinates $Z^\a$ as before
and anti-commuting coordinates $\psi^a$, $a=1,\ldots,N$.
Super-twistor space $\T_{[N]}$ is the subset $\C^{4|N}-\C^{0|N}$ on
which $Z^\a \ne 0$, and the projective super-twistor space
$\PT_{[N]} =\CP^{3|N}$ is the space of equivalence classes under
complex scalings~\cite{Ferber}:
\begin{equation*}
\PT_{[N]} =\CP^{3|N}=\{Z^I=(Z^\a,\psi^a)\in
\C^{4|N}-\C^{0|N}\}/\{Z^I\sim \lambda Z^I\, ,
\lambda\in\C^\times\}\, .
\end{equation*}
        Note that in this definition we have a
fibration $\PT_{[N]}\rightarrow \PT$ given by
$(Z^\alpha,\psi^a)\rightarrow Z^\alpha$. However, this fibration
is not preserved by the action of the superconformal group
$SL(4|N;\C)$.

The $N=4$ superspace is special for twistor theory because in that
case there is a global holomorphic volume form on the projective
super-twistor space,
\begin{equation*}
\Omega_s=\Omega\rd\psi_1\rd\psi_2\rd\psi_3\rd\psi_4\, ,
\end{equation*}
with $\Omega$ the bosonic 3-form defined in~(\ref{defmeas}). This
has weight zero, since each $\rd\psi_a$ has weight $-1$ according
to the Berezinian integration rule $\int \psi_1\rd\psi_1=1$.

Anti-chiral super-Minkowski space $\CM^-_{[N]}$ with coordinates
$x^{AA'}_+,\ti \theta^{aA'}$
        arises as the space of
$\CP^{1|0}$s in $\PT_{[N]}$ via the incidence relations
\begin{equation}\label{chiralinc}
(\omega^A, \pi_{A'},\psi^a)= (x_+^{AA'}\pi_{A'}\, , \pi_{A'}, \ti
\theta^{aA'}\pi_{A'}) ,
\end{equation}
where we have used $\pi_{A'}$ as homogeneous coordinates on
$\CP^{1|0}$. Chiral super-Minkowski space $\CM^+_{[N]}$ with
coordinates $x^{AA'}_-,\theta_a^A$ arises as the space of
$\CP^{1|N}$s in $\PT_{[N]}$ via the incidence relations
\begin{equation}\label{achiralinc}
(\omega ^A, \pi_{A'},
\psi^a)=(x_-^{AA'}\pi_{A'}+\psi^a\theta_a^A\, , \pi_{A'},
\psi^a)\, ,
\end{equation}
where now we have used $(\pi_{A'}\, , \psi^a)$ as homogeneous
coordinates on the $\CP^{1|N}$s. A point of full super-Minkowski
space $\CM_{[N]}$  with coordinates $x^{AA'},\theta_a^A,\ti
\theta^{aA'}$ arises from a choice of $\CP^{1|N}$  in $\PT_{[N]}$
together with  a choice of $\CP^{1|0}\subset \CP^{1|N}$, so that
full super-Minkowski space is the space of `flags'
$\CP^{1|0}\subset \CP^{1|N}$ in $\PT_{[N]}$~\cite{Ferber}.  Taking
(\ref{chiralinc}) and (\ref{achiralinc}) together we have
$x^{AA'}_+=x_-^{AA'}+\ti \theta^{aA'}\theta_a^A$ and it is usual to
define $x^{AA'}= \frac12(x_+^{AA'}+x_-^{AA'})$.\footnote{To obtain
standard conventions in Lorentz signature we must take
$x^{AA'}=iy^{AA'}$ for real $y^{AA'}$; our conventions are adapted
to split and Euclidean signature.}

The massless field formulae generalising~(\ref{massl1})
and~(\ref{massl2}) now give rise to superfields encoding
supermultiplets. The easiest way to see this is to expand out an
element $\mathcal F_n\in H^1(\PT_{[N]}(U), \caO(n))$ as follows:
\begin{equation*}
\mathcal F_n= f_{(n)} +
f_{(n-1)a}\psi^a+f_{(n-2)a_1a_2}\psi^{a_1}\psi^{a_2} +
f_{(n-3)a_1a_2a_3}\psi^{a_1}\psi^{a_2}\psi^{a_3} + \ldots .
\end{equation*}
Here $f_{(n-k) \ldots}$ has homogeneity degree $n-k$ so that its
Penrose transform is a massless field of helicity $-(n-k-2)$ on
space-time with skew-symmetric indices $a_1, \ldots ,a_k$, and it
transforms as a $k$-th rank anti-symmetric tensor under the
R-symmetry group.

It is possible to perform the transform on $\mathcal F_n$ to
obtain a superfield directly on $\CM^\pm$, the $\pm$ depending on
whether we integrate over $\CP^{1|0}$s or $\CP^{1|N}$ fibres.
Particularly interesting examples are furnished by the cases of
$n=\pm 2$ in the context of linearised $N=4$ Einstein
supergravity. We can define \bea\label{sugramult}
H^+(x_-,\theta_a^A) &=& \oint_{\CP^{1|4}} \mathcal
F_2(x^{AA'}_-\pi_{A'}+\psi^a\theta_a^B, \pi_{A'},\psi^b)
\pi_{A'}\rd \pi^{A'}\, \rd^4\psi \label{defH-} \eea and \bea
H^-(x_+,\tilde\theta^{aA'}) &=&\oint_{\CP^{1|0}} \mathcal F_{-2}(
x^{AA'}_+\pi_{A'},\pi_{A'},\tilde\theta^{aA'}\pi_{B'})
\pi_{A'}\rd\pi^{A'}\, . \label{sugramult2} \eea The integrand
of~(\ref{defH-}) can be expanded in $\psi^a$ using Taylor series
in the anti-commuting coordinates and the variables $\psi^a$  can
be integrated out to yield a power series in $\theta_a^B$; the
standard Penrose transform in the form (\ref{zrmpos}) can then be
applied to the coefficients to yield a superfield on chiral super
Minkowski space. Eq.~(\ref{sugramult2}) can be expanded as a
Taylor series in $\tilde \theta^{aA'}$ to obtain a series whose
coefficients can be integrated using (\ref{zrmneg}) to obtain a
superfield on anti-chiral super-Minkowski space $\CM_{[N]}^-$.
This directly gives formulae for the full chiral and anti-chiral
superfields for $N=4$ supergravity in terms of the component
fields.

In order to obtain an anti-chiral or a chiral superfield for other
values of $n$ or $N$, we need to either repeatedly differentiate
$\mathcal F_n$ with respect to $\omega^A$, or to multiply it by
enough factors of $\pi_{A'}$. In the first case,  this will reduce
the homogeneity to $-2$ and enable us to apply~(\ref{sugramult2})
to obtain an anti-chiral superfield; in the second case, we
arrange for homogeneity $N-2$ and obtain a chiral superfield by
applying~(\ref{defH-}).

As before, the space of $\CP^{1|0}$s (resp.\ $\CP^{1|N}$s or flags
$\CP^{1|0}\subset\CP^{1|N}$) in $\PT_{[N]}$ is a conformal
compactification of chiral (resp.\ anti-chiral or full) super
Minkowski space on which the superconformal group acts.  We will
wish to break conformal invariance on super-twistor space by
choosing points at infinity and a scale. There are three ways in
which we can break superconformal invariance; we can choose points
at infinity in either the chiral, anti-chiral or full Minkowski
space, and these lead to different structures.

A choice of a point at infinity in chiral super-Minkowski space
corresponds to a choice of a line $I$, a $\CP^{1|0}$, in
$\PT_{[N]}$ and coordinates $(\omega^A,\pi_{A'},\psi^a)$ can be
chosen so that $I$ is given by $\pi_{A'}=0=\psi^a$.  This
determines a projection $p_1:\PT_{[N]}-I\to \CP^{1|N}$ given in
homogeneous coordinates by
\begin{equation*}
p_1:(\omega^A,\pi_{A'},\psi^a)\to (\pi_{A'},\psi^a)\, .
\end{equation*}
The fibres of the projection are the $\CP^{2|0}$s through $I$.

If we choose a point in anti-chiral Minkowski space, then this
gives a choice of a superline $I_{[N]}=\CP^{1|N}$ and we can then
choose coordinates $(\omega^A,\pi_{A'},\psi^a)$ so that $I_{[N]}$
is the set $\pi_{A'}=0$.  This, as before, leads to a fibration
$p:\PT_{[N]}-I_{[N]}\to \CP^{1|0}$ given by
\begin{equation*}
p_1:(\omega^A,\pi_{A'},\psi^a)\to \pi_{A'}\,
\end{equation*}
with fibres the $\CP^{2|N}$s through $I_{[N]}$.

The richest structure is obtained by choosing a vertex $i$ of a
super-light-cone at infinity $\mathscr{I}$ in the full conformally
compactified super-Minkowski space (as opposed to one of its
chiral versions). This is equivalent to the choice of a `flag'
$\CP^{1|0}\subset \CP^{1|N}\subset \PT_{[N]}$, i.~e. the pair
$I\subset I_{[N]}$. These lead to corresponding projections of
$\PT'_{[N]}=\PT - I_{[N]}$
\begin{equation*}
\PT_{[N]}'\stackrel{p_1}\longrightarrow \CP^{1|N}
\stackrel{p_0}\longrightarrow \CP^{1|0}\, , \qquad
Z^I=(\omega^A,\pi_{A'},\psi^a)\rightarrow
(\pi_{A'},\psi^a)\rightarrow \pi_{A'}\, .
\end{equation*}
   We will also be
interested in the case in which there is only the projection
$p:\PT'_{[N]}\to \CP^{1|0}$.  We will see that this is a weaker
structure and there will correspondingly be a larger class of
deformations.

We can define the Poisson structure
\begin{equation*}
\{f,g\}_I:=I^{IJ}\frac{\del f}{\del Z^I} \frac{\del g}{\del Z^J} =
\epsilon^{AB} \frac{\del f}{\del \omega^A} \frac{\del g}{\omega^B}
\,
\end{equation*}
as in the bosonic case, and $p_0$ can then be used to pull back
the $1$-form
\begin{equation*}
I_{IJ} Z^I\rd Z^J=\epsilon^{A'B'}\pi_{A'}\rd\pi_{B'}\,
\end{equation*}
from $\CP^{1|0}$. These are special cases of more general
correspondences between points of chiral Minkowski space and rank
two bi-vectors $X^{IJ}=X^{[IJ]}$ up to scale, and between points
of anti-chiral Minkowski space and simple (rank two) two-forms
$X_{IJ}$ up to scale. Alternative representations can be obtained
by use of the volume form $\epsilon _{I_1\ldots I_{4+N}}$ and its
inverse on $\T_{[N]}$.

     \section{The non-linear graviton}

\label{nonlineargrav}

\subsection{The conformally anti-self-dual case}

\label{nonlinantiself}

Penrose's non-linear graviton construction provides a
correspondence between curved twistor spaces and conformally
anti-self-dual space-times, and so gives a general construction of
such space-times.  This arises from nontrivial deformations of the
flat twistor correspondence in which, on the one hand, the
space-time is deformed from flat space to one with a curved
conformal structure with anti-self-dual Weyl curvature, and, on
the other, the complex structure of a region in twistor space is
deformed away from that of a region in projective space.  One
cannot deform the complex structure of the whole of flat twistor
space as $\PT=\CP^3$ is rigid and has no continuous deformations,
so we instead consider deformations of $\PT'$, which is $\CP^3$
with a line removed.  This has topology $\R^4\times S^2$.  We will
find it convenient to   start by describing the non-projective
twistor space.

A curved twistor space $\Tc$ will be taken to be a 4-dimensional
complex manifold equipped with a vector field $\Upsilon$ and a
non-vanishing holomorphic $3$-form $\Omega$ such that
\begin{itemize}
\item $\Upsilon$ gives $\Tc$ the structure of  a
         line bundle over the space $\PTc=\Tc/\{\Upsilon\}$ of orbits of
         $\Upsilon$, for which    $\Upsilon$ is the Euler vector
         field (in local coordinates $(z, z_1,z_2,z_3)$ where
       $(z_1,z_2,z_3)$ are coordinates on $\PTc$ and $z$ is a linear
       coordinate up the fibre, $\Upsilon=z\del/\del z$).
\item $\Upsilon$ and  $\Omega$  satisfy
\begin{equation}
\mathcal L_\Upsilon \Omega=4\Omega\, , \qquad \i (\Upsilon)
\Omega=0\, .
\end{equation}
\item $\PTc$ contains a holomorphically embedded Riemann sphere that
        has the same normal bundle as a complex projective line in $\CP^3$.

\end{itemize}
The last condition is in fact rather mild and holds automatically
not only for any twistor space that is constructed as described
below from a conformally anti-self dual space-time, but also for
any twistor space that is an arbitrary small deformation of such a
twistor space. The space-time is reconstructed as the moduli space
of such Riemann spheres; given one such sphere, Kodaira theory
implies the existence of a full four-dimensional family
~\cite{Kodaira}.

The existence of the holomorphic volume form $\mathrm d \Omega$
implies that $\Tc$ is a non-compact Calabi-Yau space.\footnote{The
second condition allows us to give a construction of $\Tc$ in
terms of $\PTc$ as the total space of the line bundle
$\Tc=(\L^{(3,0)} )^{1/4}$ over $\PTc$. This definition arises by
analogy with the flat case, where $\L^{(3,0)}$ is $\caO(-4)$
because the holomorphic $(3,0)$-form $\Omega$ has weight $4$ and
so it needs to be multiplied by a weight $-4$ function to define a
$(3,0)$-form. Since $\T-\{0\}$ is the total space of the line
bundle $\caO(-1)$ minus its zero-section, it is therefore the
fourth root of $\L^{(3,0)}$.  With this definition of $\Tc$, the
existence of $\Omega$ on $\Tc$ is tautological as $\Tc$ is a
covering of the bundle of $3$-forms and so $\Omega$ is the
pull-back to $\Tc$ of the corresponding $3$-form at that point. As
the $(3,0)$-form $\Omega$ has weight $4$, it is not a $(3,0)$-form
on $\cPT$, so that $\cPT$ is not a Calabi-Yau space.}
        The global existence of
$\Upsilon$ and $\Omega$ allows us to introduce local complex
coordinates $Z^\a$ on $\Tc$ such that
\begin{equation*}
\Upsilon=Z^\a\frac{\del}{\del Z^\a}\, , \qquad \Omega=\frac16
\epsilon_{\a\b\c\d}Z^\a\rd Z^\b\rd Z^\c\rd Z^\d\,
\end{equation*}
as in the flat case, with $\epsilon_{\a\b\c\d}
=\epsilon_{[\a\b\c\d]}$, $\epsilon_{0123}=1$.

        We now turn to the relation between curved twistor space and
space-time.  For complexified Minkowski space, a twistor
corresponds to an $\alpha$-plane, i.~e.\ a totally null self-dual
two-plane. In a curved complex space-time $\CMc$, which is a
complex 4 manifold with a holomorphic metric $g$ (so that locally
the metric is $g_{\m\n}(x)dx^\m dx^\n$, depending on the complex
coordinates $x^\m$ but not their complex conjugates),
$\alpha$--plane elements in the tangent space are not generally
integrable, i.e. one cannot in general find a two surface whose
tangent planes are $\alpha$-planes. A two-surface whose tangent
plane is an $\alpha$-plane at every point is called an
$\alpha$-surface. The nececessary and sufficient condition for
there to exist $\alpha$-surfaces through each $\alpha$-plane
element at every point is that the self-dual part of the Weyl
curvature should vanish,
\begin{equation}\label{crflat}
         \tilde \psi_{A'B'C'D'} = 0 .
\end{equation}
If~(\ref{crflat}) holds, then the 3 complex dimensional curved
twistor space $\PTc$  is the space of such $\alpha$--surfaces. An
$\a$-surface through $x$ is specified by an $\a$-plane in the
tangent space at $x$, and this in turn is fixed by a choice of
primed `tangent' spinor $\p_{A'}$ at $x$, up to complex scalings,
so that the space of tangent vectors is given by
$\pi^{A'}\lambda^A$ as $\lambda^A$ varies.

A point in the non-projective twistor space $\Tc$ is determined by
an $\alpha$-surface  in $\CMc$ and a tangent spinor $\p_{A'}$ that
is parallelly propagated over the $\alpha$-surface using the
Levi-Civita connection of any metric in the conformal class. It is
a non-trivial fact that the parallel propagation of such a
`tangent' spinor over its $\alpha$-surface is independent of the
choice of conformal factor for the metric in the conformal class.
A point in the projective twistor space $\PTc$ is given by the
$\alpha$-plane together with $\pi_{A'}$ up to complex scalings of
$\p_{A'}$.

For Euclidean signature, we saw that in the flat case the twistor
space $\PT=\CP^3$ is the projective spin bundle over compactified
space-time $S^4$. This generalises, and for Euclidean signature,
the curved twistor space $\PTc$ for a conformally anti-self-dual
space ${\cal   M}$ is the projective spin bundle over ${\cal M}$,
where the fibre at a point $x$ is a $\CP^1$ with homogeneous
coordinates given by the primed spinors $\p_{A'}$ at $x$, while
$\Tc$ is the corresponding non-projective spin bundle.  In terms
of coordinates $(x,\pi_{A'})$, $\Upsilon
=\pi_{A'}\del/\del\p_{A'}$ and $\Omega=\pi^{A'}D\pi_{A'}\wedge
\pi_{B'}\pi_{C'}\epsilon_{BC}e^{BB'}\wedge e^{CC'}$ where $D$ is
the covariant exterior derivative with   the Levi-Civita
connection of some metric in the conformal class, and $e^{AA'}$
are the pull-backs from space-time to the spin bundle of the
`solder forms' $e^{AA'}_\m dx^\m$ constructed from a vielbein
$e^{AA'}_\m$. \footnote{In this form, the construction makes
         sense for compact space-times of Euclidean signature with
         complicated topology: a celebrated result of Taubes is that Euclidean
         signature anti-self-dual conformal structures can be found on
         arbitrary compact 4-manifolds, possibly after performing a connected
         sum with a finite number of $\overline{\CP^2}$s, and so there are
         many nontrivial compact examples of twistor spaces.}

       The famous result of Penrose~\cite{Penrose1} is that the space-time
       $\CMc$ together with its anti-self-dual conformal structure can be
       reconstructed from the complex structure of $\Tc$ together with
       $(\Upsilon, \Omega)$ as described above, or from $\PTc$ and its
       complex structure. The existence of the correspondence is
       preserved under small deformations, either of the complex structure
       on $\PTc$, or of the anti-self dual conformal structure on $\CMc$.
       Thus one can attempt to construct anti-self-dual space-times by
       deforming, say, $\PT'$.  The key idea is that a point $x\in\CMc$
       corresponds to a Riemann sphere $\CP^1_x$ (the Riemann sphere with
       homogenous coordinates $\p_{A'}$) in $\PT$ consisting of those
       $\a$-surfaces through $x$.  It follows from Kodaira theory that the
       moduli space of deformations of $\CP^1_x$ in $\PTc$ is necessarily
       four dimensional, and naturally contains $\CMc$ as an open set (in
       general it is some analytic continuation of $\CMc$).  Furthermore,
       this family of $\CP^1_x$s still survives after deformations of the
       complex structure of $\PTc$.

       If $\CMc$ arises as such a moduli space, an anti-self-dual conformal
       structure can be defined on $\CMc$ by declaring points $x$ and $y$ to
       be null separated if $\CP^1_x$ and $\CP^1_y$ intersect. The fact that
       the existence of such a correspondence   survives deformations of
       the complex structure on $\PTc$ means
       that, given one conformally anti-self-dual space-time, a family of new
       conformally self-dual space-times can be constructed by deforming the
       complex structure of the corresponding curved twistor space $\PTc$,
       and so the equations governing the deformation of the complex structure
       correspond to the field equations for conformal anti-self-dual
       gravity.

The data of the conformal structure on $\CMc$ is then  encoded in
the complex structure of $\PTc$. There are two standard ways to
represent the complex structure. The Dolbeault approach (cf.\ the
introduction) is to regard $\PTc$ as a real $6$-manifold with an
almost complex structure, i.~e.\ a $(1,1)$-tensor $J$ subject to
the integrability condition that its Nijenhuis tensor $N(J)$
vanishes. We can equivalently encode $J$ into a $\dbar$ operator,
the restriction of the exterior derivative to the 1-forms
$\L^{(0,1)}$ in the $-i$ eigenspace of $J$. With this restriction,
$N(J)=0$ is equivalent to $\dbar^2=0$. The \v Cech approach is to
consider $\PTc$ as a $3$ complex dimensional manifold formed by
choosing a suitable open cover $V_i$,  picking holomorphic
coordinates on each $V_i$ and then encoding the data of the
manifold in the biholomorphic patching functions defined on the
overlaps $V_i\bigcap V_j$. Both these points of view lead to a
cohomological understanding of the deformation theory, the first
via Dolbeault cohomology and the second via \v Cech cohomology. In
either approach, the   deformations of the complex structure are
parametrised by $H^1(\PTc,T^{(1,0)})$. If we consider linearised
deformations of $\PT$, we obtain the following description of
linearised conformal gravity.

We represent $f\in H^1(\PTc,T^{(1,0)})$ by a $(0,1)$-form
$f^\a(Z)=f^\a{}_{\bar \b}(Z)d\bar Z^{\bar \beta} $ taking values
in the bundle of holomorphic vector fields on $\Tc$, with the
condition that $f^\a$ has homogeneity degree 1 and is defined up
to the gauge freedom $f^\alpha\rightarrow f^\alpha + a(Z) Z^\a$
for some $(0,1)$-form $a(Z)$ of homogeneity zero. This freedom can
be fixed by the requirement that $\partial f^\a/\partial Z^\a=0$,
which is the condition that the measure $\rd \Omega $ is
holomorphic for the deformed complex structure $\bar\del +
f(Z)^\a\partial/\partial Z^\a$. This implies that
$f(Z)^\a\partial/\partial Z^\a$ is a deformation of $\T$ that
preserves both $\Omega$ and $\rd \Omega$.

The Penrose transform of $f^\a$ gives a helicity $+2$ field $\psi
_{ABCD}$ in   space-time satisfying the field equation of
linearised conformal gravity, which is  the linearised Bach
equation~\cite{RB}:
\begin{equation}
\label{Bacheq1}
        \nabla^C_{A'}\nabla^D_{B'}\psi_{ABCD}=0\, ;
\end{equation}
see \cite{Mason1, Mason2} for details.

Following~\cite{BWsc} and~\cite{Mason3}, the negative helicity
conformal graviton can be represented by an element $g\in
H^1(\PT(U), \L^1(-4))$. The pull-back of $g$ to $\T$ gives a
1-form $g_\a(Z)\rd Z^\a$ on $\T$, where $g(\Upsilon)=Z^\a g_\a=0$
and the components $g_\a$ have weight $-5$. The Penrose transform
of $g_\a$ gives  a Weyl spinor $\tilde \psi_{A'B'C'D}$, now of
helicity $-2$,
        satisfying
\begin{equation}
\nabla_B^{C'}\nabla_{A}^{D'}\tilde \psi_{A'B'C'D'}=0\, .
\label{Bach2}
\end{equation}
The Penrose transform in this case is the opposite helicity to
that of $f^\a$,
        and can be derived using   the methods of
\cite{Mason2, Mason3}; it is discussed from a different point of
view in \cite{BWsc}, where $g$  appears as the component
$\psi^1\psi^2\psi^3\psi^4 g$ of the cohomology class $b$ in
$H^1(\PT_{[4]}, T^*)$ on super-twistor space, where $T^*$ is the
cotangent bundle.

\subsubsection{Real space-times}

\label{splitsign}

The non-linear graviton construction cannot be applied to
conformally curved Lorentzian space-times, as a real Lorentzian
space-time satisfying (\ref{crflat}) is conformally flat; the
self-dual part of the Weyl curvature is the complex conjugate of
the anti-self-dual part. However, it can be applied to the other
two signatures by constructing a complex space-time and seeking a
suitable real submanifold. The specialisation to Euclidean
space-times gives the construction of general conformally
anti-self-dual spaces. In this case, the twistor space is a
$\CP^1$ bundle over space-time, so that the space-time is obtained
from the twistor space by projection~\cite{AHS}.

In split signature the non-linear graviton construction changes
character, and there are two ways of constructing self-dual
spaces~\cite{Mason5, LeBrunMason}; see also \cite{Mason4}.  For
flat space in this signature, there is a complex twistor space
$\PT=\CP^3$ and a real subspace $\PT_\R=\RP^3$ fixed by the
complex conjugation $\tau: Z\to Z^*$ inherited by twistor space
from that on complex space-time, $x^\m \to ({x^\m})^*$.  There are
two routes to the curved space generalisation.  In the first, one
deforms the complex structure of a region of the complex twistor
space $\PT=\CP^3$ to obtain a curved twistor space $\PTc$ as
before, but in such a way as to preserve the complex conjugation.
The fixed point set $\PTc_\R$ of the conjugation defines an
analogue of $\PT_\R$ in the deformed case and induces a complex
conjugation on space-time that fixes a real slice of split
signature.  In the second, the complex twistor space $\PT=\CP^3$
is kept fixed but the real subspace is deformed from $\PT_\R$ to a
subspace $\PTc_\R$.
     Both
approaches lead to considering deformations of the real twistor
space from $\PT_\R$ to $\PTc_\R $, but this is embedded in
different complex spaces in the two cases.  The two kinds of
deformations are both locally encoded in the same cohomology
classes on the real twistor space, but the second approach is
better behaved globally and does not require analyticity of the
space-time, so it is more powerful. However, it is the first
approach that has
     been used to give a non-linear interpretation of the Berkovits
string theory, in which open strings move in $\PTc$ with
boundaries lying in $\PTc_\R$.  In
\S\ref{Berkovitsstring}, we will propose a modification of the
Berkovits string theory that corresponds to the second approach,
in which there is a natural geometric interpretation of the vertex
operators.  In the first approach, points in space-time correspond
to $\CP^1$'s in $\PTc$ that are invariant under the conjugation,
while in the second they correspond to discs in $\PT$ with
boundary on $\PTc_\R$.

We now describe the two constructions in more detail.  In the
first, the twistor space $\PTc$ was the deformation of a region in
flat twistor space in such a way that the complex conjugation
$\tau:\PTc\to\PTc$ is preserved.  We can construct such a twistor
space starting with a real split signature space-time $\mathcal M$
that is real analytic.\footnote{This assumption is nontrivial as
generic solutions will be non-analytic (this can be seen to follow
from the second construction).  Nevertheless, such non-analytic
solutions can be approximated arbitrarily closely by analytic
ones, and the construction captures the full functional freedom of
these solutions.} The real analyticity can be used to find a
complexification $\CMc$ of the real split-signature space ${\cal
M}$.  This can be found locally by allowing the coordinates to
take complex values, and using the analyticity of the transition
functions for the coordinates we can extend the charts and
transition functions to construct a complex manifold ${\CMc}$
which contains ${\cal M}$ as a real slice (i.e.\ a slice fixed by
complex conjugation of the coordinates we have just constructed).
The analyticity of the metric implies that it can be extended to a
holomorphic metric on $\CMc$.  The complex non-linear graviton
construction of \S\ref{nonlinantiself} can be used locally on any
suitable open set $U\subset \CMc$ to define a twistor space
$\PTc_U$ corresponding to $U$.  The complex conjugation on
space-time again sends $\alpha$-planes to $\alpha$-planes,
inducing a complex conjugation on $\PTc_U$ that fixes a real slice
$\PTc_{U\R}$ which is a totally real 3-dimensional submanifold of
the complex twistor space. A point $x$ in the real space-time
$\Mc$ corresponds to a holomorphic Riemann sphere in the complex
twistor space that intersects $\PTc_{U\R}$ in a circle and cuts
the Riemann sphere into two discs $D^\pm_x$. In the reverse
direction, the complex twistor space can be used to reconstruct a
complex conformally anti-self-dual space as before. This
naturally has a complex conjugation that determines a real slice,
on which the complex conformal structure restricts to give a real
conformally anti-self-dual structure.  In order to construct the
global complex twistor space $\PTc$, we first need to choose a
suitable open cover $\{U_i\}$ of $\CMc$ and construct the twistor
space $\PTc_{U_i}$ for each open set; we then glue these twistor
spaces together, identifying points in $\PTc_{U_i}$ with those in
$\PTc_{U_j}$ whose corresponding $\alpha$-surfaces coincide in
$U_i\cap U_j$.  However, this natural extension gives a $\PTc$
which is a non-Hausdorff manifold~\cite{Mason5}; see the appendix
for a brief description of this space.

In the second approach, we consider general anti-self-dual
conformal structures on $S^2\times S^2$.
     Recall that the conformal compactification of split signature flat space
$\R^{2,2}$ is $S^2\times S^2/\mathbb{Z}_2$, with double cover
$S^2\times S^2$.
       It turns out that there is only the
conformally flat anti-self-dual conformal structure on $S^2\times
S^2/\mathbb{Z}_2$, while there is an infinite dimensional family
of nontrivial such conformal structures on the double cover
$S^2\times S^2$~\cite{LeBrunMason}.  Real points in $S^2\times
S^2$ correspond to Riemann spheres that intersect  the    real
subspace $\PT_\R$, dividing each sphere into  two discs $D^\pm_x$.
       The best way to understand the twistor
theory in this case is to focus on one of the two discs, say
$D^+_x$, rather than the Riemann spheres.

In Euclidean space we were able to represent the twistor space
$\T$ as the bundle of primed spinors $\mathbb{S}$ because we could
solve the incidence relation $\omega^A=x^{AA'}\pi_{A'}$ with
$x^{AA'}=(\omega^A\hat
\pi^{A'}-\hat\omega^A\pi^{A'})/(\hat\p^{B'}\pi_{B'})$ when
$x^{AA'}$ was real. Thus the coordinate transformation between
$(\omega^A,\pi_{A'})$ and $(x^{AA'},\pi_{A'})$ is locally
invertible  and in fact globally invertible if $x^{AA'}=\infty$ is
allowed.  In the context of the double fibration
(\ref{doublefibration}), when the spin bundle $\mathbb S$ is
restricted to the real slice $\M$, the projection $r$ from
$\mathbb{S}$ to $\T$
       is one-to-one and identifies the spin bundle with the
twistor space.

In split signature, with $\pi_{A'}$ complex,
$x^{AA'}=(\omega^A\bar
\pi^{A'}-\bar\omega^A\pi^{A'})/(\bar\p^{B'}\pi_{B'})$ solves the
incidence relation so that  there is locally a one-to-one
correspondence between the points in the bundle of complex spinors
on $\Mc$ and twistor space.  However, this fails where
$\bar\p^{B'}\pi_{B'}$ vanishes, i.~e. when $\pi_{A'}$ is a complex
multiple of a real spinor.  This is because at real values of $x$
and $\pi_{A'}$ there are real $\alpha$-planes, and such planes
correspond to points of $\PT_\R$.  Indeed, the bundle
$\mathbb{S}_\R$ of real spinors is foliated by the lifts of real
$\alpha$-planes to $\mathbb{S}_\R$, with the lifted $\a$-plane
through $(x,\pi_{A'})$ given by the $\a$-plane through $x$ with
tangent spinor $\pi_{A'}$, i.e. the 2-surface
      in $\mathbb{S}_\R$ of the form $(x^{AA'}+ \l^A \p^{A'},\pi_{A'})$ parameterised by $\l^A$.
        Thus, there is a one-to-one identification between
$\mathbb{PS}-\{\bar\p_{A'}\p^{A'}=0\}$ and points in $\PT-\PT_\R$,
but $\PT_\R$ itself is a quotient of $\mathbb{PS_R}$ by its
foliation by $\alpha$-planes.

The set $\mathbb S_0=\{(x,\p_{A'})\in\mathbb S :
\bar\p_{A'}\p^{A'}=0\}$ is a co-dimension-1 hypersurface in
$\mathbb S$ and divides $\mathbb S$ into two halves $\mathbb
S_\pm$ on which $\pm i\bar\p_{A'}\p^{A'} \geq 0$ with common
boundary $\mathbb S_0$. We define the corresponding bundles of
projective primed spinors $\mathbb{PS}_\pm $ and $\mathbb{PS}_0$
by the same conditions on $\bar\p_{A'}\p^{A'}$.  Working now on
$S^2\times S^2$ with a general anti-self-dual conformal structure,
it is still possible to distinguish between $\mathbb{PS}_+$ and
$\mathbb{PS}_-$ globally and we focus on one half, say
$\mathbb{PS}_+$.\footnote{On $S^2\times S^2/\mathbb{Z}_2$, it is
not possible to distinguish between $\mathbb{PS}_+$ and
$\mathbb{PS}_-$; the space-time is not simply connected and, as
one traverses a non-contractible loop, $\mathbb{PS}_\pm$
interchange.}  This is a bundle of discs over $\Mc$ with boundary
$\mathbb{PS}_0$.  It turns out that $\mathbb{PS}_+$ has an
integrable complex structure and is naturally a complex
manifold---in the conformally flat case, $\mathbb{PS}_+$ is
$\PT-\PT_\R$.  The boundary, $\mathbb{PS}_0$, is naturally
foliated by the lifts of real $\alpha$-surfaces in $\Mc$ as in the
conformally flat case and the quotient is $\PTc_\R$, the space of
real $\alpha$-planes.  There is a natural way to glue $\PTc_\R$ to
the boundary of $\mathbb{PS}_+$ to obtain a smooth compact complex
manifold which is a copy of $\CP^3$ topologically.\footnote{This
is done by considering the manifold with boundary $\mathbb{PS}_+
\cup \mathbb{PS}_0$ and compressing each horizontal lift of an
$\a$-plane to a point.}  If the original space-time is smooth, it
can be shown that this gluing can be performed in such a way that
the twistor space has a smooth complex structure.  If our
anti-self-dual conformal structure on $S^2\times S^2$ is a
continuous deformation of the standard conformal structure, then
this twistor space must be the standard $\PT$ because the complex
structure on $\CP^3$ is rigid.  However, the embedding of
$\PTc_\R$ into $\PT$ will be a deformation of the standard
embedding of the real slice $\PT_\R$ inside $\PT$.

The original space-time together with its anti-self-dual conformal
structure can be reconstructed as the moduli space of
holomorphically embedded discs in $\PT$, with boundary in
$\PTc_\R$ in the appropriate topological class~\cite{LeBrunMason}.
The central role played by discs in this approach makes open
string theory seem rather natural.

Linearised deformations of the embedding of $\PT_\R$ in $\PT$
correspond to sections of the normal bundle to $\PT_\R$ over
$\PT_\R$. These can be naturally represented as purely imaginary
tangent vector fields on $\PT_\R$; they can be represented as
vector fields on $\T_\R$ of the form $if^\a\del/\del Z^\a$, where
$f^\a$ is real with homogeneity degree 1, defined up to $f^\a\to
f^\a + Z^\a \Lambda$ for $\Lambda$ of weight $0$.  This freedom
can be fixed with the gauge choice $\del_\a f^\a=0$. The only such vector fields
that give trivial deformations are the generators of
$\mathrm{SL}(4,\C)$.

The non-linear version of this is to define a submanifold $\Tc_\R$
in $\Tc$ by the constraint
\begin{equation}\label{graph}
Z^\a =X^\alpha + iF^\a( X^\a ) ,
\end{equation}
where $ X^\a = Z^\b+\bar Z^\b$ is real and $F^\a$ is a real function of four real variables of 
homogeneity degree one.  Given $\PTc_\R\subset \PTc$, there is
some freedom in the choice of $\Tc_\R$ corresponding to the shift
\begin{equation}
Z^\alpha \rightarrow Z^\alpha = e^{i\theta (X) } \left(  X^\alpha + i F^\alpha \right)
\label{Xshift}
\end{equation}
where $\theta$ is an arbitrary function of $X^\alpha$ of weight $0$; this changes the
non-projective real slice, but not the projective one.  Infinitesimally, (\ref{Xshift})
induces 
\begin{equation}
F^\alpha \rightarrow F^\alpha + \theta (X)  X^\alpha + \ldots . 
\end{equation}
This freedom can
be fixed by imposing that $\mathrm{det}\,( \delta^\a_\b + i
\del_\a F^\b)$ be real.  This implies that
\begin{equation}\label{realdet} \del_\a F^\a = \del_\a
      F^{[\a} \del_\b F^\b \del_\c F^{\c]}, \end{equation} which is an
analogue of the Calabi-Yau condition on $\Tc$.  Clearly, this is a
non-linear generalization of the $\del_\a f^\a=0$ condition above.

Our primary interest in this paper will be in the second
construction described above, but for completeness we give a
discussion of the connection between the two approaches in an
appendix.


\subsection{The Ricci-flat case}

\label{Ricciflatcase}

We now return to complex space-time and suppose that the Ricci
tensor vanishes in addition to $\tilde\psi_{A'B'C'D'}=0$. This is
the case if and only if the full Riemann curvature is
anti-self-dual, and this is equivalent to the condition that the
primed spin connection is flat, so that there exists a two complex
dimensional vector space $\C^2$ of covariantly constant primed
spinor fields.

We saw in \S\ref{nonlinantiself} that each point in $\Tc$
corresponds to an $\alpha$-surface in space-time with a
non-vanishing parallelly propagated tangent spinor field
$\pi_{A'}(x)$ defined over it. If the full Riemann curvature is
self-dual, then a tangent spinor $\pi_{A'}(x)$ on an
$\alpha$-surface is naturally the restriction of a covariantly
constant spinor field on the whole space-time and determined by a
constant spinor $\pi_{A'}\in\C^2$, e.~g.\ the value of the
covariantly constant spinor field $\pi_{A'}(x_0)$ at some point
$x_0$.  Thus we have a projection $p:\Tc\to\C^2-\{0\}$ that takes
an $\alpha$-plane with tangent spinor $\pi_{A'}(x)$ to
$\pi_{A'}(x_0)$.

         We can use this projection to characterise the twistor space
         for a Ricci-flat space-time.  A non-projective twistor space is
         a a complex 4-manifold $\Tc$ satisfying the three conditions
         given in \S\ref{nonlinantiself}. Such a twistor space
         corresponds to a conformally anti-self-dual space-time, and for
         this to be Ricci-flat, the twistor space $\Tc$ must in addition
         have
       \begin{itemize}
     \item a projection $p:\Tc\to\mathbb \C^2-\{0\}$ such that
$p_*\Upsilon =\pi_{A'}\del/\del\p_{A'}$.
\end{itemize}
        This condition arises
because $\Upsilon$ generates scalings of the tangent spinors to
$\alpha$-planes.

       The compatibility of
$\Upsilon$ with the Euler vector field on $\C^2$ means that the
projection descends to $p:\PTc\to\CP^1$, giving a fibration  over
$\CP^1$ of the projective twistor space.\footnote{ Note that the
existence of a projective
        twistor space with a projection to $\CP^1$ is not sufficient to
        reconstruct the projection $p:\Tc\to \C^2$ as, thinking of $\C^2-0$
        as the total space of the $\C^*$ bundle $\caO(-1)$ over $\CP^1$,
        $p^*\caO(-1)$ will not in general be equivalent as a line bundle
        over $\PTc $ to $\Tc\to\PTc$.  Given $p:\PTc\to\CP^1$, in order to
        guarantee that there is a Ricci-flat metric in the conformal
        equivalence class, we need to require that $p^*\caO(-1)$ is an
        equivalent line bundle to $\Tc$ as an independent condition.}  The
fibres are two-dimensional complex manifolds (but have no linear
structure in the curved case, although, as we will see, they do
have certain symplectic and Poisson structures).

In order to clarify these conditions, we can introduce global
coordinates $\pi_{A'}$ on the base $\C^2-0$ of the fibration
$p:\Tc\to\C^2-0$ and use them to build local coordinates
$(\w^A,\pi_{A'})$ on $\Tc$.  These coordinates will be homogeneous
coordinates for $\PTc$.  As $\Tc$ is fibred over $\C^2-0$, the
pull-back of the volume form gives a globally-defined two-form
$\t$ on $\Tc$ given by
\begin{equation*}
         \tau=\frac 12 I_{\a\b}\rd Z^\a \wedge \rd Z^\b =\frac 12 \e
         ^{A'B'}\rd\pi_{A'}\wedge \rd\pi_{B'}\, ,
\end{equation*}
and a holomorphic 1-form
\begin{equation}
\label{kisss} k=I_{\a\b} Z^\a \rd Z^\b =\pi_{A'} \rd\pi^{A'}\,
\end{equation}
on $\PTc$ (and $\Tc$) given by the pull-back of the holomorphic
1-form on $\CP^1$.  We can now restrict our choice of coordinates
$\w^A$ so that
\begin{equation}\label{ }
\rd\Omega = \frac 16\epsilon_{\a\b\c\d}\rd Z^\a\wedge\rd Z^\b
\wedge \rd Z^\c
        \wedge \rd
Z^\d=2 \e_{AB}\rd\w^A\wedge\rd\w^B\wedge\tau \, .
\end{equation}
        This can be expressed as the condition that we have a
holomorphic $(2,0)$ form $\m$ on the fibres given in local
coordinates
       by
\begin{equation}
\label{defmu} \m = \frac 12 \e _{AB}\rd \w^{A}\wedge \rd\w^{B} ,
\end{equation}
where $ \e _{AB}$  is the constant alternating symbol (note that
only contractions of this form with vertical vectors up the fibres
are defined). Then   \begin{equation}\label{omegamu} \rd\Omega=4   \mu\wedge\tau \, , \qquad  \Omega=  2 \mu\wedge
k\, .
\end{equation}
Dually, there is a Poisson structure determined   by a bi-vector
$I^{\a\b}$ and this  is in turn given by $\epsilon^{AB}$, the
inverse of  $\e _{AB}$, by
\begin{equation*}
\{f,g\}_I:=I^{\a\b}\frac{\del f}{\del Z^\a}\frac{\del g}{\del
Z^\b} :=\epsilon^{AB}\frac{\del f}{\del \omega^A}\frac{\del
         g}{\del \omega^B} .
\end{equation*}

Since $\rd\Omega$ and $\tau$ are globally defined by construction,
equation (\ref{omegamu}) implies that $\mu$ is globally defined up
to the addition of multiples of $\rd\pi_{A'}$.  The Poisson
structure $I^{\a\b}$ is globally and unambiguously defined, as the
relation $I^{\a\b}=\frac12 \epsilon^{\a\b\c\d}I_{\c\d}$ determines
it uniquely. We now consider the implications of the condition
that these structures be globally defined. We introduce two
coordinate patches: $U_0$ on which $\pi_{0'}$ does not vanish, and
$U_1$ on which $\pi_{1'}$ does not vanish.  We then introduce
local coordinates `up the fibres' of $p$, $w^A_0$ on $U_0$ and
$w^A_1$ on $U_1$.  These can be elevated to homogeneous
coordinates on the respective patches by defining $\omega^A_0=
\pi_{0'}w_0^A$ and $\omega_1^A=\pi_{1'}w_1^A$. The coordinates are
related in the overlap by the patching relations
\begin{equation*}
\omega^A_0=F^A(\omega^A_1,\pi_{A'})\,
\end{equation*} for some transition function $F^A$, and these
        are required to be homogeneous: $F^A(\l\omega^A_1,\l\pi_{A'})=\l
F^A(\omega^A_1,\pi_{A'})$.
         This means that, as in the flat case, we
can define the homogeneity operator $\Upsilon=Z^\a_0\del/\del
Z_0^\a=Z^\a_1\del/\del Z_1^\a$.

The requirement that the Poisson structure be expressed in its
normal form on each patch is that
\begin{equation*}
\{f,g\}_I=I^{\a\b}\frac{\del f}{\del Z^\a_0}\frac{\del g}{\del
Z_0^\b} =\epsilon^{AB}\frac{\del f}{\del \omega^A_0}\frac{\del
         g}{\del \omega^B_0} = I^{\a\b}\frac{\del f}{\del Z^\a_1}\frac{\del g}{\del Z_1^\b}
=\epsilon^{AB}\frac{\del f}{\del
         \omega^A_1}\frac{\del g}{\del \omega^B_1}\, .
\end{equation*}
A similar condition arises for the $\mu$ and in both cases the
condition amounts to the requirement
\begin{equation}
\label{sdfhjzdsf} \epsilon^{AB}=\epsilon^{CD}\frac{\del F^A}{\del
         \omega^C_1}\frac{\del F^B}{\del \omega^D_1}\,
\end{equation}
that the patching conditions preserve $\epsilon^{AB}$.

Given a global $I^{\a\b}$, the equation
\begin{equation*}\frac12 I^{\a\b}\epsilon_{\a\b\c\d}=I_{\c\d}\end{equation*}
        determines globally the scale of
$\epsilon_{\a\b\c\d}$, and vice versa. Thus, the condition for
Ricci flatness can be expressed as the condition that we have a
global holomorphically defined simple bi-vector $I^{\a\b}$ that
determines a Poisson structure, and we will refer to this as the
infinity twistor, as in the flat case.\footnote{In fact, if we
relax the simplicity condition, we obtain the condition that the
space-time admits an Einstein metric for which the Ricci scalar
can be non-zero. }

An infinitesimal deformation $f^\a$ of the complex structure is an
element of $H^1(\PTc, T^{(1,0)})$, represented either as a \v Cech
cocycle or as a Dolbeault form.  The condition that it preserves
the Poisson structure $I^{\a\b}$ is that it is a Hamiltonian
vector field that can be expressed as
\begin{equation*}
f^\a=I^{\a\b}\frac{\del h}{\del Z^\b}\,
\end{equation*}
for some $ h\in H^1(\PTc,\caO(2))$. This is  the linearised form
of (\ref{sdfhjzdsf}). Whereas the Penrose transform of a general
$f^\a$ subject to the gauge equivalence under $f^\a\rightarrow
f^\a + a(Z) Z^\a$ gives a spin-2 field $\psi _{ABCD}$ satisfying
the higher derivative equation~(\ref{Bacheq1}), the Penrose
transform of $h$ gives a spin-2 field $\psi _{ABCD}$ satisfying
the usual spin-2 equations
\begin{equation}
\label{} \nabla ^{AA'} \psi _{ABCD} =0\, .
\end{equation}


\subsubsection{Ricci-flat case in split signature}

\label{splitsign2}

In the second of the two approaches to the split signature
non-linear graviton construction, the complex twistor space is
taken to be $\PT=\CP^3$, and conformally anti-self-dual
space-times are constructed from deformations of a real slice
$\PTc_\R$, which is itself an arbitrary small deformation of the
real subspace $\RP^3$. However, in the Ricci flat case, $\PTc_\R$
is no longer an arbitrary deformation; instead it is subject to
certain conditions as will now be explained.

Again we take $\T$ to have an infinity twistor $I^{\a\b}$ defined
on it, and this determines a projection from
$\T'=\T-\{\pi_{A'}=0\}$ to $\C^2-0$ given by $Z^\a \rightarrow
\pi_{A'}$ together with the corresponding projection
$p:\PT'\rightarrow \CP^1$. This should be compatible with the real
slice in the sense that $\PTc_\R$ should project to
$\RP^1\subset\CP^1$. Equivalently, $\PTc_\R$ should lie inside the
real codimension-1 hypersurface $\Sigma:=p^{-1}(\RP^1)\subset
\PT'$, which can also be defined by the equation
$\pi_{A'}\bar\pi^{A'}=0$ with $\bar
\pi_{A'}=(\bar\pi_{0'},\bar\pi_{1'})$ the standard complex
conjugation. This is the analogue of the existence of the
projection $p:\PTc\rightarrow\CP^1$ and we need to express the
second
part of the
condition for Ricci flatness in this context.

On $\PT'$ the line bundles $\caO(n)$ of homogeneous functions of
degree $n$ are equal to the pull-backs of the corresponding line
bundles from $\CP^1$. Thus, on $\Sigma$, the complex line bundles
$\caO(n)$ naturally have a fibrewise complex conjugation fixing
the real sub-bundles $\caO_\R(n)$, which are the pull-backs of the
corresponding real sub-bundles of $\caO(n)$ on $\RP^1$ (i.~e.\
these real line sub-bundles are spanned by homogeneous polynomials
of degree $n$ in $\pi_{A'}$ with real coefficients).

The second condition necessary in order that $\PTc_\R\subset \PT$
corresponds to a Ricci-flat anti-self-dual conformal structure is
that the $\caO(4)$-valued 3-form $\Omega$, when restricted to
$\PTc_\R$, lies in $\caO_\R(4)$, or equivalently that the
restriction to $\PTc_\R$ of the $\caO(2)$-valued 2-form $\m =
\frac 12\rd \omega^A\wedge \rd\omega_A$ up the fibres is real.
This can be stated geometrically by observing first that, on each
4 real-dimensional fibre of $p$ over $\RP^1$, the form $\m$
defines a complex symplectic form with values in $\caO(2)$, and
its imaginary part defines a real symplectic form $\varpi $ with
values in $\caO_\R(2)$. Our requirement is then that on each fibre
$p^{-1}(\pi_{A'})$ of $p$ over $\RP^1$, the intersection of
$\PTc_\R$ with $p^{-1}(\pi_{A'})$ should be Lagrangian with
respect to $\varpi $, i.e., $\varpi|_{\PTc_\R\cap
p^{-1}(\pi_{A'})} =0$ for each $\pi_{A'}$.  This will guarantee
that $\m$ is real on restriction to $\PTc_\R$, since we have
required that the restriction of its imaginary part $\varpi$
vanishes; it then follows from equation (\ref{omegamu}) that
$\Omega$ is real.

An infinitesimal deformation of $\PTc_\R$ preserving this
condition is therefore generated by a Hamiltonian vector field
preserving $\m$, and so it is determined by a Hamiltonian function
$h$ which will be a global section of $\caO_\R(2)$ defined over
$\PTc_\R$ (a finite deformation can then be obtained from a
generating function).

To be more explicit, we can decompose $\omega^A$ into its real and
imaginary parts, $\omega^A=\omega_R^A+i\omega_I^A$ where
$\omega_R^A$ and $\omega_I^A$ are real;  then $\varpi =
2\rd\omega_R^A\wedge \rd\omega_{IA}$.  Assuming the deformation to
be transverse to $\del/\del \omega_I^A$, we can express $\PTc_\R$
in $\Sigma$, on which $\pi_{A'}$ is real, as the graph
\begin{equation*}
\omega^A_I=F^A(\omega_R^A, \pi_{A'}) \, ,\end{equation*} where
$F^A$ has homogeneity degree one. Then the Lagrangian condition is
\begin{equation*}
\frac{\del}{\del \omega_R^A} F^A=0\, .
\end{equation*}
These conditions can be solved by introducing a smooth real
function $H(\omega_R^A,\pi_{A'})$ on $\T_\R$ of homogeneity degree
two and defining
\begin{equation*}
F^A(\omega_R^A, \pi_{A'})=\epsilon^{AB}\frac{\del H}{\del
\omega_R^B} \, .
\end{equation*}
It can be seen that this automatically incorporates the condition
     (\ref{realdet}).

Infinitesimally,  a deformation of   $\PT_\R$ to $\PTc_\R$ is
given by pushing $\PT_\R$ along the vector field
\begin{equation*}
if^\a(Z_R^\b)\frac{\del }{\del Z_I^B} =iI^{\a\b}\frac{\del h}{\del
Z_R^\a} \frac{\del }{\del Z_I^B} = i\epsilon^{AB}\frac{\del
h}{\del \omega_R^A} \frac{\del }{\del \omega_I^B}\, ,
\end{equation*}
where we have written $Z^\a=Z_R^\a+iZ_I^\a$ for $Z_R^\a$ and
$Z_I^\a$ real, and $h=h(Z_R^\a)$ is the infinitesimal analogue of
$H$. The vector field is understood to be a normal vector field to
the real slice, so it can be taken to be imaginary.

As a final note, we observe that the hypersurface $\Sigma$ divides
$\PT$ into two halves $\PT^\pm$ according to $\pm i
\pi_{A'}\bar\pi^{A'} >0$. The holomorphic discs in $\PT$ with
boundary on $\RP^3$ divide into those that lie entirely in
$\Sigma$, and those that lie in one of $\PT^\pm$. Those in
$\PT^\pm$ correspond to two distinct copies $\M^\pm$ of space-time
$\R^4$, whereas those in $\Sigma$ correspond to points at (null)
infinity. We will wish to work with just one copy of
       space-time,
so we discard $\PT^-$ and work only with the holomorphic discs in
$\PT^+$
and hence just the one copy $\M^+$ of space-time.

\subsubsection{Superspace, super-twistor space and anti-self-dual supergravity}

\label{supersuper}

We can consider deformations of super-twistor space $\PT'_{[N]}$ to
obtain anti-self-dual solutions to the conformal supergravity
equations. The formal definition of such a deformed complex
supermanifold has been studied in the mathematics literature
\cite{green, eastwood-lebrun}.  Here we use the
       more general physics formulation in which both fermionic
       coordinates and fermionic constants are allowed.  A
supermanifold is constructed by patching together coordinate
charts $\{U_i\}$ with coordinates $Z^I_i=(Z_i^\a,\psi_i^a)$ on
each patch, where the $Z^\a_i$ are bosonic and the $\psi^a_i$
fermionic.  On the overlaps, the coordinates are   related by
patching functions
\begin{equation*}
Z^I_i:=(Z_i^\a,\psi_i^a)=P_{ij}^I(Z_j^J):=(P_{ij}^\a(Z^J_j),
P^a_{ij}(Z_j^J))\, ,
\end{equation*}
where $P^\a_{ij}$ is an even function, and $P^a_{ij}$ is
odd.\footnote{Here fermionic  parameters are allowed in these
functions. }  We also require that the matrices $\del
P^I_{ij}/\del Z^J_j$ have non-zero super-determinant (in fact, it
must be possible to choose coordinates so that it is equal to 1 in
the $N=4$ case for which the super-twistor spaces are
super-Calabi-Yau; note that our projective twistor spaces are not
Calabi-Yau for general $N$).

A complex supermanifold, e.~g.\ $\PTc_{[N]}$, is composed of an
underlying ordinary complex manifold, $\PTc$ (the \lq body') with
patching functions $P_{ij}^\a(Z_j^\b,0)$ with all anti-commuting
coordinates and parameters set to zero, and a rank $N$ vector
bundle $E\to \PTc$ (the `soul') whose patching functions are $\del
P^a_{ij}/\del \psi_j^b|_{\psi_j^b=0}$, again with all odd
parameters set to zero. It is an important feature of generic
complex  supermanifolds that they are not in general obtained by
simply reversing the Grassmann parity of the coordinates up the
fibres of the vector bundle $E\to \PTc$ (whereas this is the case
for real  supermanifolds).  The higher derivatives of the patching
functions with respect to odd variables encode information that
cannot be gauged away.

One necessary restriction for a complex  supermanifold to be a
super-twistor space is the requirement that the $\psi^a$ have
homogeneity degree 1.  One way of expressing this is to say that
the bundle $E$ should have degree $-N$ (i.~e.\ first Chern class
$-N$). As discussed earlier, the space $\CMc$ of rational curves
in $\PTc$ in the appropriate topological class will be a
space-time with anti-self-dual conformal structure.  These
rational curves will have deformations away from the body, and
their moduli space $\CMc_{[N]}^+$ will be chiral superspace with
body $\CMc$.  The full superspace is obtained as the space of
flags $\CP^{1|0}\subset\CP^{1|N}$ in $\PTc_{[N]}$, with the chiral
and anti-chiral superspaces arising as the space of $\CP^{1|0}$s
and $\CP^{1|N}$s respectively.  We are not aware of a full
presentation of this construction in the literature, and to give
one here would take us too far afield.

An infinitesimal deformation of $\PTc_{[N]}$ can be obtained by
varying the patching functions, and such an infinitesimal
variation is given in local coordinates on the overlap of two
coordinate charts by a tangent vector $f=f^\a\del/\del Z^\a_i +
f^a\del/\del\psi_i^a$, where $f^\a$ is even and $f^a$ is odd.  To
deform the complex structure, we use such a vector field on each
overlap and a nontrivial deformation is defined modulo
infinitesimal coordinate transformations on the open sets; thus
the nontrivial deformations are parametrised by the cohomology
group $H^1(\PT'_{[N]}, T^{(1,0)})$, where $T^{(1,0)}$ is (the
sheaf of sections of) the holomorphic tangent bundle of the
 supermanifold.
       This group was studied in the case of $N=4$ in
\cite{BWsc} and the spectrum of $N=4$ conformal supergravity was
obtained (see the end of section~\ref{Berkovitsstring}). A similar
analysis can be carried out for other values of $N$.

In order to obtain an anti-self-dual version of Einstein
supergravity, we need to impose the supersymmetric analogues of
the constraints imposed on a twistor space to obtain Ricci-flat
anti-self-dual four-manifolds as described in
\S\ref{Ricciflatcase}.  There is now some ambiguity because, in
the supersymmetric case, the restriction to Poincar\'e invariance
gives a projection  to   $\CP^{1|N}$ and hence also to
$\CP^{1|0}$.  In order to obtain a straightforward supermultiplet
starting from helicity $-2$ and increasing to helicity $(N-4)/2$
in the linearised theory, we require that we have a projection
\be\label{superfibre} p_1:\PTc_{[N]}\rightarrow \CP^{1|N} \ee (and
thence a further projection $p:\PTc_{[N]}\rightarrow \CP^{1|0}$)
and a global holomorphic volume form $\Omega_s$ with values in the
pull-back of $\caO(4-N)$ from $\CP^{1|0}$.

To make this more explicit, we introduce the non-projective
super-twistor space $\Tc_{[N]}$, which as before can be defined as
the total space of the pull-back of the line bundle $\caO(-1)$
from $\CP^1$ using $p$. The projection $p_1$ then determines a
projection $p:\Tc_{[N]}\rightarrow \C^{1|N}$. We can introduce
coordinates $(\pi_{A'}, \psi^a)$, $A'=0',1'$, $a=1\ldots, N$ on
$\C^{1|N}$ and complete these to a local coordinate system $Z^I$
on $\Tc_{[N]}$ by adjoining local coordinates $\omega^A$ ($A=0,1$)
of homogeneity degree 1.

In this case we can define \lq infinity twistors' $I_{IJ}$ and
$I^{IJ}$ on the non-projective twistor space $\Tc_{[N]}$ by
setting
\begin{eqnarray*}
I_{IJ}\rd Z^I\wedge \rd Z^J
&=& \rd\pi^{A'}\wedge\rd\pi_{A'}\, ,\\
I^{IJ}(\rd \Omega_s)_{IJ K_1\ldots K_{N+2}}\rd Z^{K_1}\ldots \rd
Z^{K_{N+2}} &=& I_{IJ}\rd Z^I\wedge \rd Z^J \Pi_{a=1}^N\rd\psi^a\,
.
\end{eqnarray*}

It is now straightforward to see that deformations of super-twistor
space preserving these structures must be of the form
\begin{equation*}
f^I\frac{\partial}{\partial Z^I}= I^{IJ}\frac{\partial h}{\partial
         Z^I}\frac{\partial}{\partial Z^J} \, , \qquad h\in H^1(\PTc_{[N]},
\caO(2))\, .
\end{equation*}
Such an $h$ precisely describes an anti-self-dual supergravity
multiplet, starting with helicity $2$ and going down to helicity
$(4-N)/2$; this will be discussed in more detail in
section~\ref{sdgravN}.

It is also possible to consider deformations of $\PT'_{[N]}$ that
preserve less structure. For example, later we will consider the
case where we only preserve the projection
$p:\PTc_{[N]}\rightarrow \CP^1$. In such cases, the space of
possible deformations will be larger and correspond to more fields
on space-time.

\section{The Berkovits twistor string}

\label{Berkovitsstring}

\subsection{The Berkovits open string theory}

The Berkovits string is a theory of maps from the world-sheet
$\Sigma$ to a curved super-twistor space with coordinates
$Z^I=(\omega^A , \pi_{A'}, \psi^a )$, $\ti Z^I=(\ti \omega^A , \ti
\pi_{A'},\ti \psi^a )$.  In the following, we will find it useful
to use a notation that can handle different signatures and
different reality properties in a unified way.  There are three
different cases that we will consider:
\begin{itemize}
\item[(i)] $Z^I$ are complex coordinates on a complex super-twistor space
$\Tc$ and $ \ti Z^I$ are the complex conjugate coordinates $\tilde
Z= (Z)^*$,
\item[(ii)] $Z^I, \ti Z^I$ are independent real coordinates on a
space $\Tc_\R \times \Tc_\R$ for some real twistor space $\Tc_\R$,
\item[(iii)] $Z^I, \ti Z^I$ are independent complex coordinates on a space
$\Tc \times \Tc$ for some complex twistor space $\Tc$.
\end{itemize}
For space-times of signature $++++$ or $+++-$, the twistors are
necessarily complex, while for signature $++--$ either complex or
real twistors can be used. In the flat case, $Z^I, \ti Z^I$ are
complex conjugate coordinates on $\C ^{4|4}$, real coordinates on
$\R ^{4|4}\times \R ^{4|4}$, or complex coordinates on $\C
^{4|4}\times \C ^{4|4}$; then we write $Z^I=(\omega^A , \pi_{A'},
\psi^a )$, $\ti Z^I=(\ti \omega^A , \ti \pi_{A'},\ti \psi^a )$.
For open strings in any of the three cases, the boundary of the
world-sheet $\partial \Sigma$ is constrained to map to the
submanifold defined by $Z=\ti Z$. For case (i) with complex $Z$,
this is the real submanifold $\PTc_\R$ that arose in
\S\ref{splitsign}.

     We use
world-sheet coordinates $\s, \ti \s$ with world-sheet metric
$ds^2=2d\s d\ti \s$. For Euclidean world-sheet signature, $\s,\ti
\s$ are complex conjugate coordinates $\ti \s = \s^*$ while for
Lorentzian world-sheet signature, $\s,\ti \s$ are independent real
null coordinates.


The fields include maps $Z^I(\s,\ti \s),\ti Z^I(\s,\ti \s)$ from
the world-sheet to super-twistor space and these are world-sheet
scalar fields. The action is
\begin{equation}\label{berkoS}
        S = \int d^2 \s \left( Y_{I} \ti \partial Z^I + \ti Y_{J} \partial
\ti Z^J -\ti AJ- A\ti J \right) + S_C ,
\end{equation}
where $Y_I , \ti Y_I $ are conjugate momenta of  conformal
dimensions $(1,0) $ and $(0,1)$ respectively and $\pa= \pa/\pa
\s$, $\ti \pa= \pa/\pa \ti \s$. The world-sheet gauge fields
$A,\ti A$ couple to
       currents
\begin{equation}\label{defH}
        J = Y_{I} Z^I , \hspace{2cm} \tilde J = \tilde Y_{I} \tilde Z^I
        ,
\end{equation}
so that there is a
       local symmetry
\begin{equation}
\nonumber
        Z^I \rightarrow tZ^I, \hspace{1.5cm} Y_{I} \rightarrow \frac{1}{t}
        Y_{I}, \hspace{1.5cm} \tilde Z^I \rightarrow \ti t \tilde Z^I ,
        \hspace{1.5cm} \tilde Y_{I} \rightarrow \frac{1}{\ti t} 
        \tilde Y_{I} ,
        \end{equation}
        \begin{equation}
       \ti A \rightarrow \ti A+ \frac{1}{t}\ti \partial t ,\hspace{1.5cm}
       A \rightarrow A+ \frac{1}{\ti t} \partial \ti t .
       \label{Gl1ab}
\end{equation}
This symmetry ensures that the theory projects to one defined on a 
projective twistor space $\cPT$,  $\PTc_\R \times \PTc_\R$ or
$\PTc \times \PTc$.

The action is real for Euclidean world-sheets if one chooses case
(i) above, all variables are complex, and the tilde operation is
complex conjugation, so that for any field $\Phi$, $\ti \Phi =
\Phi^* $. For Lorentzian world-sheets the action is real if all
variables are real, requiring signature $++--$, and $\Phi ,\ti
\Phi$ are independent real variables. For Euclidean world-sheets
the parameter $t$ is complex and the gauge symmetry~(\ref{Gl1ab})
is $GL(1,\C)$ while for Lorentzian world-sheets $t,\ti t$ are
independent real parameters and the gauge group is $GL(1,\R)
\times GL(1,\R)$.  For the case of Lorentzian world-sheets in
which $\Phi ,\ti \Phi$ are independent real variables, \lq Wick
rotation' gives a theory on Euclidean world-sheets in which $\Phi
,\ti \Phi$ become independent complex variables, leading to case
(iii) above, and it is the action of this theory that is used in
the Euclidean path integral.

The term $S_C$ in~(\ref{berkoS}) is the action for an additional
matter system which is a conformal field theory with Virasoro
central charges $c_C=\ti c_C$ and currents $j^r$ and $\ti j^r$,
for $r=1,\ldots \dim G$.  Here $G$ is some group whose
Ka\v{c}-Moody algebra is generated by the currents.  The
Ka\v{c}-Moody central charges are denoted by $k=\ti k$ and the
group $G$ becomes a Yang-Mills gauge group in space-time.

Open strings are included in the model with the boundary
conditions
\begin{equation}\label{bound}
       Z^I = \tilde Z^I , \hspace{2cm} Y_{I} = \tilde Y_{I} , \hspace{2cm}
       j^r = \tilde j^r
\end{equation}
on $\del\Sigma$. For complex $Z$ with $\ti Z=Z^*$, the string
endpoints lie in a real subspace $\cT_\R$ of $\cT$, which projects
onto  a real subspace
       $\cPT_\R$ of
$\cPT$. In the flat case, this is $\RP^{3|4}\subset \CP^{3|4}$
and~(\ref{bound}) breaks the $SL(4|4;\C)$ symmetry to
$SL(4|4;\R)$. This boundary condition is natural for the case of
split space-time signature $++--$, where the real subspace plays a
natural and important role, as was discussed in~\S\ref{realspace}
and \S\ref{splitsign}. As the interpretation of the results for
other signatures is less clear, we will restrict ourselves to the
split space-time signature $++--$ in what follows. For independent
real $Z, \ti Z$ and split space-time signature, the ends of the
strings lie in the diagonal $\cPT_\R$ in $\cPT=\cPT_\R\times
\cPT_\R$. For the flat twistor space $\PT=\RP^{3|4} \times
\RP^{3|4}$, the endpoints lie in the diagonal $\RP^{3|4}$,
breaking the conformal symmetry from $SL(4|4;\R)\times SL(4|4;\R)$
to the diagonal subgroup. In either case, the boundary theory
lives on a real twistor space $\cPT_\R$ (which is $\RP^{3|4}$ in
the flat case) and the scaling symmetry is broken to $GL(1,\R)$ by
the boundary conditions.

Quantisation gives the usual conformal gauge ghosts $(b ,c )$ and
$(\tilde b , \tilde c)$ together with $GL(1)$ ghosts $(u ,v )$
       and $(\tilde u , \tilde v
)$ ($v$ and $\ti v$ have conformal dimensions $(0,0)$, while $u$
and $\ti u$ have dimensions $(1,0)$ and $(0,1)$). Variables $\ti
\f$ with a tilde are right-moving ($\pa \ti \f=0$), while those
without are left-moving ($\pat \f=0$). The matter
        stress-energy tensor is
\begin{eqnarray}\label{Tmatter}
        T^m & = & Y_{I} \del Z^I + T^C \nonumber \\
         \tilde T^m & = & \tilde
        Y_{I} \ti \del \tilde Z^I + \tilde T^C ,
\end{eqnarray}
where $T^C$ and $\tilde T^C$ are the left and right-moving
stress-energy tensors for the current algebra. The stress-energy
tensor for the ghosts is
\begin{eqnarray}\label{Tghost}
        T^{gh} & = & b \del c + \del \left( b c \right) +u \del v
        \nonumber \\
       \tilde T^{gh} & = & \tilde b \ti \del \tilde c + \ti \del
       ( \tilde b \tilde c ) +\tilde u \ti \del \tilde v
        .
\end{eqnarray}
The open string theory is defined by the boundary
conditions~(\ref{bound}) on the twistor variables, together with
additional boundary conditions on the ghosts:
\begin{equation}\label{boundgh1}
        c = \tilde c , \hspace{1cm} b = \tilde b , \hspace{1cm} v = \tilde v ,
        \hspace{1cm} u = \tilde u .
\end{equation}

The BRST charges are
\begin{eqnarray}\label{defBRS1}
        Q & = & \oint d\sigma \left( c T +v J + c u \del v + c b \del c
        \right) \nonumber \\
        \tilde Q & = & \oint d\ti \sigma \left( \tilde c
        \tilde T +\tilde v \tilde J + \tilde c \tilde u \ti \del \tilde v +
        \tilde c \tilde b
         \ti \del
          \tilde c
        \right) \nonumber \\
\end{eqnarray}
and they are nilpotent provided the additional matter system has
Virasoro central charge $c_C=28$; this value cancels the
contributions $c=-26$ of the $(b ,c)$ system and $c=-2$ of the $(u
,v )$ system to the conformal anomaly. There is no $GL(1)\times
GL(1)$ anomaly because of cancellation between bosons and
fermions.

The physical open string states are BRST cohomology classes
represented by vertex operators that are $GL(1)$ neutral and are
dimension one primary fields with respect to the Virasoro and
Ka\v{c}-Moody generators~(\ref{Tmatter}), (\ref{Tghost})
and~(\ref{defH}). The super-Yang-Mills vertex operators are the
dimension one operators constructed with Ka\v{c}-Moody currents of
the auxiliary matter system~\cite{NB}:
\begin{equation}\label{VYM}
    V_\phi = j_r \phi^r (Z) ,
\end{equation}
where the $\phi ^r(Z) $ are any Lie-algebra-valued functions that
are invariant under scalings of $Z^I$ (i.~e.\ any
Lie-algebra-valued functions on $\RP^{3\mid 4}$) and have
conformal weight zero. The dimension one vertex
operators~\cite{BWsc}
\begin{equation}\label{Vgrav}
        V_f = Y_I f^I (Z), \hspace{2cm} V_g = g_I (Z) \del Z^I
\end{equation}
        are $GL(1)$-invariant provided the functions $f^I$ carries $GL(1)$ charge
$+1$ (i.~e.\ it is in $\caO (1)$) and $g_I$ carries $GL(1)$ charge
$-1$ (i.~e.\ it is in $\caO (-1)$). They will be physical if the
$f^I$ and $g_I$ satisfy
\begin{equation}\label{fgprim}
        \del_I f^I = 0 , \hspace{2cm} Z^I g_I = 0.
\end{equation}
Changing $f^I,g_I$ by
\begin{equation}\label{gaugefg}
        \delta f^I = Z^I \Lambda , \hspace{2cm} \delta g_I = \del_I \chi ,
\end{equation}
gives operators in the same BRST cohomology class as those given
in ~(\ref{Vgrav}), so that (\ref{gaugefg}) are gauge invariances
giving physically equivalent states~\cite{NB,BWsc}. The vertex
operators (\ref{Vgrav}) give the states of
       conformal supergravity~\cite{BWsc}.

Since $f^I$ has $GL(1)$ charge 1, the vector field
\begin{equation}\label{Upf}
        f=f^I \frac{\del}{\del Z^I}
\end{equation}
on $\cT$ is invariant under scaling, and the first equivalence
relation in~(\ref{gaugefg}) means that $f$ can be interpreted as a
vector field on $\cPT$~\cite{BWsc}. The first constraint
in~(\ref{fgprim}) means that $f$  is a volume-preserving vector
field. The second constraint in~(\ref{fgprim}) means that the
one-form
\begin{equation}\label{defThet}
        g= g_I dZ^I
\end{equation}
is well-defined on $\cPT$~\cite{BWsc}. The second gauge
equivalence in~(\ref{gaugefg}) means that $g$ is an abelian gauge
field.

The functions $\phi^r (Z)$ in~(\ref{VYM}) are superfields which
can be expanded in terms of ordinary functions on twistor space
with values in the line bundles $\caO (0),\caO (-1),\caO
(-2)$, $\caO (-3)$, $\caO (-4)$. By the Penrose transform, these
represent fields of helicities $(1, \frac{1}{2} ,0, -\frac{1}{2} ,
-1)$ with the correct R-symmetry representations to describe $N=4$
super-Yang-Mills states~\cite{Witten2003,NB}. Likewise, the
spectrum of Minkowski space helicity states associated with the
vertex operators~(\ref{Vgrav}) follows from the
       expansions of the superfields $f^I (Z)$ and
$g_I (Z)$ in powers of $\psi$~\cite{BWsc}. The analysis
of~\cite{BWsc} shows that, taking (\ref{fgprim},\ref{gaugefg})
into account, $f^A(Z)$ and $f^{A'}(Z)$ each describe the helicity
states $(+2,+\frac{3}{2},+1,+\frac{1}{2},0)$ of an $N=4$
supergravity multiplet (with the correct R-symmetry
representations) while $f^a(Z)$ describe the helicity states $(
+\frac{3}{2},+1,+\frac{1}{2},0, -\frac{1}{2})$ of (four) gravitino
multiplets. Similarly, $g_A, g_{A'}$ give two supergravity
multiplets with negative helicities $ (0, -\frac{1}{2},
-1,-\frac{3}{2},-2)$ and $g_a$ give (four) gravitino multiplets
       $(+\frac{1}{2},0, -\frac{1}{2}, -1,-\frac{3}{2})$.
Taken together, the space-time fields described by the vertex
operators $V_f$ and $V_g$ given in~(\ref{Vgrav}) can   be
identified with the physical states of $N=4$ conformal
supergravity.

\subsection{Generalised boundary conditions}

In split signature, the non-linear graviton can be constructed
from deformations of a real subspace $\PTc_\R$ in a fixed flat
twistor space $\PT$, as was reviewed in \S\ref{splitsign}. This
suggests a modification of the Berkovits string model in which,
for the case (i) of complex $Z$, the strings live in $\PT$  and
the open string boundaries are constrained to lie in the general
subspace $\PTc_\R$ defined in terms of functions $F^\a$ by
(\ref{graph}) instead of the real subspace defined by the
condition $Z= Z^*$.  We then consider a string theory
      in which the boundary condition $Z^I=\tilde Z^I$ is
replaced with
\begin{equation}\label{sgraph}
Z^I-\tilde Z^I=\hat F^I(Z^J+\tilde Z^J)\,
\end{equation}
for some function $\hat F^I$ of homogeneity degree one. There is a gauge freedom in the definition of $F$, which can be
multiplied by a function of homogeneity degree $0$ (see also the
discussion following equation (\ref{graph})).  This can be fixed
by
       imposing  the condition that $\mathrm{sdet} (\delta_J^I + \del_J
       \hat F^I)=
        \mathrm{sdet}  (\delta_J^I - \del_J \hat F^I)$ where sdet
        denotes the super-determinant.
This is the condition that the Calabi-Yau forms $\rd \Omega$ in
$Z^\a$ and in $\tilde Z^\a$ agree.
       The corresponding boundary
          conditions for $Y$ are found by requiring the surface term in the
          variation of the action to vanish.  Varying the
          action~(\ref{berkoS}) gives terms proportional to the field
          equations together with a surface term
\begin{equation}
\label{} \int_{\del \Sigma} (Y_I\delta Z^I-\tilde Y_I\delta\tilde
Z^I) =\frac 12\int_{\del \Sigma} \left[ (Y_I-\tilde Y_I)(\delta
Z^I+\delta \tilde Z^I)+(Y_I+ \tilde
       Y_I)(\delta Z^I-\delta\tilde Z^I) \right]\, ,
\end{equation}
where the boundary  $\del \Sigma$ is specified by $\s +\ti \s =0$.
Using equation (\ref{sgraph}),  this will vanish if
       the boundary conditions for $Y$ are modified to become
       \begin{equation} \label{genboundY1}
       Y_J-\tilde
Y_J=-\hat F^I{}_{,J}(Y_I+\tilde Y_I)\, .
\end{equation}

In the cases (i) or (iii) above in which $\tilde Z^\a$ and $Z^\a$
are independent quantities,
       the deformation of the boundary condition
amounts to a deformation of the location of the diagonal subspace
inside $\PT_\R\times \PT_\R$ or $\PT\times \PT $ where the
world-sheet boundary is constrained to lie. In the complex case
(ii) in which $Z$ is complex and $\tilde Z =(Z)^*$ and the
boundary is the real axis $\s =  \s ^*$, it is useful to write
$\hat F =i F$ so that (\ref{sgraph}) becomes
\begin{equation}\label{sgraph1}
Z^I-\bar Z^I=iF^I(Z^J+\bar Z^J)\, ,
\end{equation}
where $\mathrm{sdet} (\delta_J^I + i\del_J F^I)$ is constrained to
be
        real (in order to fix the gauge freedom). This is a
        supersymmetric version of (\ref{graph}), and the boundary condition
        (\ref{genboundY1}) becomes
       \begin{equation} \label{genboundY11}
       Y_J-\bar
Y_J=-iF^I{}_{,J}(Y_I+\bar Y_I)\, .
\end{equation}

With these boundary conditions, the  worldsheets of degree $1$
        correspond to points of the compactified space-time $S^2\times S^2$, and this  has the non-trivial
split signature anti-self-dual conformal structure determined by
$F^I$.  The construction of \S\ref{splitsign} then suggests that
the geometric interpretation of the vertex operator $V_f = Y_If^I$
should be that $f^I$ determines an infinitesimal variation in
$F^I$, and so deforms the  boundary conditions.

Next we turn to the interpretation of the vertex operator $V_g =
g_I\del Z^I$. If one adds a boundary term
\begin{equation}
\label{} \int_{\del \Sigma} G_I(Z^J+\tilde Z^J) \del (Z^I+\tilde
Z^I)
\end{equation}
to the action (\ref{berkoS}), for some  $G_I=G_I(Z^J+\tilde Z^J)$,
then the condition that the surface term in the variation of the
action vanishes is now
\begin{equation}
\label{genboundY2}
       Y_J-\tilde Y_J=-\hat F^I{}_{,J}(Y_I+\tilde Y_I) +
2G_{[I,J]}\del (Z^J+\tilde Z^J)\, ,
\end{equation}
so that the surface term leads to a modification of the boundary
conditions for $Y$.
       Then the vertex
operator $g_I\del Z^I$ corresponds to a deformation of $G_I$.

The quantisation of the string models based on the generalised
boundary conditions~(\ref{sgraph}) and~(\ref{genboundY2}) will be
discussed elsewhere.

\section{Gauged $\beta$-$ \gamma$ systems}

\label{betagamma}

\subsection{1-form symmetries}

\label{1formsymm}

The system (sometimes referred to as a $\beta$-$ \gamma$ system)
\begin{equation}\label{bg}
        S = \int d^2 \s Y_{I} \ti \partial Z^I ,
\end{equation}where the $Z^I$ are coordinates on some manifold (or supermanifold)
$M$, has recently been discussed in~\cite{Wittenbetgamm,Nekrasov}.
The Berkovits twistor string has kinetic terms of this form, with
super-twistor space as the target space. If $k^i = k^i_IdZ^I$ are
1-forms on $M$ labeled by an index $i$, $i=1,\ldots p$, then the
chiral currents
\begin{equation}
\label{kcur} K^i = k^i_I\partial Z^I
\end{equation}
are conserved:
\begin{equation}
\label{kcurfeq} \ti \partial K^i =0 
\end{equation}
and generate a symmetry with parameters $\a_i(\s)$ satisfying $\ti
\partial \a_i =0$,
\begin{equation}
\label{varal} \d Z^I=0, \qquad \d Y_I = k^i_I \partial \a_i+ 2 \a
_i k^i_{[I,J]} \partial Z^J .
\end{equation}
The rigid symmetry with constant parameters was discussed
in~\cite{Wittenbetgamm}. Both bosonic and fermionic local
symmetries can be considered, and below we consider models
       with $d$ bosonic currents and $n$ fermionic currents and $p=d+n$. The currents $K^i$ commute, so they
satisfy an abelian Ka\v{c}-Moody algebra with vanishing central
charge:
\begin{equation}
\label{kcuralg} [K^i(\s), K^j(\s')]=0 .
\end{equation}

This can be promoted to a local symmetry by coupling to gauge
fields $\ti B_i$ to give the action
\begin{equation}\label{bgB1}
        S = \int d^2 \s \,  \left( Y_{I} \ti \partial Z^I - \ti B_i K^i \right) ,
\end{equation}
which is invariant under (\ref{varal}) and
\begin{equation}
\label{varb} \d \ti B_i= \ti \partial \a_i
\end{equation}
for general local parameters $\a_i (\s, \ti \s)$. Gauge-fixing and
introducing ghosts $s_i$ and anti-ghosts $r^i$ gives the action
\begin{equation}\label{bggf}
        S = \int d^2 \s \left( Y_{I} \ti \partial Z^I + r^i \ti \partial
        s_i\right)
        ,
\end{equation}
and the BRST charge
\begin{equation}
\label{} Q = \oint d\sigma \, s_i K ^i\end{equation} is nilpotent.

For the vertex operator $V_f=f^I Y_I$,
\begin{equation}
\label{} [Q,V_f]=(\partial s_i)f^I k^i_I+2 s_i f^I k^i_{[I,J]}
\partial Z^J =\partial (s_if^I k^i_I) -s_i[ {\cal L}_f k^i]
_I\partial Z^I
\end{equation}
and so $f^IY_I$ is BRST invariant provided
\begin{equation}
\label{} f^I k^i_I=0, \qquad f^I k^i_{[I,J]}=0 ,
\end{equation}
while the integrated vertex operator $\int V_f$ is invariant (up
to a surface term) provided the Lie derivative of $k^i$ with
respect to the vector field $f$ vanishes,
\begin{equation}
\label{}
       {\cal L}_f k^i=0 .
\end{equation}
Changing the vertex operator $g_I \partial Z^I$ by a BRST exact
term leads to the symmetry
\begin{equation}
\label{} \d g_I = \h_i k_I^i
\end{equation}
for any $\h_i(Z)$, since $\h_i k_I^i \del Z^I = \{ Q , \h_ir^i
\}$.

This can be generalised to the case in which the one-forms $k^i$
are not globally-defined\footnote{As emphasised by E.\ Witten,  a geometrically clearer formulation of the construction and of its generalisation can be 
given
in terms of the distribution (i.~e.\ the sub-bundle of the cotangent bundle $T^*M$ of $M$) generated by the $k^i$. In particular, the distribution 
does not depend on the choice of basis for the one-forms $k^i$.} but are local sections of a bundle
\cite{Hull:2006qs}. For example, the $k^i$ might be
       a local section of the co-frame bundle,
        i.e.
a local basis for the cotangent bundle $T^*M$. If $M$ is a bundle
over some $E$, the $k_i$ could be a local section of the co-frame
bundle of $E$ (or rather the pull-back of this co-frame bundle).
We will  be interested mainly in the case in which $M$ is
projective (super-)twistor space, and is a bundle over $E$ where
$E$ is $\CP^1$ or $\CP^{1|N}$. Given an open cover $\{ U_r \}$ of
$M$, suppose there is a set of 1-forms $k^i_r$ in each patch
$U_r$, with
\begin{equation}
\label{} k^i_r = (L_{rs})^i{}_jk^j_s
\end{equation}
in the overlaps $U_r\cap U_s$, and transition functions $L_{rs}$ in
$GL(d|n)$ if the $k^i_r$ consist of $d$ bosonic one-forms and $n$
fermionic ones. The $k^i_r$ are then sections of a bundle  $X$
over $M$, and we can introduce a connection one-form $(\hat B_r )_i=(\hat B_r )_{iI }d Z^I$ with transition functions
\begin{equation}
\label{}
(\hat B_r)_{iI } = (L^{-1}_{rs})_i{}^j(\hat B_s)_{jI }+ \partial _I \hat \alpha _i
\end{equation}
 for the bundle $\hat X$ whose structure group is
the group of fibre translations (with parameters $ \hat \alpha _i
$). Then
 the gauged theory is well-defined provided the gauge
fields $\ti B_i$ are taken to be connections on the pull-back of
$\hat X$ to a bundle over the world-sheet, by a similar construction to that given in~\cite{Hull:2006qs}. The
theory is locally the same as that described above.

\subsection{1-form symmetries and scale symmetry}

\label{1formscalesymm}

A natural generalisation of the construction of the last section
would be to consider a set of vector fields $V_j= V_j^I(Z) \pa
/\pa Z^I$ on $M$, and construct the currents $ V_j^I Y_I$. A
necessary condition for the current algebra to close is that the
$V_j$ are closed under the Lie bracket, so that they generate the
action of a group $L$ on $M$. In certain circumstances, the
corresponding symmetries can be gauged, resulting in a theory on
the quotient space $M/L$. Thus the gauging leads to replacing $M$
with $M/L$, and gauging symmetries from vectors and 1-forms on $M$
is equivalent to gauging symmetries from 1-forms only on $M/L$.
There is then no loss of generality in considering general $M$
without gauging the symmetries generated by vector fields on $M$.
However, it will be useful to consider the case of the Euler
vector field
\begin{equation}
\Upsilon = Z^I \frac{\del}{\del Z^I} \label{Eulervect}
\end{equation}
generating the one-dimensional group $L_S$ of scale
transformations. Gauging the symmetries from 1-forms and
$\Upsilon$ on $M$ is then the same as gauging 1-forms alone on the
projective space $PM=M/L_S$, but using the formulation on $M$ will
be useful for the Berkovits twistor string.

Suppose the one-forms $k^i$ have scaling weights $h_i$ under the
action of  (\ref{Eulervect}), so that for each $i$
\begin{equation}
\label{weights} {\cal L}_\Upsilon k^i= h_i k^i
\end{equation}
where ${\cal L}_\Upsilon $ is the Lie derivative with respect to
$\Upsilon $,
       and have constant vertical projections, so that
$\i (\Upsilon ) k^i=e^i$ for some constants $e^i$, i.e.
\begin{equation}
\label{zki1} Z^Ik_I^i=e^i .
\end{equation}
If $h_i=0, e^i=0$, then $k^i$ is horizontal and is the pull-back
of a form on $PM$, the projective space given by taking the
quotient by the action of the scalings generated by $\Upsilon$.
Then the current $J= Y_I Z^I$ has the commutation relations
\begin{equation}
\label{kmalg1} [J(\s) , K^i (\s') ]= h_i K^i (\s ) \d (\s -\s
')+e^i \d ' (\s -\s ')
\end{equation}
for each $i$. If $Z^I=(Z^\a,Z^a)$ and $Y_I=(Y_\a, Y_a)$ where
$Y_\a, Z^\a$ with $\a=1,..., D$ are bosonic $\beta$-$ \gamma$
systems and $Y_a, Z^a$ with $a=1,..., N$ are fermionic $b$-$c$
systems, then
\begin{equation}
\label{kmalg2} [J(\s) ,J(\s') ]= \d ' (\s -\s ') (D-N) .
\end{equation}
Then the currents $J, K^i$ generate a Ka\v{c}-Moody algebra which
is non-abelian if the weights $h_i$ are not all zero and which has
central charges $e^i$, $D-N$. If the $e^i$ were not constant, the
algebra would not close and one would need to introduce the $e^i$
as extra generators.

This symmetry can be gauged by introducing gauge fields $\ti A,
\ti B_i$ only if $e^i=0$, so that the $k^i$ are all horizontal; it
will now be assumed that this is the case. The gauged action is
\begin{equation}\label{bgB2}
        S = \int d^2 \s \, \left(
        Y_{I} \ti \partial Z^I -\ti A J-\ti B_i K^i \right) ,
\end{equation}
which is invariant under the gauge transformations given by
(\ref{varal}) together with
\begin{equation}
\label{vartia}
       \d \ti A =0
\end{equation}
       and
\begin{equation}
\label{varba} \d \ti B_i= \ti \partial \a_i - h_i \ti A \a _i .
\end{equation}
It is also invariant under the scaling symmetry
\begin{equation}\label{Gl1}
        Z^I \rightarrow tZ^I, \hspace{1.5cm} Y_{I} \rightarrow \frac{1}{t}
        Y_{I}, \hspace{1.5cm} \ti A \rightarrow \ti A+ \frac{1}{t}\ti \partial t \hspace{1.5cm} \ti B_i \rightarrow
        t^{-h_i} \ti B_i .
        \end{equation}

Introducing ghosts $v, s_i$ and anti-ghosts $u, r^i$, the BRST
charge is now
\begin{equation}
\label{brstk} Q = \oint d\sigma \left( vJ+\sum _i [s_i K ^i - vh_i
s_i r^i] \right) .
\end{equation}
The ghost $s_i$ is a world-sheet scalar with scaling weight $-h_i$
(transforming as $s_i \rightarrow
        t^{-h_i} s_i$ under $GL(1)$)
while the antighost $r^i$ has world-sheet conformal dimension one
and
        scaling weight $h_i$.
Then $Q^2$ is proportional to $\int \k v \partial v$, where
\begin{equation}
\label{kappbefW} \k= D-N - \sum _i \e_i (h_i)^2
\end{equation}
with $\e_i=1$ for bosonic symmetries (with $\a_i $ a bosonic
parameter) and $\e_i=-1$ for fermionic symmetries (with $\a_i $ a
fermionic parameter). The constant $\k$ is the central charge for
the Ka\v{c}-Moody algebra generated by the currents
\begin{equation}
\label{} J_{gf}= J- \sum _i h_i s_i r^i
\end{equation}
which generate scalings of the gauge-fixed action, and quantum
consistency (cancellation of the anomaly in the scaling symmetry)
requires $\k=0$~\footnote{It was pointed out to us by E.\ Witten that, if a global and everywhere nonzero function $w$ exists on $M$ then the last term
(involving the scaling weigths $h_i$) in the anomaly~(\ref{kappbefW}) can be eliminated by adding to the BRST operator $Q$ a term proportional to $\oint \partial
v \log w $. This is natural in the formulation in terms of the distribution generated by the one-forms $k^i$ rather than that using
a specific choice of $k^i$ adopted here.}.

\section{Gauging the Berkovits twistor string}

\label{GaugedBerkovits}

The formalism of the previous section will now be applied to the
Berkovits twistor string, generalised to a target space $\cT$ that
is a supermanifold with $D$ bosonic dimensions and $N$ fermionic
ones; the flat twistor space is $\C ^{D|N}$, $\R ^{D|N}\times \R
^{D|N}$ or $\C ^{D|N}\times \C ^{D|N}$. The case of physical
interest is $D=4$, and we will see that, remarkably, this value is
selected by anomaly cancellation in some of the models.

We saw in \S\ref{Ricciflatcase} that the twistor space $\Tc$ for a
Ricci-flat space-time is fibred over $\C^2-0$, so that $\PTc$ is
fibred over $\CP^1$, and this in particular implies the existence
of the 1-form $k$ given by (\ref{kisss}), corresponding to an
infinity twistor. In the flat case, this requires working with
$\PT '= \CP^3-\CP^1$, which has such a fibration, whereas the full
twistor space $\CP^3$ does not.  In the supersymmetric case,
$\PTc$ is fibred over $\CP^{1|0}$ or $\CP^{1|N}$, and in the
latter case a local basis of $N$ fermionic 1-forms on $\CP^{1|N}$
pull back to $N$ locally defined fermionic 1-forms $k^a$ on
super-twistor space.  In this section we will assume that the
target space $\cT$ is equipped with a set of 1-forms $k^i$ and
gauge the corresponding symmetries. In the following sections, we
will suppose that these 1-forms arise from a fibration of the
super-twistor space that follows from the condition for a
Ricci-flat space-time, and find that the gauging restricts the
physical states of the string theory so that they can be
associated with deformations of the super-twistor space preserving
the fibration structure, and hence the Ricci-flatness.

Given a set of 1-forms $k^i=k^i_I(Z)dZ^I$ and $\ti k^i=\ti
k^i_I(\ti Z)d\ti Z^I$ of weights $h_i, \ti h_i$ there are currents
\begin{equation}
\label{kcurt} K^i = k^i_I\partial Z^I, \qquad \ti K^i = \ti
k^i_I\ti \partial \ti Z^I .
\end{equation}
These are conserved Ka\v{c}-Moody currents for the free theory
given by~(\ref{berkoS}) with $A=\ti A=0$. For the case of
Euclidean world-sheets, in which $\ti \s = \s ^*$ and $\ti Z=
Z^*$, the currents $\ti K^i$ are the complex conjugates of the
$K^i$. For the other cases, the $\ti K^i$ and the $K^i$ are
independent currents satisfying $K^i =\ti K^i$ on the boundary as a
result of the boundary conditions~(\ref{bound}).

We assume that the 1-forms satisfy
       \begin{equation}
\label{} Z^I k_I^i=0 \, , \qquad \ti Z^I \ti k_I^i=0
\end{equation}
       so that the central charges $e^i, \ti e^i$  vanish and gauging is possible.
Then gauging the symmetries
       generated by $K^i ,\ti K^i $ gives the action
       \begin{equation}\label{berkoSK}
        S = \int d^2 \s \left( Y_{I} \ti \partial Z^I + \ti Y_{J} \partial
\ti Z^J -\ti AJ- A\ti J -B_i\ti K^i-\ti B_iK^i \right) + S_C ,
\end{equation}
and this is invariant under~(\ref{varal}),
(\ref{vartia}),(\ref{varba}) together with the corresponding
symmetries with parameter $\ti \a, \ti t$. For open strings, the
boundary conditions~(\ref{bound}) are imposed as before.

Under the symmetries with parameter $\a, \ti \a$, the action
changes by a total derivative term
\begin{equation}
\label{totderr} \d S= \int d^2 \s (\partial- \ti  \partial) \left(
\a K- \ti \a \ti K
       \right)
\end{equation}
       and with the boundary conditions~(\ref{bound}), this vanishes for gauge transformations in which
       the parameters satisfy
       \begin{equation}
\label{dsfghs} \a= \ti \a
\end{equation}
on the boundary.

Gauge-fixing by choosing conformal gauge and requiring all gauge
fields to vanish introduces
       the ghosts $(u,v)$ and $(\ti u , \ti v )$ of the Berkovits string, together with the
       ghost system $(r^i, s_i )$ of the last section and its conjugate system
        $(\ti r^i, \ti s_i )$.
        The open string theory is defined by the boundary
conditions~(\ref{bound}) on the twistor variables and
\begin{equation}\label{boundgh2}
        c = \tilde c , \hspace{1cm} b = \tilde b , \hspace{1cm} v = \tilde v ,
        \hspace{1cm} u = \tilde u , \hspace{1cm} r ^i= \tilde r ^i , \hspace{1cm}
        s _i= \tilde s _i
\end{equation}
on the ghosts.

The BRST operators are
\begin{eqnarray}\label{defBRS2}
        Q & = & \oint d\sigma \left( c T +v J +s_i K^i + c u \del v + c b \del c + c r^i \del s_i
        - \sum _i vh_i s_i r^i
        \right) \nonumber \\
         \tilde Q & = & \oint d\ti \sigma \left( \tilde c \tilde T +\tilde v \tilde
         J +\tilde s_i \tilde K^i + \tilde c \tilde u \tilde \del
        \tilde v + \tilde c \tilde b \ti \del
          \tilde c + \tilde c \tilde r ^i\ti \del \tilde s_i
- \sum _i \ti v\ti h_i\ti s_i  \ti r^i
        \right) . \nonumber \\
\end{eqnarray}
In $Q^2$, there are two potentially non-zero terms: a conformal
anomaly term proportional to $C \int  c\pa ^3c$, where $C$ is the
Virasoro central charge, and a gauge anomaly term proportional to
$k \int  v \partial v$, where  $k$ is the  Ka\v{c}-Moody central
charge. The Virasoro central charge is
\begin{equation}
\label{Vircharge} C=D-N+ c_C - 28 -2(d-n) ,
\end{equation}
where $D-N$ comes from the $YZ$ system, $c_C$ is the central
charge of the auxiliary matter system $S_C$, the contribution
$-28=-26-2$ comes from the $bc$ and $uv$ systems, and $-2(d-n)$
comes from the $(r^i ,s_i )$ system consisting of $d$ fermionic
ghosts and $n$ bosonic ones. The Ka\v{c}-Moody central charge is
\begin{equation}
\label{KMcharge} k= D-N - \sum _i \e_i (h_i)^2 ,
\end{equation}
where $\e_i=1$ for bosonic symmetries (with $\a_i $ bosonic) and
$\e_i=-1$ for fermionic symmetries (with $\a_i $ fermionic).

The gauge anomaly cancels if $\k=0$. If $\k \ne 0$, one might
attempt to cancel the anomaly against
        a contribution from the matter system $S_C$.
If the matter system $S_C$ has a current $J_C$ generating a
$GL(1)$ Ka\v{c}-Moody symmetry with central charge $\k_C$, and
$S_C$ is chosen to contain the coupling $\ti A J_C$, then
\begin{equation} \label{} k= D-N +\k_C - \sum _i \e_i (h_i)^2 .
\end{equation}
However, this is likely to lead to problems from mixing between
the
       auxiliary matter system and the twistor space sector, and its most natural
       interpretation would be as a change in the definition of the twistor space.
       We therefore restrict ourselves  to solutions with
\begin{equation}
\label{}
       D-N - \sum _i \e_i (h_i)^2 =0 ,
\end{equation}
so that no resort to such a compensating coupling is needed.

There will be similar anomalies with coefficients $\ti C, \ti k$
from $\ti Q$. Quantum consistency requires $C=\ti C=0$ and $k
=\ti k=0$. In the next section, some string theories in which
these anomalies cancel will be considered.

\section{World-sheet anomaly cancellation in twistor strings}

\label{Anomalies}

\subsection{No supersymmetry}

\label{nosusy}

Consider first the bosonic case in which $N=0, n=0$, so that the
twistor space  $\cPT$ is an ordinary (bosonic)   complex manifold
of dimension $D-1$. The Penrose construction  of the non-linear
graviton  for $D=4$ requires the projective twistor space $\cPT$
to be fibred over $\CP^1$.
      We then restrict ourselves to twistor spaces  in which  $\cPT$ is fibred over $\CP^1$
      (or in the real case, to spaces $\cPT_\R \times \cPT_\R$ with $\cPT_\R$ fibred over $\RP^1$). Then
      there is a
holomorphic 1-form on $\CP ^1$, given by $\e^{A'B'} \p_{A'} \wedge
d\p_{B'}$ where $\p_{A'}$ are homogeneous coordinates on $\CP^1$,
and its pull-back to $\cPT$ is
\begin{equation}
\label{kinf} k = I_{\a\b}Z^\a d Z^\b
\end{equation}
with $I_{\a\b}$ the dual of the infinity twistor. This in turn
pulls back to a 1-form on (non-projective) twistor space $\cT$,
again given by (\ref{kinf}). This 1-form has weight $h=2$. Gauging
the symmetry generated by this 1-form then gives  the
Ka\v{c}-Moody central charge  $k = D-h^2=D-4 $, which vanishes
precisely when $D$ takes the value $D=4$ needed for Penrose's
twistor space, and no $\k_C$ is needed. Then
from~(\ref{Vircharge}) with $D=4, d=1$, we find
\begin{equation}
\label{} C= c_C-26
\end{equation}
so the matter system can be taken to be a critical bosonic string
with $c_C=26$.

\subsection{$N$ supersymmetries, $\PTc$ fibred over $\CP^{1|N}$}

\label{Nsusys}

Suppose now that there are $N$ fermionic dimensions, and the
projective twistor space is fibred over $\CP^{1|N}$ (or
$\RP^{1|N}\times \RP^{1|N}$). On $\CP^{1|N}$, a section of the
co-frame bundle gives one bosonic one-form and $N$ fermionic ones.
The bosonic 1-form is the globally-defined $k$ given
in~(\ref{kinf}), while the $N$ locally-defined fermionic one-forms
$k^a$ are of the form
\begin{equation}
\label{1formka} k^a=d \psi ^a +e^a_{A'} d \p ^{A'}
\end{equation}
and are of weight $h_a=1$. Here $e^a_{A'}$ satisfies
\begin{equation}
\label{piepsi} \p^{A'}e^a_{A'}=- \psi^a ,
\end{equation}
so that the $k^a$ satisfy $\i(\U) k^a=0$. In a patch where
$\p^{A'}\r_{A'}\ne 0$ for some fixed spinor
   $\r_{A'}$,
this can be solved by
\begin{equation}
\label{} e^a_{A'}=-  \, \frac { \psi^a \r_{A'}} {\p^{B'}\r_{B'}}
\end{equation}
so that
\begin{equation}
\label{piouu1} k^a = {\p^{A'}\r_{A'}} \, d \left( \frac { \psi^a }
{\p^{B'}\r_{B'}} \right)\, .
\end{equation}

These forms pull back to one-forms $(k, k^a)$ on $\PTc$ and $\cT$,
so they can be used in the construction of the last section. The
$k^a$ are only locally-defined, but the gauging is still defined
globally, as discussed at the end of \S\ref{1formsymm}. Now
from~(\ref{KMcharge}), the Ka\v{c}-Moody central charge $k$ is
independent of $N$ and
\begin{equation}
\label{} \k= D-4 ,
\end{equation}
so that anomaly cancellation again  selects $D=4$.
Then~(\ref{Vircharge}) gives
\begin{equation}
\label{} C= c_C-(26-N) ,
\end{equation}
so that the matter system should be chosen to have $c_C=26-N$.

\subsection{General weights}

The form (\ref{kinf}) is of weight $h=2$, but a 1-form of general
weight $h$ can be made by multiplying by a function $w(Z)$ of
weight $h-2$ (so that $w$ is a section of $\caO(h-2)$) to give
\begin{equation}
\label{kinfw} \hat k =w(Z) I_{IJ}Z^I d Z^J .
\end{equation}
Similarly, multiplying~(\ref{1formka}) by a $w^a(Z) $ that is a
section of $\caO(h_a-1)$ gives for each $a$
\begin{equation}
\label{kinfwa1} \hat k^a=w^a(Z)(d \psi ^a -e^a_{A'} d \p ^{A'})
\end{equation}
which is of weight $h_a$.

Introducing such factors gives many formal anomaly-free solutions
for which the central charges~(\ref{Vircharge})
and~(\ref{KMcharge}) vanish. For example, choosing all $\hat k^a$
to be of equal weights $h'$, the conditions are
\begin{eqnarray}
0 & = & D-N+c_C -30 \nonumber \\ 0 & = & D-N -h^2 +N(h')^2 .
\label{cond0hh}
\end{eqnarray}
In the bosonic case $N=0$, the only solution with $D=4$ is the
model with $h=2$ and matter central charge $c_C = 26$ discussed
in~\S\ref{nosusy}; however, formally there are higher dimensional
solutions of~(\ref{cond0hh}) with
\begin{equation}
       h^2=D , \hspace{2cm} c_C = 30-D .
\end{equation}
For the case $D=4$ with $N$ fermionic currents,
\begin{eqnarray}
c_C & = & 26+N \nonumber \\ h^2 - N(h')^2 & = & 4-N .
\label{D4andNcurr}
\end{eqnarray}
For $h' =1$, there are solutions with $h=2$ and $c_C = 26+N$ (including an $N=4$ model which is distinct from the $N=4$ model with $c_C = 22$ discussed in~\S\ref{Nsusys}), and there are additional solutions of~(\ref{D4andNcurr})  with
$h'>1$. It is straightforward to find further anomaly-free
solutions corresponding to currents of general weights $h,h_a$.

\subsection{Weightless forms}

\label{weight0}

An important special case of the construction with general $w,w^a$
consists in choosing $w$ of weight $-2$ and all the $w^a$ of weight
$-1$, which gives forms $\hat k, \hat k^a$ all with weights $0$.
Then~(\ref{KMcharge}) gives the same constraint $D=N$ as for the
Berkovits string, and with $D=4$ this selects $N=4$. If one gauges
$\hat k$ and $n$ of the $\hat k^a$ with $0\le n\le N$, then the
central charge is
\begin{equation}
\label{} C=c_C-30+2n .
\end{equation}
There are two models of particular interest with $D=N=4$, that
with $n=0$ and that with $n=4$.

If $w$ is chosen to depend on $\p_{A'}$ only, then the one-form
$\hat k$ is closed, $d\hat k=0$.
       In a patch where $\p^{A'}\r_{A'}\ne 0$ for some fixed spinor $\r_{A'}$
with $k^a$ given by (\ref{piouu1}), choosing $w^a =(
{\p^{A'}\r_{A'}})^{-1}$ for each $a$ gives
\begin{equation}
\label{piouu2} \hat k^a =  d \left( \frac { \psi^a }
{\p^{B'}\r_{B'}} \right)
\end{equation}
which automatically satisfies $d\hat k^a = 0$. More generally, for
any $w^a (\p)$ on $\CP^1$ of weight $-1$, we can choose $\hat k^a =
d (  { \psi^a } w^a)$ (with no sum over $a$).

A potential problem with this construction is that functions
$w(Z), w^a (Z)$ of negative weights can have singularities. For
example,  for  weight $-1$, $w' =( {\p^{A'}\r_{A'}})^{-1}$ is
singular on the surface $ {\p^{A'}\r_{A'}}=0$ on which ${\p^{A'}=
\l \r^{A'}}$ for arbitrary parameter $\l$.  A function $w(Z)$ of
weight $h$ on $\CP^1$ will have $-h$ singularities if $h<0$, and
it is not clear how to define the construction at these
singularities.

For the case of real twistor space with $Z,\ti Z$ independent and
real, there are non-singular functions of negative weights. For
example, a function of weight $-2$ on $\RP^1$ is given by
\begin{equation}
\label{mpost} w(\p)= \frac {1}{M^{A'B'}\p_{A'}\p_{B'}}
\end{equation}
where $\p_{A'}$ are real homogeneous coordinates for $\RP^1$, and
this is non-singular if the constant symmetric real matrix
$M_{A'B'}$ is positive definite, since the point $\p^{A'}= 0$ is
excluded. This can then be pulled back to a non-singular function
of weight $-2$ on any  space that is fibred over $\RP^1$. For a
real twistor space given by a region of $\RP^{3|4}\times
\RP^{3|4}$, or more generally one that is of the form
$\PTc_\R\times \PTc_\R$ for some real $\PTc_\R$ that is fibred
over $\RP^1\times \RP^1$,  non-singular functions $w(\p), \ti
w(\ti \p)$ can be constructed in this way, and they can be used to
construct well-defined one-forms $\hat k(Z), \hat{\ti k} (\ti Z)$
of weight $h=\ti h=0$. A function $w'$ of weight $-1$ can be
defined as $w' =\sqrt {w}$   as $w$ is positive.

For the complex case, $w(Z)$ can be chosen to be non-singular in a
holomorphic disc with boundary on the real subspace, so that it is
non-singular on the embedding of the open string world-sheet in
super-twistor space.
       For   a twistor space $\PT$ fibred over $ \CP^1$, $w$ can be chosen as
\begin{equation}
\label{hdhddhd} w=\frac {1}{( \r^{A'} _1\p_{A'})( \r^{B'}_2
\p_{B'})}
\end{equation}
for some fixed complex spinors $ \r^{A'} _1, \r^{A'} _2$. Then
each singularity lies in a    plane  $\r^{A'} \p_{A'}=0$. Recall
that twistor space divides into two parts $\PT^\pm $ with $\pm
i\pi_{A'}\bar\pi^{A'}\geq 0$ and that these two parts correspond
to two copies of space-time. To obtain just one copy of
space-time, we choose  $\PT^+$, say, as the twistor space, and the
space of holomorphic discs in this part of twistor space with
boundary on $\PT_\R$ gives a complete copy of space-time. If we
take both $ \r^{A'} _1, \r^{A'} _2$ to lie in $\PT^-$, then $w(Z)$
is non-singular on $\PT^+$ and the gauging of the twistor string
is well-defined for world-sheets that are discs in $\PT^+$.

In the complex case with $\ti Z = Z^*$, the cancellation of the
surface term in the variation (\ref{totderr}) requires that $w\a =\ti
w \ti \a = (w\a)^*$ on the boundary. If $w(Z)$ is real on the real
axis $Z=Z^*$, this gives the boundary condition $\a=\ti \a$ as before,
but if $w$ is a complex function on the real axis, then the boundary
conditions of $\a$ and hence of the ghosts $s$ are modified.  However,
in the case of Euclidean world-sheet, in which $Z$ and $\ti Z$ are
independent complex variables, the boundary condition is $Z=\ti Z$ and
it is possible that $w(Z), \ti w (\ti Z)$ can be chosen so that $w(Z)=
\ti w (\ti Z)$ on the boundary with $w(Z)$ non-singular on the
holomorphic disc, and the boundary condition on $\a$ is $\a=\ti \a$.

The models in which the zero-weight one-form (\ref{kinfw}) or the
one-forms (\ref{kinfw}), (\ref{kinfwa1}) are gauged are then
well-defined both for the real case, and for the complex case with
     independent complex coordinates $Z, \ti Z$.
     The models depend on an arbitrary function $w$, or on the
functions $w$ and $w^a$, but these only enter into the BRST charge.
It will be seen in the next section that the spectrum is
independent of $w,w^a$, provided these functions are chosen to have
no zeroes or poles; tree-level amplitudes at degree zero are also
independent of the choice of $w,w^a$, as will be checked explicitly
in an example in~\S\ref{3pointN4}.

\section{Spectra of the twistor string theories}

\label{spectra}

\subsection{Physical vertex operators}

\label{vertices}

In this section, we will investigate the      constraints and
gauge invariances for the  vertex operators $V_f,V_g,V_\phi$ for
each of the anomaly-free theories of the last section, and obtain
the ghost-independent part of the BRST cohomology. We will discuss
the ghost-dependent vertex operators elsewhere.

The gauged twistor string is constructed on a twistor space with a
set of 1-forms $k^i = k^i_IdZ^I$ with weights $h_i$ defined by
(\ref{weights}) and satisfying
\begin{equation}
\label{zki2} Z^Ik_I^i=0 .
\end{equation}
The vertex operator $ V_f = Y_I f^I (Z)$ is physical provided
\begin{equation}\label{fgprimf}
        \del_I f^I = 0 , \qquad f^I k^i_I=0, \qquad f^I k^i_{[I,J]}=0
\end{equation}
for each $i$. However, the gauge invariance (\ref{gaugefg})
       is now modified, as
       \begin{equation}
\label{qurs} \{ Q, u\} = J+ \sum _i h_i r^is_i .
\end{equation}
If all the weights $h_i$ vanish, then $\L J$ is BRST trivial for
any $\L(Z)$ of zero weight, and
\begin{equation}\label{gaugeHTf}
        \delta f^I = Z^I \Lambda \end{equation}
changes $V_f$ by a BRST trivial term. However, if any of the
weights $h_i$ are non-zero, then the extra ghost terms in
(\ref{qurs}) mean that (\ref{gaugeHTf}) is not a symmetry. This is
just as well, as the constraints (\ref{fgprimf}) are only
invariant under (\ref{gaugeHTf}) if all the $h_i$ are zero.

The vertex operator $V_g = g_I (Z) \del Z^I $ is physical provided
\begin{equation}\label{fgprimg}
Z^I g_I = 0,
\end{equation}
and it has the gauge invariances
\begin{equation}\label{gaugeHT}
       \delta g_I = \del_I \chi, \qquad \d g_I = \h _ik_I^i
        \end{equation}
for any $\chi (Z )$ and any $\h_i (Z)$ of weights $-h_i$.

The Yang-Mills vertex operator $ V_\phi = j_r \phi^r (Z)$ receives
no further constraints from the gauging. In the following the
spectrum will be analysed for the anomaly-free strings of the last
section in the flat case. The twistor space is $\PT _{[N]}' = \PT
_{[N]}-I$ and results from removing the appropriate (super)line
$I$ (which is $I=\CP^{1|0}$ or $I_{[N]}=\CP^{1|N}$ in the complex
case, and  $\RP^{1|0}\times \RP ^{1|0}$ or $\RP^{1|N}\times \RP
^{1|N}$ in the real case) from $\CP^{3|N}$ or $\RP^{3|N}\times \RP
^{3|N}$. The vertex operators live on the boundary of the
world-sheet, which in turn lies in $\RP^{3|N}$.

\subsection{Self-dual gravity without supersymmetry}

\label{sdgrav0}

Consider first the bosonic $N=0$  theory of~\S\ref{nosusy} with
the one-form
\begin{equation}
\label{kinfq} k = I_{\a\b}Z^\a d Z^\b
\end{equation}
on the twistor space $\PT'=\CP^3-\CP^1$ (or $\PT'_\R =\RP^3-\RP^1$
in the real case), so that
\begin{equation}
\label{} k_\a =-I_{\a\b}Z^\b , \qquad k_{[\a ,\b ]}=-I_{\a\b} .
\end{equation}
The coordinates on twistor space are $Z^\a =(\omega^A,\pi_{A'})$
and
\begin{equation}
\label{kinfp} k = \e^{A'B'} \p_{A'} d \p_{B'} .
\end{equation}

Then $f^I= (f^A,f_{A'})$ are of degree one and the constraints
(\ref{fgprimf}) imply
\begin{equation}
\label{fdet1} \frac{\del f^A}{\del\omega^A} = 0, \qquad f_{A'}=0 ,
\end{equation}
which in turn imply
\begin{equation}\label{solvef1}
        f^A = \epsilon^{AB} \frac{\del h}{\del \omega^B}
\end{equation}
for some twistor function $h(Z)$ homogeneous of degree 2. Via the
twistor transform, this corresponds to a space-time field of
helicity $2$ satisfying the field equations of linearised Einstein
gravity~\cite{Mason1}.

The 1-form $g=g_\a\rd Z^\a$ in the vertex operator $g_\a \del
Z^\a$ satisfies $Z^\a g_\a =0$, which means that $g_\a$ is defined
on the projective twistor space, and moreover it follows
from~(\ref{gaugeHT}) that it is defined up to two gauge freedoms:
\begin{equation}
\label{twogaugetr} g_\a \rightarrow g_\a +\del_\a \chi \, , \qquad
g_\a \rightarrow g_\a +I_{\a\b}Z^\b \eta\, .
\end{equation}
The four components  of $g_\a $ are subject to  one constraint and
two gauge invariances, and the remaining degree of freedom is
conveniently represented by a
       function $\tilde f$ of homogeneity degree $-2$ defined  by
\begin{equation}\label{plusmult}
\tilde h = I^{\a\b}\del_\a g_\b = \e^{AB} \del_A g_B ,
\end{equation}
which is   invariant under the two gauge transformations given
in~(\ref{twogaugetr}). This function of degree $-2$ corresponds to
a space-time scalar field. Finally, the Yang-Mills vertex operator
with functions $\phi_r(Z)$ of degree zero gives states of helicity
$+1$ in the adjoint of the gauge group $G$.

Thus the spectrum of this theory consists of a state of helicity
$+2 $, a scalar state of spin $0$ and $dim(G)$ states of helicity
$+1$.  Note that the state of spin zero could come from a scalar
field or a 2-form gauge field. An interacting theory with this
spectrum is self-dual gravity coupled to self-dual Yang-Mills and
a scalar (or 2-form gauge field), and this has covariant field
equations but no covariant action. In the absence of the scalar,
the field equations would be
\begin{equation}
\label{sdfhjzfs} R=*R, \qquad F=*F ,
\end{equation}
where $R$ is the curvature 2-form, $F$ is the Yang-Mills field
strength and $*$ denotes the Hodge duality operation. Finding out
whether this interacting theory arises, and finding the form of
the scalar coupling, requires investigating the interactions
arising from string amplitudes. This will be discussed elsewhere.

\subsection{Supergravity with $N$ supersymmetries}

\label{sdgravN}

       Consider next the case of~\S\ref{Nsusys}, with
projective twistor space $\PT'_{[N]}$ of dimension $3|N$ (given by
$\CP^{3|N}-\CP^{1|0}$, or $\RP^{3|N}-\RP^{1|0}$ in the real case)
that is fibred over $\CP^{1|N}$, and the gauging associated with
the bosonic one-form (\ref{kinfp}) and the $N$ fermionic one-forms
\begin{equation}
\label{} k^a=d \psi ^a -e^a_{A'} d \p ^{A'} .
\end{equation}
The vector field $f^I$ decomposes as
$f^I=(f^\a,f^a)=(f^A,f_{A'},f^a)$ and the conditions
       (\ref{fgprimf}) imply
\begin{equation}
\label{} \frac{\del f^A}{\del\omega^A} = 0, \qquad f^{A'}=0
,\qquad f^{a}=0 ,
\end{equation}
and again
\begin{equation}\label{solvef2}
        f^A = \epsilon^{AB} \frac{\del h}{\del \omega^B}
\end{equation}
for some super-twistor function $h (Z)$ homogeneous of degree 2.

Consider first the case $N=4$. Then $h(Z)$ has an expansion
\begin{equation}\label{negspec}
h(Z^I)= g(Z^\a)+\chi _a (Z^\a) \psi ^a + A_{ab}(Z^\a) \psi ^a \psi
^b+ \Lambda^d (Z^\a) \e _{abcd}\psi ^a \psi ^b \psi ^c+ \varphi
(Z^\a) \e _{abcd} \psi ^a \psi ^b \psi ^c \psi ^d ,
\end{equation}
where $Z^\a =(\omega^A,\pi_{A'})$ are the coordinates on bosonic
twistor space. This gives twistor fields $g,\chi _a,A_{ab},
\Lambda _{abc},\varphi$ in
$\caO(2),\caO(1),\caO(0),\caO(-1),\caO(-2)$ respectively. Via the
twistor transform, these correspond to space-time fields of
helicities $2,3/2,1,1/2,0$ in the $SL(4,\R)$ representations
$({\bf 1},{\bf 4},{\bf 6},{\bf 4'},{\bf 1})$ respectively. We then
obtain the following positive helicity fields in space-time: a
graviton $g_{\m\n}$, four gravitini $\chi _a^\m$, six helicity one
fields $A_{ab}^\m$, four helicity half fields $\Lambda _{abc}$ and
a scalar $\varphi$. These satisfy the field equations of
linearised $N=4$ supergravity.

For general $N$, one again  has an expansion
\begin{equation}\label{negspecads}
h(Z^I)= g(Z^\a)+\chi _a (Z^\a) \psi ^a + A_{ab}(Z^\a) \psi ^a \psi
^b+...\end{equation} terminating with a term of order $\psi ^N$,
giving twistor fields  in $\caO(2),\caO(1),..,\caO(2-N)$
corresponding to space-time fields of helicities
$2,3/2....,2-(N/2)$ in the $SL(N,\R)$ representations $({\bf
1},{\bf N}, {\bf N(N-1)/2},...,{\bf N'},{\bf 1})$ respectively.

For the vertex operator $g_I \del Z^I$, $g_I=(g_A, g_{A'},g_a)$
and the symmetry~(\ref{gaugeHT}) with the one-forms $k^a$ can be
used to set $g_a=0$. Then (\ref{plusmult}) again defines a
function of homogeneity degree $-2$ that is invariant under the
remaining symmetries, and gives rise to the conjugate multiplet to
the one obtained from $f$. For $N=4$, this is
\begin{equation}
\tilde h(Z^I)= \tilde g(Z^\a) \e _{abcd} \psi ^a \psi ^b \psi ^c
\psi ^d +\tilde\chi^d (Z^\a) \e _{abcd}\psi ^a \psi ^b \psi ^c +
\tilde A_{ab}(Z^\a) \psi ^a \psi ^b+\tilde\Lambda_a (Z^\a) \psi ^a
+ \tilde \varphi (Z^\a) ,
\end{equation}
giving twistor functions $\tilde g,\tilde \chi _a,\tilde A_{ab},
\tilde \Lambda _{abc},\tilde \varphi$ in
$\caO(-6),\caO(-5),\caO(-4),\caO(-3),\caO(-2)$ corresponding to
helicities $-2,-3/2,-1,-1/2,0$ with multiplicities $1,4,6,4,1$
respectively. For general $N$, this gives twistor fields in
$\caO(-2-N),..,\caO(-3),\caO(-2)$ corresponding to helicities
$-N/2,...,-1/2,0$.

Finally, the Yang-Mills sector is represented by a function of
degree zero in super-twistor space, corresponding to helicities
$1,1/2,...,-N/2$ in the $SL(N,\R)$ representations $({\bf 1},{\bf
N}, {\bf N(N-1)/2},...,{\bf N'},{\bf 1})$, and for $N>4$, there
are higher-spin fields with helicities less than $-1$.

For $N=4$ this is the spectrum of $N=4$ supergravity coupled to
$N=4$ super-Yang-Mills. For $N<4$, this is a self-dual
supergravity theory coupled to self-dual Yang-Mills. Interacting
self-dual supergravity theories in 2+2 dimensions have been
discussed in~\cite{KNG1,KNG2,KNG3,KNG92,BS,Siegel2}. For $N>4$, we find
multiplets with spins greater than two, and with more than one
state of helicity $-2$. Free theories can be written down for all
these spectra, but the possibilities for interactions are more
limited. However, there is the intriguing possibility of self-dual
interactions for these theories, as the usual higher-spin
inconsistencies are absent for certain self-dual theories. The
possibility of interactions will be discussed   in section
\ref{Discuss}.

\subsection{$N=8$ supergravity}

\label{N8grav}

Consider the theory of \S\ref{weight0} formulated in $N=4$
super-twistor space with the gauging for the single weightless
1-form
\begin{equation}
\label{kinfws1} \hat k =w(Z) I_{IJ}Z^I d Z^J ,
\end{equation}
where $w$ is of degree $-2$. We need  only assume a fibration over
$\CP^{1|0}$, so that the flat twistor space can be taken to be
$\PT '_{[4]}= \CP^{3|4}-  \CP^{1|4}$ (or the real analogue
thereof). We choose $w=w(\p)$ so that $\hat k$ is closed, $d\hat
k=0$.  For real twistors $Z,\ti Z$, the function $w$ could be chosen as
in~(\ref{mpost}), and for complex ones as in (\ref{hdhddhd}).

Starting with the vector field $f^I$, we work through the various
conditions and gauge equivalences as follows. In this case, the
constraints~(\ref{fgprimf})  are weaker than in \S\ref{sdgravN} as
$d\hat k=0$, but there is now a gauge invariance of the
type~(\ref{gaugeHTf}) since the form has weight $h=0$. We set
$f^I=(f^\a,f^a)=(f^A,f_{A'},f^a)$. We fix the gauge freedom
$f^I\rightarrow f^I + Z^I \Lambda$ from equation (\ref{gaugeHTf})
by requiring that
\begin{equation}
\label{} \frac{\del f^A}{\del\omega^A} = 0 ,
\end{equation}
which in turn implies
\begin{equation}\label{solvef3}
        f^A = \epsilon^{AB} \frac{\del h}{\del \omega^B}
\end{equation}
for some twistor function $h(Z)$ homogeneous of degree 2. This has
the expansion (\ref{negspec}) and gives the space-time fields of
helicities $2,3/2,1,1/2,0$ of the positive helicity $N=4$
supergravity multiplet.

For $w=w(\p)$, (\ref{kinfws1}) implies $\hat k_{[I,J]}=0$, so that
the constraints~(\ref{fgprimf}) give
\begin{equation}\label{fgprimfs}
        \del_I f^I = 0 \qquad w(\p) f^{A'} \p_{A'}=0 ,
\end{equation}
       implying that $f_{A'}=\pi_{A'}\lambda$ for
some $\lambda$ of homogeneity degree $-1$. The function $\lambda$
can be understood to be determined in terms of the $f^a$ by the
condition  $\del_I f^I=0$ (cf.\ eq.~(\ref{fgprimf})) and so $\l$
does not represent any independent degrees of freedom.  We
expand the $f^a$ to obtain
\begin{equation}
\label{formforfe} f^e= \chi^e(Z^\a) + A^e_{a}(Z^\a) \psi ^a +
\Lambda^e_{ab} (Z^\a) \psi ^a \psi ^b + \varphi^{ea} (Z^\a) \e
_{abcd} \psi ^b \psi ^c \psi ^d + \tilde \Lambda^e (Z^\a) \e _{abcd} \psi
^a \psi ^b \psi ^c \psi ^d .\\
\end{equation}
We have used the same symbol as in equation (\ref{negspec}) to
denote fields of the same helicity. Eq.~(\ref{formforfe}) gives
four gravitino multiplets, each with states of helicities $
3/2,1,1/2,0,-1/2$, and so leads to a further four gravitini,
sixteen helicity one fields, twenty four helicity one half fields,
sixteen scalars and four helicity minus one half fields.

The 1-form $g=g_I\rd Z^I$ in the vertex operator $g_I \del Z^I$
satisfies $Z^Ig_I=0$, which means that $g_I$ is defined on the
projective twistor space; moreover $g_I$ is defined up to two
gauge freedoms:
\begin{equation}
\label{} g_I\rightarrow g_I +\del_I\chi \, , \qquad g_I\rightarrow
g_I+wI_{IJ}Z^J\eta\, . \label{gaugegag}
\end{equation}
We  define a gauge-invariant function $\tilde h$ of homogeneity
degree $-2$ by (\ref{plusmult}) and this again gives rise to the
conjugate supergravity multiplet
      with helicities
$-2$,$-3/2$,$-1$,$-1/2$,$0$ and multiplicities $1,4,6,4,1$
respectively.

The fermionic components $g_a$ contribute further states to the
spectrum. In order to  see this and find the full spectrum, we
write $g_I=(g_\a,g_a)=(g_A,g^{A'}, g_a)$. The gauge freedom
$g_I\rightarrow g_I + \partial_I \chi$ can be fixed by imposing the
gauge condition $g^{A'}\pi_{A'}=0$. This implies
$g^{A'}=\pi^{A'}\xi$ for some $\xi$ which can then be set
to zero by use of the gauge freedom $\delta g_I =  I_{IJ}Z^J
\eta$. Consider next the  two degrees of freedom in $g_A$. One is
the component $\omega^A g_A$, which is determined in terms of the
$g_a$ by the final constraint $Z^I g_I=0$ (cf.~(\ref{fgprimg}))
and  so is not an independent degree of freedom. This leaves one
degree of freedom represented by  the gauge-invariant function
$\tilde h$
       given  by (\ref{plusmult}),
       corresponding to the negative helicity  $N=4$ supergravity multiplet.

The remaining components $g_a$ are unconstrained and, together
with $\tilde h$, determine the gauge fixed conponents of $g_I$.
The $g_a$ can be expanded as
\begin{equation}
g_e=\tilde\chi_e (Z^\a) \e _{abcd}\psi ^a \psi ^b \psi ^c\psi^d +
\tilde A_e^d(Z^\a) \e_{abcd}\psi ^a \psi ^b\psi^c+\tilde\Lambda
_{eab} (Z^\a) \psi ^a\psi^b + \tilde \varphi (Z^\a)_{ea}\psi^a
+\Lambda(Z^\a)_e\, .
\end{equation}
This gives four negative helicity gravitino multiplets,
       conjugate to those from $f^a$.

Note that the spectrum is independent of the choice of $w(\p)$.
Combining all the positive and negative helicity states, we obtain
a spectrum consisting of a graviton $h_{\m\n}$, 8 gravitini, 22
vector fields, 32 spin-half fields $\Lambda _{abc}$ and 34
scalars. This is six $N=4$ vector multiplets short of the full
$N=8$ supergravity spectrum.  In addition, the Yang-Mills vertex
operator gives vector multiplets in the adjoint of some group $G$.
If $G$ is six-dimensional, then the spectrum of $N=8$ supergravity
is obtained.

\subsection{$N=4$ supergravity coupled to super-Yang-Mills} \label{N4grav}

Consider the theory of \S\ref{weight0} formulated in $N=4$
super-twistor space with the gauging for the    weightless 1-form
\begin{equation}
\label{kinfws2} \hat k =w(Z) I_{IJ}Z^I d Z^J ,
\end{equation}
where $w$ is of degree $-2$, and the four weightless 1-forms
\begin{equation}
\label{kinfwa2} \hat k^a=w'(Z)(d \psi ^a -e^a_{A'} d \p ^{A'})
\end{equation}
where $w'$ is of degree $-1$. We  assume a fibration over
$\CP^{1|4}$, so that the flat twistor space can be taken to be
$\PT '_{[4]}= \CP^{3|4}-  \CP^{1|0}$ (or the real analogue
thereof). It will be assumed that $w,w'$ are chosen so that $\hat k,\hat
k^a$ are closed, and  that they  have no zeroes or poles on  the
boundary space defined by the boundary condition $Z=\ti Z$ (which
is $\RP^{3|4}$ for the Lorentzian world-sheet theory). It was
shown in the previous subsection that the constraints from $\hat
k$ imply that the vertex operator $V_f$ is determined by a
function $h(Z)$ of degree $2$ and   four functions $f^a$ of degree
$1$, while $V_g$ is given in terms of a function $\tilde h(Z)$ of
degree $-2$ and four functions $g_a$ of degree $-1$. The
constraints $f^I \hat k_I^a=0$ from the fermionic 1-forms give
\begin{equation}
\label{} w' f^a=0 ;
\end{equation}
this implies that $f^a=0$ as $w'$ is chosen to have no zeroes on
$Z=\ti Z$, while the symmetry $\d g_I= \eta_a \hat k_I^a$ can be
used to set $g_a=0$. In this way the gravitino multiplets are
eliminated, leaving the twistor functions $h(Z)$ of degree $2$ and
$\tilde h(Z)$ of degree $-2$, and this gives the spectrum of $N=4$
supergravity. In addition, the vertex operators $V_\phi$ give the
spectrum of $N=4$ super-Yang-Mills with gauge group $G$, so the
spectrum of $N=4$ supergravity coupled to $N=4$ super-Yang-Mills
is obtained.

\section{Amplitudes for $N=8$ and $N=4$ supergravity}

\label{3pointN4}

The scattering amplitudes for the Berkovits string, calculated
from open string correlation functions with vertex operators
$V_f,V_g, V_\f$ inserted on the world-sheet boundary, give rise to
nontrivial scattering amplitudes and hence to interactions for the
space-time fields~\cite{NB,BM,BWsc}.  The $n$-point tree-level
amplitude is given by the formula \cite{NB,BM}
\begin{equation}
\label{dfggfds} \sum_d \left\langle   c V_1(\s_1) c V_2(\s_2)c
V_3(\s_3)\int  d\s_4V_4(\s_4)\dots \int  d\s_nV_n(\s_n)R
\right\rangle _d
\end{equation}
where $V_i$ are any of the vertex operators $V_f,V_g, V_\f$ and
$\langle \dots \rangle _d$ is the correlation function on a disc
of degree $d$, corresponding to a gauge instanton on the disc with
a topologically non-trivial configuration for the gauge field $A$
characterised by the integer $d$ \cite{BM}.  The coordinates are
written as $Z^I= \rho \hat Z^I$, where $\rho $ is a scale factor
(which is complex for complex $Z$), and a BRST-invariant operator
$R$ is
\begin{equation}
\label{} R= \d (\rho -1) v +\dots
\end{equation}
This has the property that it gives an insertion of the zero-mode
of the ghost $v$, so that the integration over $v$ is non-zero,
and regulates the integral over $\rho$. (Changing the insertion
point $\s_0$ changes  $R(\s_0)$ by a BRST exact term, so that the
amplitude is independent of  $\s_0$.) Integrating out $\rho, v$
leaves an amplitude defined on a \lq small Hilbert space'  of
$GL(1)$-neutral states independent of the $v$ zero-mode, giving
results defined on the projective twistor space \cite{BM}.

Consider now the new  theories based on weightless forms of
\S\ref{weight0}, \S\ref{N8grav}, \S\ref{N4grav}, corresponding to
$N=8$ supergravity or  $N=4$ supergravity coupled to
super-Yang-Mills. These new string theories   are similar to the
Berkovits string, and the twistor fields $Y,Z$ have the same
world-sheet dynamics and the same vertex operators. However, there
is an additional ghost sector and the extra terms in the BRST
operator give extra constraints and extra gauge invariances for
the twistor wave-functions $f^I, g_I$, while there are no further
constraints or invariances for the Yang-Mills wave-functions
$\f_r$. In the $N=8$ theory, there is an extra anti-commuting
ghost $s$ of conformal weight zero, which has one zero mode on the
disc, so that one insertion of the $s$ zero-mode is needed to
obtain a non-zero amplitude. For any BRST-invariant vertex
operator $cV$, $scV$ is also BRST-invariant, so that a non-zero
amplitude is given by replacing e.g.  $c V_1(\s_1)$ with $s c
V_1(\s_1)$ in (\ref{dfggfds}). Upon integrating over the $s$
zero-mode,  the amplitude (\ref{dfggfds}) is recovered. For the
$N=4$ theories of section  \ref{N4grav}, there is in  addition one
zero-mode for each of the four commuting ghosts $s^a$, and the
integral over these can be handled by choosing appropriate
pictures for the vertex operators $V_i$. A convenient choice is to
replace $c V_1(\s_1)$ with $s \d ^4(s^a) c V_1(\s_1)$ in
(\ref{dfggfds}). Again, on integrating out the ghost zero modes
$s,s^a$, the formula  (\ref{dfggfds}) is recovered.

       As a result, after integrating out the zero-modes of the new ghosts, the tree-level correlation
functions for the $N=4$ and $N=8$ theories of \S\ref{N8grav} and
\S\ref{N4grav} have the same form as   for the Berkovits string
in~\cite{NB,BM,BWsc} when written in terms of $f^I, g_I, \f_r$.
However, in our case these wave-functions are subject to further
constraints and have further gauge invariances. As we have seen,
these can be used to
        write $f^I, g_I$ in terms of the
unconstrained wave-functions  $h,\ti h$ (defined by
(\ref{solvef2}),(\ref{plusmult})) for the $N=4$ theory, or $h,\ti
h,f^a,g_a$ for the $N=8$ theory.
       These are wave-functions
for supergravity and matter systems whose field equations are of
2nd order in space-time derivatives for bosons (1st order for
fermions), not  those for conformal supergravity with 4th order
equations for bosons. When written in terms of $h,\ti h$ or $h,\ti
h,f^a,g_a$, the scattering amplitudes of the new twistor strings
should then give interactions for Einstein gravitons and matter.
These will be systematically investigated and compared with known
gravity amplitudes elsewhere, but it is straightforward to see
that non-vanishing amplitudes are obtained in certain examples,
confirming that these theories have non-trivial interactions, and
moreover we can compare these with the known MHV gravity
amplitudes.

We now check this for tree-level amplitudes
       at degree zero
       by first calculating
amplitudes   in terms of
       $f^I, g_I$ using the procedure described in \cite{BWsc,BM}, and
       then writing these in terms of the $h,\ti h$
       defined by (\ref{solvef2}) and (\ref{plusmult}).
       The Yang-Mills amplitudes are the same as for the Berkovits string.
At degree zero, the amplitudes $\langle V_g V_g V_g\rangle$,
$\langle V_f V_g V_g\rangle$ vanish automatically. Now consider
the amplitude $\langle V_{f_1} V_{f_2} V_{g_3}\rangle$.
Following the procedure given in \cite{BWsc}, we obtain the
formula
       \begin{equation}
       \langle V_{f_1} V_{f_2} V_{g_3}\rangle=
\int_{\RP^{3|4}} \Omega_s\, f^I_1f^J_2
\partial_{[I}g_{3J]}  ,
\label{3ptform1}
\end{equation}
       where $\Omega_s$ is the volume form on $\RP^{3|4}$.
Briefly, this formula follows upon identifying the open string
worldsheet with the upper-half complex plane, inserting open
string vertex operators on the real axis, and  evaluating the
correlation function $\langle V_{f_1} (\sigma_1 )  V_{f_2}
(\sigma_2 ) V_{g_3} (\sigma_3 ) \rangle$ of three vertex operators
given in terms of the $f^I$ and $g_I$ by $V_f = Y_I f^I (Z)$ and
$V_g = \del Z^I g_I (Z)$. This correlation function is computed by
taking contractions and using the OPE
\begin{equation}
        Z^I(\sigma_1) Y_J(\sigma_2 )\sim
\frac {\delta^I_J}{\sigma_1-\sigma_2 } . \end{equation} The
contractions give rise to a factor of
$(\sigma_1-\sigma_2)(\sigma_2-\sigma_3)(\sigma_3-\sigma_1)$ in the
denominator that cancels an identical factor in the numerator
coming from  the integral over zero-modes of the conformal ghost
$c$. The result is then integrated over the space of zero-modes of
the fields $Z^I(\s)$, which are just constant maps from the disc
to twistor space, giving an integral over $\RP^{3|4}$.  To obtain
the formula~(\ref{3ptform1}), one also needs to integrate certain
terms by parts and use the fact that $\del_I f^I=0$. Furthermore, it
can be checked that, for our vertex $V_f$ with
\begin{equation}
\label{} f_i^I=(\epsilon^{AB}\frac {\partial h_i}{\partial
\omega^B},0,0) , \qquad i= 1,2,3 ,
\end{equation}
the  formula for the remaining amplitude $\langle V_{f_1} V_{f_2} V_{f_3}\rangle$ given in~\cite{BWsc}
(eq.~(5.10) of that paper) yields
\begin{eqnarray}
       \langle V_{f_1} V_{f_2} V_{f_3}\rangle & = & \frac{1}{(\sigma_1-\sigma_2)(\sigma_2-\sigma_3)(\sigma_3-\sigma_1)} \times \nonumber \\ & & 
\int_{\RP^{3|4}} \Omega_s\, \left( \epsilon^{AB} \epsilon^{CD} \epsilon^{EF} - \epsilon^{CB} \epsilon^{ED} \epsilon^{AF} \right)  \frac {\partial h_1}{\partial
\omega^E \partial \omega^B} \frac {\partial h_2}{\partial
\omega^A \partial \omega^D}\frac {\partial h_3}{\partial
\omega^C \partial \omega^F}
  . \nonumber\\
\label{3Vfs}
\end{eqnarray}

We now focus on the amplitudes between two positive helicity and
one negative helicity graviton states so we consider the case in
which the wave functions are given in terms of functions $h,\tilde
h$. We choose
\begin{equation}
\label{} f_1^I=(\epsilon^{AB}\frac {\partial h_1}{\partial
\omega^B},0,0) \, , \qquad f_2^I=(\epsilon^{AB}\frac {\partial
h_2}{\partial
        \omega^B},0,0) \,  , \qquad g_{3I}=(
g_{3A}\Pi_{a=1}^4\psi^a ,0,0)\, ,
\end{equation}
where $h_1$, $h_2$ and $g_{3A}$ are functions of the bosonic
twistor coordinates $Z^\a$ alone, $g_{3A}$ has weight $-5$ and
\begin{equation}
\label{} \epsilon^{AB}\frac{\partial }{\partial \omega^A}
g_{3B}=\tilde h_3\, ,
\end{equation}
where $\tilde h_3$ has homogeneity degree $-6$. Performing the
integrals over the odd variables, the integral~(\ref{3ptform1})
now becomes
\begin{equation}
\langle V_{f_1} V_{f_2} V_{g_3}\rangle = \int_{\RP^{3}} \Omega_s\,
\epsilon^{AB} \left(\frac{\partial }{\partial
          \omega^A}h_1\right)\left(\frac{\partial }{\partial
          \omega^B}h_2\right) \tilde h_3 \, 
          \label{3ptform2}
\end{equation}
where $\Omega$ is the volume form on $\RP^{3}$. We now take $h_1$,
$h_2$, and $\tilde h_3$ to be momentum eigenstates with momenta
$P_i^{AA'}=p_i^Ap_i^{A'}$, $i=1,2,3$:
\begin{equation}
h_i=\exp \left(
\frac{\omega^AP_{iAA'}\alpha^{A'}}{\pi_{B'}\alpha^{B'}}\right)
\left(\frac{\pi_{A'}\alpha^{A'}} {p_{1B'}\alpha^{B'}}\right)^{3}
\delta
         (\pi_{A'}p_1^{A'}) \,
         \label{repr1}
\end{equation}
for $i=1,2$ and
\begin{equation}
       \tilde h_3=\exp \left(
\frac{\omega^AP_{iAA'}\alpha^{A'}}{\pi_{B'}\alpha^{B'}}\right)
\left( \frac{\pi_{A'}\alpha^{A'}} {p_{1B'}\alpha^{B'}}\right)^{-5}
\delta
         (\pi_{A'}p_1^{A'}) \, .
\label{repr2}
\end{equation}
Here $\alpha_{A'}$ is a fixed spinor on which the
representatives~(\ref{repr1}) and~(\ref{repr2}) in fact do not
         depend (see e.~g.~\cite{Wittenparity,Witten2003}).  The
         integral~(\ref{3ptform2}) can now be done; after some
         delta-function manipulations, this yields the standard formula for the three
         point MHV amplitude for gravity in split signature (or in Lorentz
         signature with complex
         momenta)~\cite{Gio,Nair2,Bjeretal}:
\begin{equation}
\langle V_{f_1} V_{f_2} V_{g_3}\rangle =
\delta^4(P_1+P_2+P_3)
\frac{\left( p_{1A} p_2^A \right)^6 }{\left( p_{3B}p^{B}_{1}\right)^2 \left( p_{2C}p^{C}_{3}\right)^2 } \, .
\end{equation}
Thus the new $N=4$ and $N=8 $ twistor string theories each have at
least one non-trivial interaction, and this gives precisely the
helicity $(++-)$ 3-graviton interaction of Einstein gravity.

Under scaling the infinity twistor $I^{IJ} \to R I^{IJ}$, $\e^{AB}
\to R \e^{AB} $, so that if $f^I, g_I$ are kept fixed, then $h \to
R^{-1}h $ and $\ti h \to R\ti h $. Then the amplitude scales as
$R^{-1}$, so that $R^{-1}$ sets the strength of  the gravitational
coupling.

\section{Discussion}  \label{Discuss}

In this paper, a number of new twistor string theories have been
constructed. They were shown to be free from perturbative
world-sheet anomalies, and the ghost-independent part of the
spectra in space-time have   been found. The full BRST cohomology
including ghost-dependent vertex operators will be discussed
elsewhere. The key questions that remain are whether these give
fully consistent quantum theories, and whether they have
non-trivial interactions.
      We have
      seen in section~\ref{3pointN4}
      that non-vanishing 3-point supergravity  amplitudes are obtained in the $N=4$ and $N=8$ cases, so
         these theories have   non-trivial interactions.  Other amplitudes for these theories, and those for the other theories, will be discussed elsewhere.

The string theories giving the  $N=4$ and $N=8$ theories involve
arbitrary functions $w,w'$ of homogeneity $-2$ and $-1$
respectively. These can be chosen to be non-singular for the
theory with Lorentzian world-sheet and independent real
coordinates $Z, \ti Z$ (with target space $\RP^{3|4}\times
\RP^{3|4}$ in the flat case) and for the Wick-rotated version of
this with Euclidean world-sheet and independent complex
coordinates $Z, \ti Z$  (with target space $\CP^{3|4}\times
\CP^{3|4}$ in the flat case). There is also a theory with
Euclidean world-sheet obtained from this by setting $ \ti Z = Z^*$
     (with target space $\CP^{3|4}$ in the flat case); in this case, we can   choose
      $w,w'$ to be non-singular on the disc  but complex on the boundary, resulting in a modification of the boundary conditions for the ghosts, or we can choose  $w,w'$ to be   real on the boundary but
singular on the disc. With the latter choice, however, the gauging
of the weightless one-forms may be problematic. The $N=4$ and
$N=8$ theories then arise from the real theory with Lorentzian
world-sheet and real $Z,\ti Z$, while the amplitudes are
calculated using the Euclidean version of this.

The Berkovits twistor string gives a theory of $N=4$
superconformal gravity coupled to $N=4$ super-Yang-Mills for any
gauge group  that can arise as a current algebra of a $c=28$
conformal field theory. However, it is known that $N=4$
superconformal gravity coupled to $N=4$ super-Yang-Mills has an
$SU(4)$ (or $SL(4,\R)$ in split signature) R-symmetry anomaly that
cancels only if $G$ is 4-dimensional~\cite{FradTsey,RomPvN}, so
$G=SU(2) \times U(1)$ or $U(1)^4$. This is so for the theory with
minimal kinetic term $\int W^2$, but a similar result is expected
to apply for the theory with non-minimal kinetic term $\int
e^{-2\F}e^{2W}$ arising from the twistor string~\cite{BWsc}. This
suggests that the Berkovits string may only be consistent at loops
for special gauge groups, and that there are constraints and
potential inconsistencies that have not yet been found.
In~\cite{BWsc}, it was suggested that these may come from open
string tadpole cancellation. At loops, there may be interactions
with a closed string sector, and further issues could arise from
closed strings. (Closed string vertex operators are constructed
from products of left-moving and right-moving vertex operators, so
that one might expect the closed string spectrum to be related to
the tensor product of the open string spectrum with itself. The
twistor space spectrum appears to be the tensor product of that
for open strings, but it is not clear what this means for the
space-time spectrum, as the conventional Penrose transform does
not apply to non-holomorphic fields $\F (Z,\ti Z)$.)

The new string theories described here have the same form as the
Berkovits string, but with extra terms in the BRST operator. It is
therefore to be expected that for these theories, too, there will
be further constraints that will eliminate some models. We do not
understand these constraints from the string theory perspective,
but some clues might be obtained from the corresponding space-time
theories. The new theories have different symmetries from those of
conformal supergravity (for example, they do not have a gauged
R-symmetry or a conformal symmetry) and so they will have
different anomalies, and different constraints from anomaly
cancellation. Interestingly, there are supersymmetric theories
which can be defined in 2+2 dimensions that have no analogue in
3+1 dimensional space-time, and the spectra of some of these arise
here.

First, the theory of section~\ref{sdgrav0} has the spectrum of
self-dual gravity coupled to self-dual Yang-Mills and a scalar (or
2-form gauge field). Consistent non-linear interactions are
possible classically for this theory, with field equations given
by some scalar-dependent modification of (\ref{sdfhjzfs}). There
is no covariant action for such field equations, but there are
non-covariant actions of the type proposed by Plebanski~\cite{JP}.
The theory is  a chiral one in $2+2$ dimensional space-time, and
so it is prone to potential anomalies. An interacting theory of
self-dual gravity coupled to self-dual Yang-Mills in $2+2$
dimensions arises from the ${\cal N}=2$ string~\cite{OV}, and this is
believed to be a consistent quantum theory (however,
see~\cite{MM,CS2}). This suggests the intriguing possibility that
the $N=0$ twistor string found here could be dual to an ${\cal N}=2$
string theory. A string theory with the spectrum of self-dual
gravity coupled to self-dual Yang-Mills and a   2-form gauge field
is given by the   ${\cal N}=2$ string whose target space is generalised
K\" ahler~\cite{Hull:1996zt}; this is obtained by coupling the
$(2,2)$ supersymmetric sigma-model with torsion
\cite{Gates:1984nk} to ${\cal N}=2$ world-sheet supergravity. The
theories of section~\ref{sdgravN} with $N<4$ give supersymmetric
extensions of this bosonic theory with self-dual supergravity
coupled to self-dual super-Yang-Mills and $N$ supersymmetries, and
these could be consistent non-trivial theories if the $N=0$ theory
is.

For $N=4$, we have two twistor theories, both of which have the
spectrum of $N=4$ supergravity coupled to $N=4$ super-Yang-Mills.
One is the    theory of~\S\ref{sdgravN} with $N=4$   (for any
gauge group   that can arise as a current algebra of a $c=22$
conformal field theory) and the other is the theory
of~\S\ref{N4grav}. However, there are a number of different
supersymmetric theories with this spectrum, and the question we
now turn to is which of these arises in the twistor string.
Consider first the Yang-Mills sector, for which there is the free
theory and  two possible interacting supersymmetric theories. For
$N=4$ Yang-Mills, there is the standard non-chiral theory, which
can be rewritten in the Chalmers-Siegel form~\cite{CS} with
Yang-Mills kinetic
      term $\int EF +E^2$ where
$E$ is a self-dual 2-form and $F=dA+A^2$ is the usual Yang-Mills
field strength. There is also Siegel's chiral theory with
Yang-Mills kinetic term $\int EF$~\cite{Siegel1}.  This is
sometimes called a self-dual theory, but it has the same spectrum
as the usual super-Yang-Mills theory. It differs from the usual
theory in that the interactions are chiral, i.e. they are not
symmetric under the parity transformation interchanging positive
and negative helicities, and the action is  linear in the
negative helicity fields (such as $E$). The full non-chiral $N=4$
super-Yang-Mills theory is obtained in the Berkovits string, and
the same is true for our $N=4$ theory as it  is the same as that
of Berkovits in the Yang-Mills sector.

The supergravity sector has the  spectrum of $N=4$ Einstein
supergravity, and we have seen that it has at least one
non-trivial interaction. Just as for Yang-Mills, there is the
possibility of either the standard non-chiral theory or of one
with chiral interactions. A formulation of Einstein gravity with
chiral interactions was discussed in~\cite{Siegel2,AH1}. The
fields consist of a vierbein $e_{\mu}{}^a$ (the analogue of the
Yang-Mills connection $A$) and an independent Lagrange multiplier
field $\omega _\mu ^{ab}$ which is anti-self-dual in the Lorentz
indices $ab$ (the analogue of the anti-self-dual Lagrange
multiplier field $E$). The multiplier  $\omega _\mu ^{ab}$ imposes
the constraint that the anti-self-dual part of the Levi-Civita
spin-connection $\Omega (e)$ constructed from $e$ vanishes, so
that the corresponding curvature is self-dual. An $N=4$
supersymmetric version of this theory was given by
Siegel~\cite{Siegel2}, with component action given  by truncating
the
     $N=8$ component action of ref.~\cite{Siegel2}.

To determine whether
     the free,  chiral or
the non-chiral interacting  $N=4$ supergravity arises from the two
$N=4$ string theories
     requires further
analysis of the scattering amplitudes, and we will return to this
elsewhere.  However, the theory of~\S\ref{N4grav} has the usual
non-chiral Yang-Mills interactions
     and has a non-trivial cubic gravitational coupling, so it is presumably
     the full Yang-Mills theory coupled to either chiral or non-chiral $N=4$ supergravity.
     The usual non-chiral interacting $N=4$ supergravity coupled to Yang-Mills theory has no
anomalies, but it is expected to have ultra-violet divergences.
Nonetheless, it has a limit in which gravity decouples to leave
$N=4$ super-Yang-Mills, and this is believed to be a consistent
      ultra-violet finite field theory.
      The theory of chiral $N=4$ supergravity coupled to
      $N=4$ super-Yang-Mills is likely to have better ultra-violet behaviour than the full supergravity (and might conceivably be finite) and
     it has a similar decoupling limit so that, whichever supergravity
      theory arises, there should be a decoupling limit giving pure $N=4$ super-Yang-Mills amplitudes.
This  limit in the twistor theory is given by scaling the infinity
twistor so that $I^{IJ}\to 0$. Then from (\ref{trans1}), for any
supergravity wave-function $h$, the corresponding $f^\a$ will
vanish and so any amplitude involving $h$ will vanish. It will be
interesting
      to check that this   leads  to a full decoupling of gravity at all
      orders in perturbation theory.
There is then the intriguing possibility that this twistor string
can give $N=4$
      super-Yang-Mills in this limit.

For the $N=4$ supergravity and Yang-Mills theories, a relation
with ${\cal N}=2$ strings has also been suggested
in~\cite{Siegel1,Siegel3}, and again there is the possibility of a
link between our twistor strings and an ${\cal N}=2$ string theory. A
relation between Siegel's $N=4$ supersymmetric ${\cal N}=2$ string and a
different twistor string theory was suggested
in~\cite{NeitzkeVafa}.

Next, consider the theory of section~\ref{N8grav}, giving the
spectrum of $N=4$ supergravity plus four $N=4$ gravitino
multiplets, together with super-Yang-Mills (for any gauge group
      that can arise as a current algebra of a $c=26$ conformal
field theory). There are then $8$ gravitini of helicity $+3/2$ and
$8$ gravitini of helicity $-3/2$, so that the theory should be an
$N=8$ supergravity theory. Again, there is the possibility of
either a theory with chiral interactions, or a non-chiral one.
(There is also the possibility of a free theory.) If it is a
standard non-chiral $N=8$ supergravity, the total number of vector
fields should be 28 and this requires the number of Yang-Mills
multiplets to be six. This suggests that, if the twistor string
gives a consistent non-chiral theory, there must be a constraint
fixing the number of vector multiplets to be 6. The Berkovits
string is expected to have a constraint fixing the number of
vector multiplets to be $4$, to cancel the  anomalies of conformal
supergravity,
    and
both constraints could arise in the same, as yet unknown, way.
Alternatively, the theory arising could be Siegel's  chiral $N=8$
supergravity~\cite{Siegel2}, in which the negative helicity fields
appear linearly. In~\cite{Siegel2}, Siegel argued that the ${\cal N}=2$
string gives $N=4$ chiral Yang-Mills from the open string sector
and $N=8$ chiral supergravity from the closed string sector, and
that the chirality of the interactions implied that the
supergravity and super-Yang-Mills fields do not couple, so that
one can consistently have  $N=8$ chiral supergravity and an
arbitrary number of $N=4$ chiral Yang-Mills multiplets. It will be
interesting to see whether either of these interacting $N=8$
supergravity theories arise here. If the space-time theories
arising from the perturbative string theory are chiral
supergravities, then it is possible that non-perturbative effects
could give rise to the non-chiral interactions,   as they do for
Yang-Mills in Witten's topological twistor string
\cite{Witten2003}.

Finally, for the models of section~\ref{sdgravN} with $N>4$, the
spectrum is chiral with states of spin greater than $2$, and with
more than one state of spin 2. It is believed that there are no
chirally-symmetric theories with spins higher than 2 or with more
than one graviton which have non-trivial interactions, but the
no-go theorems do not apply
      to chiral theories.
Consider first the $N>4$ Yang-Mills theories, with
       helicities
$1,1/2,...,-N/2$ in the $SL(N,\R)$ representations $({\bf 1},{\bf
N}, {\bf N(N-1)/2},...,{\bf N'},{\bf 1})$, and all in the adjoint
of the Yang-Mills gauge group, so that for $N>4$ there are
negative helicity states of spin greater than one. The field
equation for a  free massless field $\Phi_{A'_1A'_2...A'_n}$ of
helicity $-n/2$ is
\begin{equation}
\label{feqn} \nabla ^{BA'_1} \Phi_{A'_1A'_2...A'_n} = 0 .
\end{equation}
For a field in a representation of the gauge group, the
corresponding field equation is (\ref{feqn}) where $\nabla $ is
the Yang-Mills covariant derivative. For $n\ge 2$ this is
consistent only if the Yang-Mills connection is self-dual,
\begin{equation}
\label{asds} F_{A'B'}=0 .
\end{equation}
The chiral $N=4$ theory is of this type, with self-dual Yang-Mills
coupled to a field $E_{A'B'}$ with field equation of the form
(\ref{feqn}). There are then  consistent chiral  interactions for
the $N>4$ Yang-Mills  multiplets
      of this type provided the  Yang-Mills equation is the self-duality condition (\ref{asds}).
It remains to investigate whether such interactions can be
supersymmetric, and we will return to this elsewhere. For $N>4$,
the chirality of the spectrum will mean that it is unlikely that
there will be a covariant action.

Similar considerations apply to the $N>4$ supergravities arising
from the twistor strings, in which there are negative helicity
states of spin greater than two. The field equation for a  free
massless field   of helicity $-n/2$ is again (\ref{feqn}), but
with $\nabla $ denoting  the gravitational covariant derivative.
In curved space, this has an integrability condition for $n>2$
(the Buchdahl constraint) given by
\begin{equation}
\label{} \ti \psi _{A'B'C'D'}=0
\end{equation}
where $\ti \psi _{A'B'C'D'}$ is the anti-self-dual part of the
Weyl-curvature. For Lorentzian signature, this would imply that
space-time is conformally flat, but for Euclidean or split
signatures, non-trivial conformally self-dual spaces are possible.
A free field of helicity $-n/2$ can then be consistently coupled
to conformally self-dual gravity. Self-dual supergravities for
$N\le 8$ have been given in~\cite{Siegel2}, and it is to be
expected that these can be coupled to the free supermultiplet with
helicities $0,-1/2,...,-N/2$. Such theories could provide
consistent interactions for the space-time theory arising from the
$N\le 8$ twistor strings, with the self-dual supergravity fields
arising from the twistor field $f$ and the negative helicity
multiplet from the twistor field $g$. For $N>8$   supergravity,
just as for  $N>4$ Yang-Mills, there are consistent interactions
that can be written down and it remains to be seen whether these
can be supersymmetric.

Much remains to be done to investigate the interactions of the
theories presented in this paper.
     It would be interesting to find and analyse super-twistor space actions, following~\cite{Mason3, BMS}, and to seek 
     corresponding modifications of Witten's topological
     twistor string that gave similar results.
       It is conceivable that some of the strings found here  give   free theories, and that others may be inconsistent.
       However, it is encouraging
that suitable interacting supersymmetric space-time theories
       exist for many of the cases,  and interesting that the interactions are typically
       chiral for $N\ne 4,8$.
       However, the most promising theories  are the $N=4$ theory giving an interacting theory
       of supergravity coupled to super-Yang-Mills, and the one giving $N=8$ supergravity. The $N=4$ theory has a decoupling limit
giving pure Yang-Mills, opening the prospect of a twistor string
formulation of super-Yang-Mills loop amplitudes.


\section*{Acknowledgements}

We would like to thank Nathan Berkovits, Edward Witten, Parameswaran Nair, Roger
Penrose, Simon Salomon, David Skinner, Kellogg Stelle and Arkady
Tseytlin for helpful discussions; we are especially grateful to Edward Witten for his advice and remarks. MA thanks the Theory Division at
CERN for hospitality and financial support. MA and LJM also thank
the Institute for Mathematical Sciences at Imperial College London
for hospitality and financial support. The work of MA was
also supported in part by a PPARC Postdoctoral Research Fellowship
with grant reference PPA/P/S/2000/00402, by the \lq FWO-Vlaandere'
through projects G.0034.02 and G.0428.06, by the Belgian Federal
Science Policy Office through the Interuniversity Attraction Pole
P5/27, and by the European Union FP6 RTN programme
MRTN-CT-2004-005104. The work of LJM was also partially supported
by the European Union through the FP6 Marie Curie RTN {\em Enigma}
(contract number MRTN-CT-2004-5652) and by a Leverhulme Senior
Research Fellowship.

\appendix
\section{Appendix: relation between split signature constructions}

In this appendix, we continue our discussion in \S\ref{splitsign}
of two distinct twistor constructions for space-times of split
signature. In the first construction, we obtained a deformed
twistor space $\PTc$ with a complex conjugation $\tau:\PTc\to\PTc$
whose fixed point set defined a real slice $\PTc_\R$, whereas in
the second we considered a deformation $\PTc_\R$ of the real slice
$\PT_\R$ inside $\PT$. Although the first construction is perhaps
more intuitive, the second is more powerful and has a better
conceptual fit with the Berkovits open twistor string model, so we
will derive the first construction from the second. We will assume
that we have obtained a twistor space  $\PTc$ by suitably gluing
together the  twistor spaces for small open sets in space-time,
with the assumption that the space-time is  $S^2\times S^2$
globally and admits an analytic conformal structure. This space is
non-Hausdorff, and we give a brief description of it here.

The second construction starts from the data of $\PTc_\R\subset
\PT$ determined by equation (\ref{graph}):
\begin{equation}\label{graphagain}
Z^\a-\bar Z^\a=iF^\a (Z^\b+\bar Z^\b)\, .
\end{equation}
With the assumption of analyticity, $F^\a$ can be analytically
continued to become a holomorphic function $F^\a(Z^\b)$ on a
neighbourhood containing $\T_\R$ (initially, $F^\a (Z^\b)$ was
defined only for real values of $Z^\a$). Thus equation
(\ref{graphagain}) will make sense when $\bar Z^\a$ is replaced by
$\tilde Z^\a$ where $\tilde Z^\a$ is close to, but not necessarily
equal to $\bar Z^\a$.  This gives the equation
\begin{equation}\label{cgraph}
Z^\a-\tilde Z^\a=iF^\a (Z^\b+\tilde Z^\b) ,
\end{equation}
where now $\tilde Z^\a$ is an independent variable that is no
longer the complex conjugate of $Z^\a$.   For $F^\a$ sufficiently
small, this equation can be solved for $\tilde Z^\a$ in terms of
$Z^\b$ as
\begin{equation}\label{glue}
\tilde Z^\a=P^\a (Z^\b )
\end{equation}
for some invertible functions $P^\a$.  Since (\ref{graph}) was
defined for $Z^\a \in \Tc_\R$ and $\tilde Z^\a\in \overline
{\Tc_\R}$, the analytic continuation (\ref{cgraph}) will be
defined for $Z^\a$ in some neighbourhood $V$ of
$\Tc_\R\subset\PT_+$ and, from the reality properties of
(\ref{graph}), the $P^\a$  will map $V$ holomorphically onto the
complex conjugate set $\bar V \subset \PT_-$. It follows from this
definition that the real slice $\PTc_\R$ is given by the subset of
$V$ on which $\tilde Z^\a= \bar Z^\a$, since (\ref{cgraph}) then
reduces to (\ref{graph}).

We will construct $\PTc$ by gluing together two copies of $\CP^3$
using $P^\a(Z^\b)$.  We now take $Z^\a$ to be holomorphic
coordinates on one copy of $\CP^3$, denoted $\PT_+$, and $\tilde
Z^\a$ to be coordinates on another copy denoted $\PT_-$.  We
construct $\PTc$ by interpreting equation (\ref{glue}) as a
patching relation for constructing a complex manifold by gluing
the neighbourhood $V\subset \PT_+$ to $\bar V\subset \PT_-$.  We
note, however, that this global description is not Hausdorff.
Furthermore, the full space $\PTc$ admits a complex conjugation
$\tau$ which interchanges $\PT_+$ and $\PT_-$ so that $\tau$ maps
the point $Z^\a\in\PT_+$ to the point $\tilde Z^\a=\bar Z^\a\in
\PT_-$ and vice-versa.  In order to see that this is well defined,
we need to check that it is compatible with the patching
(\ref{glue}); if $Z^\a \in V$ then $\tau(Z^\a)$ is the point in
$\PT_-$ with $\tilde Z^\a=\bar Z^\a$, but $Z^\a$ is identified
with $\tilde Z^\a=P^\a(Z^\b)$ in $\PT_-$ whose conjugate point is
$Z^\a=\overline{ P^\a(Z^\b)}$ in $\PT_+$.  For $\tau$ to give the
same point in each case, we need to see that $\bar Z^\a=
P^\a(\overline{P^\b(Z^\c)})$.  This follows from the fact that
(\ref{glue}) is equivalent to (\ref{cgraph}) and $F^\a$ is a real
function for real values of its argument, so that its analytic
continuation satisfies $\overline{F^\b(Z^\a+\tilde
Z^\a)}=F^\b(\bar Z^\a+\bar{\tilde Z}^\a)$.  Thus (\ref{cgraph})
implies
\begin{equation*}
\bar{\tilde Z}^\a-\bar Z^\a=iF^\a(\bar Z^\b+\bar{\tilde Z}^\b)
\end{equation*}
and this equation is the same as (\ref{cgraph}) except that the
role of $Z^\a$ has been taken by $\bar{\tilde Z}^\a$ and that of
$\tilde Z^\a$ by $\bar Z^\a$. Thus we have  $\bar
Z^\a=P^\a(\bar{\tilde
      Z}^\b)=P^\a(\overline{P^\b(Z^\c)})$ as desired.

Given a holomorphic disc $D_x$ in $\PT_+$ with boundary on
$\PTc_\R$, we can define the Riemann sphere $\CP^1_x=D_x\cup
\tau(D_x)$ in $\PTc$ since $\tau$ fixes $\PTc_\R$ and hence glues
the boundary of $D_x$ to that of $\tau(D_x)$.  It is a standard
theorem in complex analysis that this embedding will actually be
holomorphic along $\del D_x$ as well as over the interiors of
$D_x$ and $\tau(D_x)$.

We can carry out the non-linear graviton construction on $\PTc$
and construct the space $\CMc $ of Riemann spheres in $\PTc$ in
the same family as $\CP^1_x$.  This will be four complex
dimensional as before, and admit a holomorphic conformal structure
that is anti-self-dual.  The anti-holomorphic involution $\tau$ on
$\PTc$ takes Riemann spheres to Riemann spheres, and so it induces
a complex conjugation on $\CMc$ that preserves the conformal
structure; thus it fixes a real slice ${\mathcal
      M}\subset \CMc$ on which the conformal structure is real.  The
points of the real slice correspond to Riemann spheres in $\PTc$
that are mapped to themselves by the anti-holomorphic involution.
Such Riemann spheres contain an equatorial circle that is fixed by
the involution, and  which must lie in the fixed points $\PTc_\R$
in $\PTc$.  Thus such a Riemann sphere corresponds to a pair of
holomorphic discs in $\PTc$ with common boundary on $\PTc_\R$ and
conversely a disc $D$ gives rise to the Riemann sphere $D\cup
\tau(D)$ as described above.

\end{document}